\documentclass{article}
\usepackage{epstopdf}
\usepackage{tikz}
\usepackage[mathscr]{eucal}
\usepackage{amsthm}
\usepackage{amsmath}
\usepackage{amssymb}
\usepackage{mathrsfs}
\usepackage{subcaption}
\usepackage{marginnote}
\usepackage{float}
\usepackage{graphicx}
\usepackage[utf8]{inputenc}
\usepackage[export]{adjustbox}
\usepackage{newunicodechar}
\usepackage[stable]{footmisc} 
\newunicodechar{≤}{$\le$}
\newunicodechar{ê}{\^{e}}

\usepackage{color}
\newcommand{\aSt}{\mathcal{W}}

\newcommand{\IrrJTL}[2]{\mathcal{X}_{#1,#2}}
\newcommand{\AStTL}[2]{\aSt_{#1,#2}}

\newcommand{\bAStTL}[2]{\overline{\mathcal{W}}_{#1,#2}}

\newcommand{\q}{\mathfrak{q}}

\newcommand{\aX}{\mathcal{X}}
\newcommand{\G}{\mathrm{G}}

\usepackage{subeqnarray,amsmath,amsfonts,amssymb}
\usepackage{bm}    
\usepackage{cite}  
\usepackage{epic,eepic}
\usepackage{epsf}
\usepackage{url}

\usepackage{float}

\usepackage{tikz}
\usepackage{psfrag}
\usetikzlibrary{snakes}
\usetikzlibrary{patterns}
\usetikzlibrary{calc}

\usepackage{array}
\newlength\squareheight
\newenvironment{absolutelynopagebreak}
{\par\nobreak\vfil\penalty0\vfilneg
	\vtop\bgroup}
{\par\xdef\tpd{\the\prevdepth}\egroup
	\prevdepth=\tpd}

\newcommand{\appropto}{\mathrel{\vcenter{
			\offinterlineskip\halign{\hfil$##$\cr
				\propto\cr\noalign{\kern2pt}\sim\cr\noalign{\kern-2pt}}}}}

\newfloat{rules}{thp}{foobar}
\floatname{rules}{List}
	
	\newcommand{\be}{\begin{equation}}
	\newcommand{\ee}{\end{equation}}

	\def\reff#1{(\protect\ref{#1})}
	\def\spose#1{\hbox to 0pt{#1\hss}}
	\def\ltapprox{\mathrel{\spose{\lower 3pt\hbox{$\mathchar"218$}}
			\raise 2.0pt\hbox{$\mathchar"13C$}}}
	\def\gtapprox{\mathrel{\spose{\lower 3pt\hbox{$\mathchar"218$}}
			\raise 2.0pt\hbox{$\mathchar"13E$}}}
	
	\def\proof{\par\medskip\noindent{\sc Proof.\ }}
	\def\qed{ $\square$ \bigskip}

	\def\scra{\mathcal{A}}
	
	\def\scrc{\mathcal{C}}
	
	\def\scre{\mathcal{E}}
	\def\scrf{\mathcal{F}}

	\def\scrv{\mathcal{V}}

	\newcommand{\bv}{ {\bf v} }

	\def\R{{\mathbb R}}

	\newtheorem{theorem}{Theorem}[section]

	\newtheorem{lemma}[theorem]{Lemma}

	\newenvironment{scarray}{
		\textfont0=\scriptfont0
		\scriptfont0=\scriptscriptfont0
		\textfont1=\scriptfont1
		\scriptfont1=\scriptscriptfont1
		\textfont2=\scriptfont2
		\scriptfont2=\scriptscriptfont2
		\textfont3=\scriptfont3
		\scriptfont3=\scriptscriptfont3
		\renewcommand{\arraystretch}{0.7}
		\begin{array}{c}}{\end{array}}

\begin{document}

\title{Four-point geometrical correlation functions in the two-dimensional $Q$-state Potts model: connections with the RSOS models}

\author{Yifei He$^{1}$, Linnea Grans-Samuelsson$^{1}$, Jesper Lykke Jacobsen$^{2,3,1}$, Hubert Saleur$^{1,4}$ \\
[2.0mm]
 ${}^1$Universit\'e Paris-Saclay, CNRS, CEA, Institut de physique th\'eorique,\\ 
91191, Gif-sur-Yvette, France\\
 ${}^2$Laboratoire de Physique de l'Ecole Normale Sup\'erieure, ENS, Universit\'e PSL,\\ CNRS, Sorbonne Universit\'e, Universit\'e de Paris, F-75005 Paris, France \\
 ${}^3$Sorbonne Universit\'e, \'Ecole Normale Sup\'erieure, CNRS, \\
Laboratoire de Physique (LPENS), 75005 Paris, France \\
${}^4$Department of Physics,
  University of Southern California, Los Angeles, CA 90089, USA }


\maketitle

\begin{abstract}
	The ``bootstrap determination'' of the geometrical correlation functions in the two-dimensional Potts model proposed in  a paper by Picco et al. \cite{Picco:2016ilr} was later shown in  \cite{Jacobsen:2018pti} to be incorrect, the actual spectrum of the model being considerably more complex than initially conjectured. We provide in this paper a geometrical interpretation of the four-point functions built in \cite{Picco:2016ilr}, and explain why the results obtained by these authors, albeit incorrect,  appeared so close to those of their numerical simulations of the Potts model. Our strategy is based on a cluster expansion of correlation functions in RSOS minimal models, and a subsequent numerical and algebraic analysis of the corresponding $s$-channel spectrum, in full analogy with our early work on the Potts model  \cite{Jacobsen:2018pti}. Remarkable properties of the lattice amplitudes are uncovered, which explain in particular the truncation of the spectrum of  \cite{Jacobsen:2018pti} to the much simpler one of the RSOS models, and which will be used in a forthcoming paper to finally determine the geometric four-point functions of the Potts model itself.

\end{abstract}

\tableofcontents

\section{Introduction}

The determination of correlation functions in two-dimensional critical geometrical models is a difficult and important problem that has stymied the community since the early days of conformal field theory. 

In the recent couple of years, a new approach to this problem has been proposed based on potential connections  with Liouville theory and a systematic use of the crossing symmetry constraints. Attention focussed on the $Q$-state Potts model \cite{potts_1952},  which, for $Q$ generic, can be formulated geometrically in terms of clusters (via the well-known Fortuin-Kasteleyn expansion \cite{FORTUIN1972536}). The limit $Q\to 1$ is then particularly interesting, since it describes the  percolation problem. 


In the pioneering work  \cite{Picco:2016ilr}, a simple crossing symmetric spectrum was proposed to describe some of the four-point functions of the order operator in the Potts model; in terms of clusters, these correspond to  probabilities  of having four points connected in different ways, as we shall discuss below.  The proposal was checked using Monte-Carlo simulation, and reasonable agreement was found.  A later work \cite{Jacobsen:2018pti} based on a combination of algebraic and numerical techniques however showed that the speculated spectrum cannot be the true spectrum for the Potts model: it misses an infinite number of states which, despite having small amplitudes, are essential to cancel the unwanted singularities in $Q$ appearing in the four-point function of \cite{Picco:2016ilr}. To this day, the full four-point functions  remain therefore unknown. \footnote{See also \cite{Dotsenko:2019dcu} for some recent study of the four-spin correlations using the Coulomb Gas approach.}

Interestingly, it  was found afterwards in \cite{Ribault:2018jdv} that the spectrum of \cite{Picco:2016ilr} could be obtained from a certain limit of minimal models when the central charge is  taken to be an irrational number (see section~\ref{sec2-2} below). The corresponding CFT was further elucidated analytically  in \cite{Migliaccio:2017dch,Ribault:2019qrz}. The main question  however remains: what statistical physics model does the spectrum in \cite{Picco:2016ilr} actually describe, if it is not the Potts model, and why does it give results apparently so close numerically \cite{Picco:2019dkm} to those of the Potts model?  The goal of the present  paper is to answer this question. Remarkably, we shall also obtain results of considerable importance for the solution of the Potts problem itself, as will be described in a subsequent work \cite{bootstrappaper}.

Our approach follows our general philosophy of analysing lattice models in as much detail as possible. In the present case, we will focus on the geometrical interpretation of correlation functions in restricted solid-on-solid (RSOS) models, following and extending the early work of, in particular, V. Pasquier \cite{Pasquier:1986jc,Pasquier:1987xj,Pasquier_1987} and I. Kostov \cite{Kostov:1989eg}. We shall find that the lattice correlation functions  of certain operators in these models have graphical expansions that are very similar to---albeit slightly different from---those occurring in the Potts model. The main difference between the two models is, perhaps not surprisingly, the weight given to  clusters with non-trivial topologies.  The fine structure of these weights allows for  intricate cancellations of the Potts spectrum given in \cite{Jacobsen:2018pti}, leading to the spectrum of (unitary or non-unitary) minimal models in the corresponding limits. By following the logic in  \cite{Ribault:2018jdv}, and taking appropriate limits of the lattice model, we are then able to provide a geometrical interpretation of the correlation functions  proposed in  \cite{Picco:2016ilr}, and explain why---and by how much---they differ from the true Potts model ones.  


The paper is organized as follows. In the next section, we briefly review the geometrical correlations in the Potts model, and provide further motivations to study the relation with minimal models. In section \ref{PottsRSOS}, we describe the general strategy for comparing the Potts and RSOS correlations, and we state in particular the main results about the RSOS lattice model relevant for establishing its connection with the Potts model. These will be used in the following section to study in detail the geometrical formulation of four-point functions in minimal models of type $A$ and $D$. There, we define the relevant geometrical quantities---the ``pseudo-probabilities" in the RSOS minimal models, which are to be compared with the true probabilities in the Potts model. In section \ref{pseudoATL}, we turn to the $s$-channel spectra involved in these two quantities, which we exhibit in terms of the affine Temperley-Lieb algebra as studied in \cite{Jacobsen:2018pti}. The properties of the spectra in the two cases are characterised by several striking facts about the ratios between certain amplitudes entering the $s$-channel expressions of the probabilities. The amplitude ratios are exact expressions (ratios of integer-coefficient polynomials in $Q$), which we obtain here conjecturally based on numerical observations, deferring the task of proving them to a future publication. These facts are then used in section \ref{MM} to recover the minimal models spectra. In section \ref{limit}, we discuss the limit when the central charge goes to an irrational number and compare with the CFT results. The last section contains our conclusions. 

To focus on the comparison with the Potts model, we only state relevant results on the RSOS model in the main text, but also provide a more systematic formulation in the appendices. In particular, in appendix \ref{Z_Proof} we give a proof of the identity of the RSOS and Potts partition functions.%
\footnote{This appendix is adapted from an unpublished work by A.D.\ Sokal and one of the authors \cite{Sokal:2008}.}
In appendix \ref{NptRSOS}, we state the rules for computing the RSOS $N$-point functions. Appendix \ref{3ptC} gives the results on $3$-point couplings in the type $A$ and $D$ RSOS model which are used in the main text for the geometrical formulation of the minimal models four-point functions.
Finally, appendix \ref{app:num} explain the numerical methods (beyond those already described extensively in the appendices of \cite{Jacobsen:2018pti}) used
for extracting the exact amplitude ratios.

\section{Correlation functions in the Potts model}\label{geometricalproblem}

\subsection{Lattice model}\label{Pottslatticemodel}
Let us briefly recall the geometrical problem of interest. The lattice $Q$-state Potts model \cite{potts_1952} is defined on a graph $G=(V,E)$ with vertices $V$ and edges $E$. A spin variable $\sigma_i = 1,2,\ldots,Q$ is attached to each vertex $i \in V$ with the interaction energy $-K \delta_{\sigma_i,\sigma_j}$ associated with each edge $(ij) \in E$. The partition function is given by
\begin{equation}
 Z = \sum_{\{ \sigma \}} \prod_{(ij) \in E} {\rm e}^{K \delta_{\sigma_i,\sigma_j}} ,
\end{equation}
where we have absorbed the temperature into the definition of interaction energy $K$. While this initial formulation requires $Q$ to be a positive integer, $Q \in \mathbb{N}$,
it is easy to rewrite $Z$ more generically in terms of the cluster formulation due to Fortuin and Kasteleyn (FK) \cite{FORTUIN1972536}. Setting  $v = {\rm e}^K - 1$,
one finds
\begin{equation}
 Z = \sum_{A \subseteq E} v^{|A|} Q^{\kappa(A)} \,, \label{Z_FK}
\end{equation}
with the sum going over all $2^{|E|}$ subsets of $E$, and $|A|$ denoting the number of edges in the subset. The partition function is now defined for real values of $Q$ where $\kappa(A)$ indicates the number of connected components---the so-called FK clusters---in the subgraph $G_A = (V,A)$. We will take $G$ to be the two-dimensional square lattice and temperature parameter to be its critical value $v_{\rm c} = \sqrt{Q}$ \cite{potts_1952,Baxter_1973} such that in the continuum limit the model is conformally invariant. In this limit, we consider the geometry of the infinite plane, so that boundary effects are immaterial. 

The partition function \eqref{Z_FK} can be equivalently formulated \cite{Baxter_1976} in terms of the loop model on the medial lattice ${\cal M}(G) = (V_{\cal M},E_{\cal M})$. The vertices $V_{\cal M}$ stand at the mid-points of the original edges $E$, and are connected by an edge $E_{\cal M}$ whenever the corresponding edges in $E$ are incident on a common vertex of $V$. In particular, for the two-dimensional square lattice $G$ we consider, ${\cal M}(G)$
is just another square lattice, rotated by $\frac{\pi}{4}$ and scaled down by a factor of $\sqrt{2}$.
There is a bijection between $A\subseteq E$ in the partition function \eqref{Z_FK} and completely-packed loops on ${\cal M}(G)$. The loops are defined such that they turn around the FK clusters and their internal cycles, and in this way, they separate the FK clusters and their dual clusters. (See figures \ref{plat} and \ref{platdual} below for an example.) The partition function is then written as
\cite{Baxter_1976}
\begin{equation}
 Z = Q^{|V|/2} \sum_{A \subseteq E} \left( \frac{v}{n} \right)^{|A|} n^{\ell(A)} \,, \label{Z_loop}
\end{equation}
where $\ell(A)$ denotes the number of loops. The loop weight is given by
\begin{equation}
 n = \sqrt{Q} = \mathfrak{q} + \mathfrak{q}^{-1} \,, \label{loop-fugacity}
\end{equation}
where $\mathfrak{q}$ is a quantum group related parameter. Notice that on a square lattice, we have simply $\frac{v_{\rm c}}{n} = 1$, i.e., at the critical point, \eqref{Z_loop} depends only on $\ell(A)$.

\subsection{Correlation functions on the lattice}

On the lattice, it is natural to consider the correlation functions of the order parameter (spin) operator
\begin{equation}
 {\cal O}_a(\sigma_i) \equiv Q \delta_{\sigma_i,a} - 1. \, \label{spin-op}
\end{equation}
One can however define more general correlation functions of a geometrical type by switching to the cluster or loop formulations. We are mainly interested in the geometrical correlation functions defined in terms of the FK clusters as following. Consider a number of distinct marked vertices $i_1, i_2,\ldots, i_N \in V$, and let ${\cal P}$ be a partition of a set of $N$ elements. One can then define the probabilities
\begin{equation}
 P_{\cal P} = \frac{1}{Z} \sum_{A \subseteq E} v^{|A|} Q^{\kappa(A)} {\cal I}_{\cal P}(i_1,i_2,\ldots,i_N | A) \,, \label{P_corr_def}
\end{equation}
where $Z$ is given by (\ref{Z_FK}), and ${\cal I}_{\cal P}(i_1,i_2,\ldots,i_N | A)$ is the indicator function that, $\forall k,l \in \{1,\ldots,N\}$ belong to the same block of the partition ${\cal P}$ if and only if
vertices $i_k$ and $i_l$ belong to
the same connected component in $A$. We will denote
${\cal P}$ by an ordered list of $N$ symbols ($a,b,c,\ldots$) where identical symbols refer to the same block. Taking $N=2$ for instance,
$P_{aa}$ is the probability that vertices $i_1,i_2$ belong to the same FK cluster, whereas $P_{ab} = 1 - P_{aa}$ is the probability that $i_1,i_2$
belong to two distinct FK clusters. 

The probabilities $P_{\cal P}$ can be related to the correlation functions of the spin operator 
\begin{equation}
G_{a_1,a_2,\ldots,a_N} = \left \langle {\cal O}_{a_1}(\sigma_{i_1}) {\cal O}_{a_2}(\sigma_{i_2}) \cdots {\cal O}_{a_N}(\sigma_{i_N}) \right \rangle, \,
\label{order-param-corr}
\end{equation}
where the expectation value is defined with respect to the normalization $Z$. Here $a_1,a_2,\ldots,a_N$ is a list of (identical or different) symbols defining a partition ${\cal P}$. To evaluate the expectation value
of a product of Kronecker deltas, one initially supposes that $Q$ is integer, and uses that spins on the same FK cluster are equal, while spins on different
clusters are statistically independent. This leads to $Q$-dependent relations, which can be analytically continued to real values of $Q$.
In the case of $N=2$, one finds that
\begin{equation}
 G_{a_1,a_2} =  \left(Q \delta_{a_1,a_2} -1\right) P_{aa}, \label{G_2-point}
\end{equation}
i.e., the two-point function of the spin operator is proportional to the probability that the two points belong to the same FK cluster.
Therefore ${\cal O}_a(\sigma_i)$ effectively ``inserts'' an FK cluster at $i \in V$ and ensures its propagation until it is ``taken out'' by another spin operator.

In the context of four-point functions, there are 15 probabilities
$P_{aaaa}, P_{aabb},\ldots,P_{abcd}$ whose combinatorial properties were discussed in \cite{Delfino:2011sc}. We will focus on the same subset of four-point functions as studied in \cite{Picco:2016ilr} which are the probabilities of the four points belonging to one or two clusters, namely: $P_{aaaa}$, $P_{aabb}$, $P_{abba}$ and $P_{abab}$.
The relation with the corresponding $G_{\cal P}$ reads \cite{Delfino:2011sc}
\begin{subequations} \label{sumrules}
	\begin{eqnarray}
	G_{aaaa} &=&(Q-1)(Q^2-3Q+3) P_{aaaa}+(Q-1)^2(P_{aabb}+P_{abba}+P_{abab}) \,, \label{Gaaaa}\\
	G_{aabb} &=&(2Q-3) P_{aaaa}+(Q-1)^2P_{aabb}+P_{abba}+P_{abab} \,, \\
	G_{abba} &=&(2Q-3)P_{aaaa}+P_{aabb}+(Q-1)^2P_{abba}+P_{abab} \,, \\
	G_{abab} &=&(2Q-3)P_{aaaa}+P_{aabb}+P_{abba}+(Q-1)^2P_{abab} \,.
	\end{eqnarray}
\end{subequations}
As stated before, for arbitrary real values of $Q$, the left-hand sides of these equations are only formally defined: it is in fact the right-hand sides that give them a meaning.
Notice that the linear system has determinant $Q^4 (Q-1)(Q-2)^3 (Q-3)$ and therefore cannot be fully inverted for $Q=0,1,2,3$. 

\subsection{Continuum limit}

In the continuum limit, and at the critical point, the Potts model is conformally invariant for $0 \le Q \le 4$. One then expects that the correlation functions \eqref{sumrules} are given by the spin correlation functions in the corresponding CFT. Parametrizing
\begin{equation}
 \sqrt{Q}=2\cos \left( {\pi\over x+1} \right) \,, \mbox{ with } x \in [1,\infty] \,,
\end{equation}
we can then write the central charge as
\begin{equation} \label{cx}
c=1-{6\over x(x+1)} \,.
\end{equation}
Note that the quantum-group related parameter $\mathfrak{q} = {\rm e}^{\frac{i \pi}{x+1}}$ is not a root of unit in this generic case, i.e., we do {\em not} restrict $x$ to be integer, as would be the case for the minimal models. 
We also use the Kac table parametrization of conformal weights\footnote{To compare with \cite{Picco:2016ilr} one must identify $\beta^2={x\over x+1}$ (so that ${1\over 2}\leq \beta^2\leq 1$). Moreover, the conventions used in their paper for the exponents are switched with respect to ours: they call $\Delta_{sr}$ what we call $h_{rs}$.\label{paraid}}
\begin{equation}
h_{r,s}={[(x+1)r-xs]^2-1\over 4x(x+1)} \,.
\end{equation}
Usually,  the labels $(r,s)$ are positive integers, but---like for the parameter $x$---we shall here allow them to take more general values. Of course, when $(r,s)$ are not integer, the corresponding conformal weight is not degenerate. It is well known in particular that the order parameter operator has conformal weight $h_{1/2,0}$ \cite{denNijs:1983zz,Nienhuis:1984wm}. Part of the challenge  since the early days of CFT has been to understand what such weight exactly means---in particular, what are the OPEs of the field with itself, and how they control the four-point functions. 

\subsection{A potential relationship with  minimal models}
\label{sec2-2}

It so happens that when
%
\begin{equation} \label{p_gcd_q}
  x={q\over p-q}, \mbox{ with } p>q \mbox{ and } p\wedge q=1 \,,
\end{equation}
for $p$ even and $q$ odd, the conformal weight $h_{1/2,0}$ belongs to the Kac table
\begin{equation} \label{Kactable}
 h_{m,n} = \frac{(p m - q n)^2 - (p - q)^2}{4 p q}
\end{equation}
of the minimal models ${\cal M}(p,q)$ with central charge
\begin{equation}\label{MMc}
{\cal M}(p,q):~c=1-6{(p-q)^2\over pq},
\end{equation}
where the cases $p-q=1$ correspond to unitary minimal models, and $p-q > 1$ are non-unitary.
Using the parametrization
\begin{equation} \label{param_n_m}
 p=2n, \quad  q=2m+1 \,,
\end{equation}
with non-negative integers $n > m$ (and $n-m=1$ corresponding to unitary cases), it is easy to see from \eqref{Kactable} that indeed
$h_{1/2,0}=h_{m,n}$ (since $p/2=n$, while $pm-qn=-n$). The question then arises, as to whether (some of) the geometrical correlations of interest for the corresponding value of $Q$ with
\begin{equation}
 \sqrt{Q}=2\cos  \frac{\pi(p-q)}{p} 
\end{equation}
could conceivably be obtained from the four-point functions of the field with $h=h_{m,n}$, for positive integer Kac labels $m,n$, in a
minimal model%
\footnote{{\em A} rather than {\em the} minimal model, as there might be several modular invariants.}
CFT with the same central charge. 
\bigskip
%


In \cite{Picco:2016ilr}, the authors first conjectured CFT four-point functions describing the Potts probabilities:
\begin{eqnarray}
&&\hbox{\underline{Conjecture in \cite{Picco:2016ilr}:}}\nonumber\\
&&\langle V^DV^NV^DV^N\rangle\propto P_{aaaa}+\mu P_{abab}\label{PRSfunc}
\end{eqnarray}
where $\mu$ is a constant, and similarly for $P_{abba}$ and $P_{aabb}$ with the left hand side replaced by $\langle V^DV^NV^NV^D\rangle$ and $\langle V^DV^DV^NV^N\rangle$. The $V^D$ and $V^N$ here have conformal dimension $h_{1/2,0}=\bar{h}_{1/2,0}$ and were later found in \cite{Migliaccio:2017dch} to originate from the diagonal and non-diagonal sectors respectively of the type $D$ minimal models. While the central charge in the minimal models is rational, the following limit of the minimal models spectrum was taken \cite{Migliaccio:2017dch} to provide an extension to the irrational cases:\footnote{With the identification of the parameters as explained in footnote \ref{paraid}, the $p,q$ in \cite{Ribault:2018jdv} is also switched with respect to ours and the limit \eqref{PRS4pt} correspond to the limit $\frac{p}{q}\to\beta^2$ in \cite{Ribault:2018jdv}.}
\begin{equation}\label{PRS4pt}
p,q\to\infty,\;\;\frac{q}{p-q}\to x \,,
\end{equation}
where $x$ is a finite number. In such  a limit, it was argued in \cite{Ribault:2018jdv,Ribault:2019qrz} that the levels of the null vectors, which are removed in irreducible modules of minimal models, go to infinity, and one obtains Verma modules with the same conformal dimensions: the non-diagonal sector contains fields with conformal dimensions $(h_{r,s},h_{r,-s})$ where $r\in \mathbb{Z}+\frac{1}{2},s\in 2\mathbb{Z}$, and the spectrum in the diagonal sector becomes continuous. The limit spectrum was then used in a  conformal block expansion for the numerical bootstrap of the four-point function \eqref{PRSfunc}, and the results obtained were found to be in reasonable agreement  with Monte-Carlo simulations \cite{Picco:2019dkm}. The corresponding structure constants were later obtained and shown to match  \cite{Migliaccio:2017dch} with a non-diagonal generalization \cite{Estienne:2015sua} of the Liouville DOZZ formula \cite{Zamolodchikov:1995aa}.

This  elegant and tempting procedure does not, however, give the true Potts probabilities. In particular, the latter are expected to be smooth functions in $Q$ (as already argued in \cite{Jacobsen:2018pti}), while there are poles in the four-point functions \eqref{PRSfunc} at rational values of $x$ when \cite{Ribault:2018jdv}:
\begin{equation}\label{polesra0}
p\equiv0\;\text{mod}\;4 \,,
\end{equation}
corresponding to the values of $Q$:
\begin{equation}\label{polesra}
Q=4\cos^2 \left(\frac{\pi}{4}\right),\;4\cos^2 \left(\frac{\pi}{8}\right),\;4\cos^2 \left(\frac{3\pi}{8}\right),\; \ldots \,.
\end{equation}
The authors of \cite{Picco:2016ilr} then further conjectured the following relation in \cite{Picco:2019dkm} (hence, proposing an formula for their $\mu$ parameter, which was initially adjusted numerically):
\begin{eqnarray}
&&\hbox{\underline{Conjecture in \cite{Picco:2019dkm}:}}\nonumber\\
&&\langle V^DV^NV^DV^N\rangle\appropto P_{aaaa}+\frac{2}{Q-2} P_{abab}.\label{PRSfuncmod}
\end{eqnarray}
This expression accommodates the first pole of \eqref{polesra} at $Q=2$ in the four-point function, and was observed using Monte-Carlo simulations \cite{Picco:2019dkm} to be approximately correct. It also becomes exact for $Q=0,3,4$. {\it A priori}, there is no reason why such a combination of the geometric quantities should enter the four-point function in the CFT. In addition, it is unclear how the other poles in $Q$ given by \eqref{polesra}---which were truncated out in the conformal block expansion in \cite{Picco:2019dkm}---could be accounted for in the four-point function \eqref{PRSfuncmod} in terms of the geometric quantities. 

Despite these issues, it is fascinating to see that the four-point functions of minimal models (and their irrational limits) do indeed seem to provide some insights on the geometrical problem of the Potts model. The question is why, and whether this is useful.

An important motivation for this paper is to clarify  this matter, and to establish in particular that the geometrical four-point functions \eqref{P_corr_def}
cannot be obtained by analytic continuation of the minimal models results in this way. It will turn out that the difference between the two types of
correlation functions is numerically small, and probably indiscernible by Monte-Carlo methods \cite{Picco:2019dkm}, although they are certainly detectable
by the transfer matrix techniques developed in \cite{Jacobsen:2018pti} and used in the present paper. The quantities defined and
studied in \cite{Picco:2016ilr,Migliaccio:2017dch,Ribault:2018jdv,Picco:2019dkm,Ribault:2019qrz} will prove to be skewed versions of the true correlation
functions \eqref{P_corr_def}, as we shall explain in detail in section~\ref{limit}.

\bigskip
To make progress, we shall  again follow a direct approach, and   study the geometrical correlation functions of minimal models on the lattice. Setting aside the CFT aspects for a moment, let us recall that minimal models can in fact be obtained as a continuum limit of well-defined RSOS lattice models associated with Dynkin diagrams of the ADE type \cite{PhysRevB.30.3908, Pasquier:1986jc,Pasquier:1987xj}. In this formalism, the correlation functions of the order parameters on the lattice become, in the continuum limit, (some of) the correlation functions of minimal models. In particular, certain order parameter(s) in the RSOS lattice model give rise to the field with conformal weight $h_{m,n}$ in the Kac table and thus coincide with the Potts order parameter at the same central charge---recall the relation (\ref{param_n_m}). On the other hand, the RSOS lattice model has a natural formulation in terms of clusters and loops \cite{Pasquier:1986jc,Pasquier:1987xj,Pasquier_1987,Kostov:1989eg}, somewhat similar to the one in the Potts model, and therefore the correlation functions acquire a geometrical interpretation which can be compared with that of the Potts model. In the following sections, we will study the RSOS four-point functions and their geometrical content, with focus on the operator whose conformal weight coincides with the one of the Potts order parameter, in order to understand the relation and differences with geometrically defined correlation functions of the Potts model. We will use the main results of RSOS correlations functions without detailed proofs,  and leave this  to the appendices.

\section{Comparing Potts and RSOS correlations: general strategy}\label{PottsRSOS}

\begin{figure}[t]
	\begin{centering}
	\begin{subfigure}[h]{0.32\textwidth}
			\includegraphics[width=0.9\textwidth]{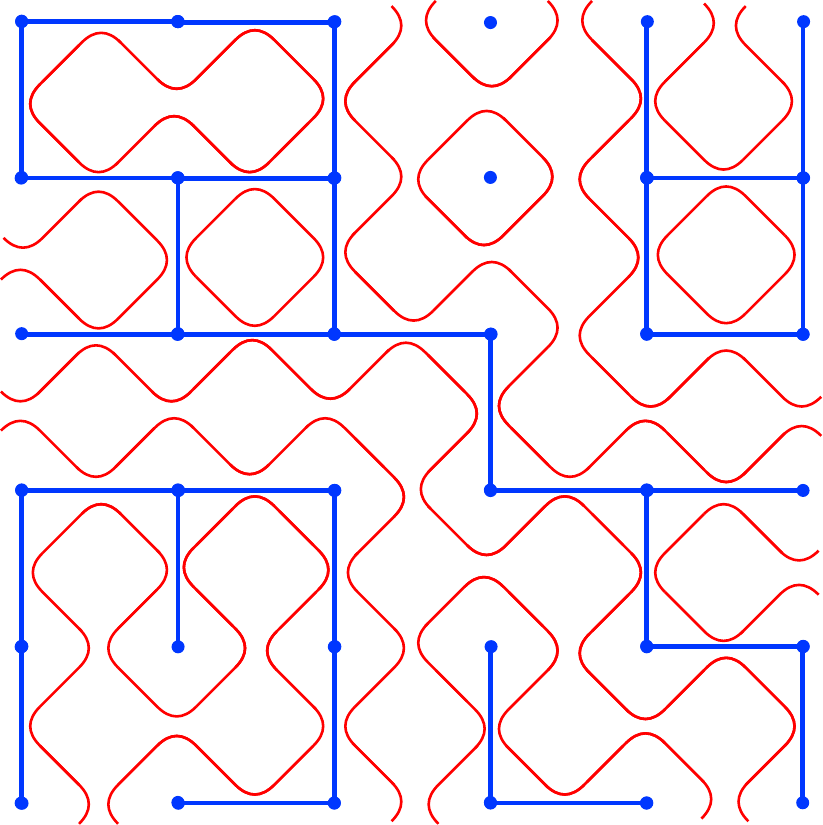}
			\caption{}
			\label{plat}
	\end{subfigure}
	\begin{subfigure}[h]{0.32\textwidth}
			\includegraphics[width=0.9\textwidth]{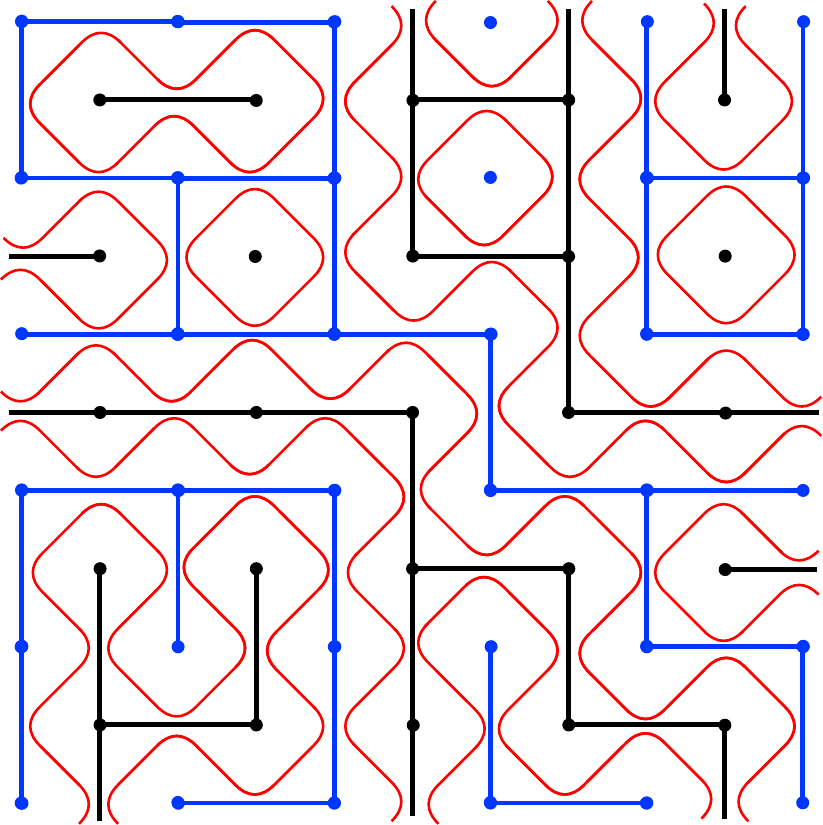}
			\caption{}
			\label{platdual}
	\end{subfigure}
	\begin{subfigure}[h]{0.32\textwidth}
		\includegraphics[width=0.9\textwidth]{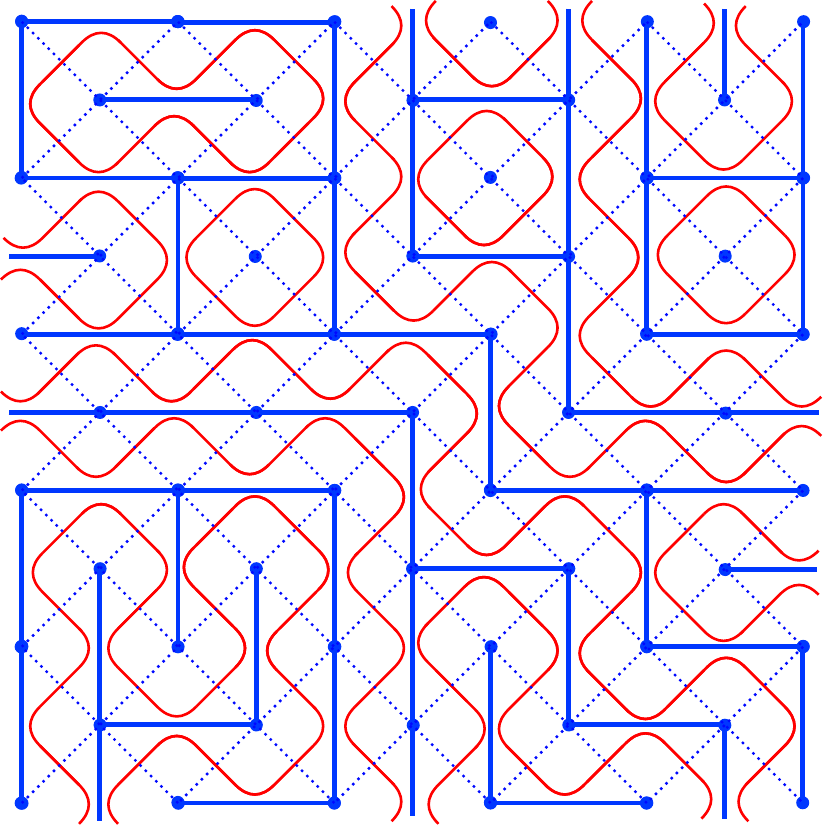}
		\caption{}
		\label{RSOSlat}
    \end{subfigure}
	\caption{In \ref{plat}, we show a cluster configurations on the Potts lattice (blue) and the corresponding loops (red). The clusters are separated by even number of loops. From here, one can draw the clusters on the dual lattice (black) as shown in \ref{platdual}. An RSOS clusters/loops configuration \ref{RSOSlat} is equivalent to this Potts clusters/loops configuration. The links forming the clusters are on the diagonals of the plaquettes, the latter being indicated with dashed lines. The loop configuration is the same as in the Potts case; however, the weights of the loops are different.}
	\end{centering}
\end{figure}

Let us take a more detailed look at the formulation of the Potts model in terms of clusters and loops. 
Consider a Potts cluster configuration given by the subgraph $G_A=(V, A)$ , where the loops are formulated in the usual way as described in section \ref{Pottslatticemodel}. Taking the centers of each plaquette (i.e., lattice face), and defining them as the vertices $V^*$ of another lattice, we obtain the dual Potts model on the graph $G^*=(V^*,E^*)$. The previous loop configuration in fact predetermines the clusters on the dual lattice given by subgraph $G^*_{A^*}=(V^*,A^*)$, where the $A^*$ are all the edges in $E^*$ which do not cross the loops. There is thus a one-to-one map between the Potts cluster configurations $G_A$ and its dual $G^*_{A^*}$. As shown in figures \ref{plat} and \ref{platdual}, we see that when put together, the loops separate the Potts clusters from their dual clusters. A consequence of this mapping, which will turn out particularly important in the following, is that going from one Potts cluster to another
one requires traversing an {\em even} number $k = 2 l$ of loops, with $l \ge 1$ integer. Namely, when the first Potts cluster is separated from the second
one by $l-1$ distinct surrounding clusters, there will also be $l$ surrounding dual clusters, and since clusters and dual clusters alternate each time we traverse
a loop, the number of surrounding loops will be $2l$ indeed. In this paper we are only interested in correlation functions in which all the marked clusters
reside on the direct (not dual) lattice.

The RSOS model, on the other hand, is defined through a map from the lattice to a finite graph $H$ \cite{Andrews:1984af,1985JSP....38..435F}. The nodes on $H$ are taken as the possible values of a ``height" variable $\sigma_i$ associated with site $i$, while neighbouring sites are constrained to have heights which are neighbours on $H$. As a result, the clusters on the RSOS lattice are formed by linking the diagonals of the plaquettes, and each plaquette takes one of the diagonal links:
\begin{equation}
\vcenter{\hbox{
\begin{tikzpicture}
\coordinate (A) at (-0.3, -0.3) {};
\coordinate (B) at (-0.3, 0.3) {};
\coordinate (C) at (0.3, -0.3) {};
\coordinate (D) at (0.3, 0.3) {};
\filldraw (A) circle (0.5 pt);
\filldraw (B) circle (0.5 pt);
\filldraw (C) circle (0.5 pt);
\filldraw (D) circle (0.5 pt);

\node(A2) at ([shift=(-180:8pt)]A) {$\sigma_1$};
\node(B2) at ([shift=(-180:8pt)]B) {$\sigma_4$};
\node(C2) at ([shift=(0:8pt)]C) {$\sigma_2$};
\node(D2) at ([shift=(0:8pt)]D) {$\sigma_3$};

\draw[densely dotted] (A)--(B);
\draw[densely dotted] (B)--(D);
\draw[densely dotted] (C)--(D);
\draw[densely dotted] (C)--(A);
\end{tikzpicture}}}
\vcenter{\hbox{\hspace*{0cm}$=$\hspace*{0.3cm}$\delta_{\sigma_{1}\sigma_{3}}$}}
\vcenter{\hbox{
		\begin{tikzpicture}
		\coordinate (A) at (-0.3, -0.3) {};
		\coordinate (B) at (-0.3, 0.3) {};
		\coordinate (C) at (0.3, -0.3) {};
		\coordinate (D) at (0.3, 0.3) {};
		\filldraw (A) circle (0.5 pt);
		\filldraw (B) circle (0.5 pt);
		\filldraw (C) circle (0.5 pt);
		\filldraw (D) circle (0.5 pt);
		
		\node(A2) at ([shift=(-180:8pt)]A) {$\sigma_1$};
		\node(B2) at ([shift=(-180:8pt)]B) {$\sigma_4$};
		\node(C2) at ([shift=(0:8pt)]C) {$\sigma_2$};
		\node(D2) at ([shift=(0:8pt)]D) {$\sigma_3$};
		
		\draw[densely dotted] (A)--(B);
		\draw[densely dotted] (B)--(D);
		\draw[densely dotted] (C)--(D);
		\draw[densely dotted] (C)--(A);
		\draw[] (A)--(D);
		\end{tikzpicture}}}
\vcenter{\hbox{\hspace*{0.1cm}$+$\hspace*{0.3cm}$\delta_{\sigma_2\sigma_4}$}}
\vcenter{\hbox{
		\begin{tikzpicture}
		\coordinate (A) at (-0.3, -0.3) {};
		\coordinate (B) at (-0.3, 0.3) {};
		\coordinate (C) at (0.3, -0.3) {};
		\coordinate (D) at (0.3, 0.3) {};
		\filldraw (A) circle (0.5 pt);
		\filldraw (B) circle (0.5 pt);
		\filldraw (C) circle (0.5 pt);
		\filldraw (D) circle (0.5 pt);
		
		\node(A2) at ([shift=(-180:8pt)]A) {$\sigma_1$};
		\node(B2) at ([shift=(-180:8pt)]B) {$\sigma_4$};
		\node(C2) at ([shift=(0:8pt)]C) {$\sigma_2$};
		\node(D2) at ([shift=(0:8pt)]D) {$\sigma_3$};
		
		\draw[densely dotted] (A)--(B);
		\draw[densely dotted] (B)--(D);
		\draw[densely dotted] (C)--(D);
		\draw[densely dotted] (C)--(A);
		\draw[] (B)--(C);
		\end{tikzpicture}}}
\end{equation}
where the choice of the local weights multiplying each term will be deferred to the next paragraph.
It is straightforward to see that there is an equivalence between the RSOS clusters/loops configurations and the ones in the Potts model, as shown in figure \ref{platdual} and \ref{RSOSlat}. This is discussed in more details in appendix \ref{Z_Proof}, where we also give a proof of the equivalence between the partition functions of the two models. Notice, however, that two distinct clusters on the RSOS lattice are mapped to Potts clusters only when separated by an even number of loops (otherwise one is mapped to a Potts cluster and the other one to a dual Potts cluster). This will play a role when we consider the geometric four-point functions of the two models.

\bigskip

Taking the graph $H$ to be a Dynkin diagram $\mathcal{D}$ of the ADE type with Coxeter number $p$ and introducing its adjacency matrix $\mathcal{A}$, the eigenvalues $\lambda_{(r)}$ take the form
\begin{equation}\label{lambdadef}
 \lambda_{(r)}=2\cos \frac{r\pi}{p} \,,
\end{equation}
and the normalized eigenvectors are denoted $S^{\sigma_i}_{(r)}$, where $r$ takes values in the set ${\cal D}^*$ of exponents of the algebra. (See figure \ref{ADE} for the diagrams $\mathcal{D}$ and their corresponding exponents ${\cal D}^*$ to be considered in this paper.) They enter the definition of the Boltzmann weight of a certain configuration, as we discuss in details in appendices \ref{Z_Proof} and \ref{NptRSOS}. Choosing a special eigenvector
\begin{equation}\label{speigen}
S_{\sigma_i}\equiv S^{\sigma_i}_{(p-q)}, 
\end{equation}
where $p>q$ and $p\wedge q=1$ as in \eqref{p_gcd_q}, a representation of the Temperley-Lieb (TL) algebra \cite{TL71} is defined by the basic action of the generator $e$ on a face:%
\footnote{Since we shall consider non-unitary cases in which some of the components $S_\sigma$ are non-positive, we stress that one should use the determination of the square root satisfying always $(S_\sigma S_\sigma)^{1/2} = S_\sigma$.}
%
\begin{equation}
e_i := e(\sigma_{i-1} \sigma_i \sigma_{i+1} | \sigma_{i-1} \sigma'_{i} \sigma_{i+1})= \includegraphics[width=0.2\textwidth,valign=c]{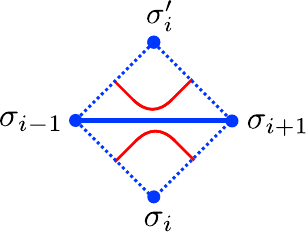}=
		\delta_{\sigma_{i-1},\sigma_{i+1}}{ \left(S_{\sigma_{i}} S_{\sigma'_{i}}\right)^{1/2}\over S_{\sigma_{i-1}}} \,.
\end{equation}
Here the label $i$ refers to the spatial position of the face. These generators satisfy the relations
\begin{subequations}
\begin{eqnarray}
 e_i^2 &=& \lambda_{(p-q)} e_i \,, \\
 e_i e_{i \pm 1} e_i &=& e_i \,, \\
 e_i e_j &=& e_j e_i  \,, \mbox{ when } |i-j| \ge 2
\end{eqnarray}
\end{subequations}
defining the TL algebra \cite{TL71}. The continuum limit of the RSOS model thus defined is known to be given by the ADE minimal models with central charge \eqref{MMc}\cite{RIGGS1989673,Nakanishi:1989cv}.

Like for the Potts model, the torus partition function of the RSOS model can be expanded into configurations of clusters/loops \cite{Pasquier:1986jc,Pasquier:1987xj,Pasquier_1987,Kostov:1989eg}.
For each configuration, the contractible loops get the weight $\lambda_{(p-q)}$, while the situation for the non-contractible loops is more complicated: one must sum over terms for $r\in {\cal D}^*$ \cite{DiFrancesco:1988xz}, in each of which the non-contractible loops get the weight $\lambda_{(r)}$. This is in contrast with the Potts model \cite{DiFrancesco:1987gwq} where one sums over only two terms: one where non-contractible loops get the same weight as contractible ones, and one where they get the weight zero (this last term comes formally with multiplicity $Q-1$). As a result, the operator content of the two models is profoundly different: the minimal models are rational, while the Potts model is irrational.

In general, the operators whose two-point function is defined by assigning to non-contractible loops (on the twice punctured sphere) the weight $\lambda_{(r)}$ have conformal weight \footnote{While not explicitly appearing in the literature as far as we know, this equation follows the mapping of the non-unitary minimal models onto a Coulomb gas---see, e.g., \cite{DiFrancesco:1988xz}.}
\begin{equation}\label{hdef}
h_r=\bar{h}_r={r^2-(p-q)^2\over 4pq} \,.
\end{equation}
The difference between the minimal models and Potts spectra thus becomes particularly important in the non-unitary case, $p-q>1$. In this case, the minimal models always contain an operator of negative conformal weight associated with the term for which non-contractible loops get the weight $\lambda_{(1)}$. This operator leads thus to an effective central charge $c_{\rm eff}=c-24h_1 =1-{6\over pq}$. Meanwhile, in the Potts model, all conformal weights are positive, and $c_{\rm eff}=c$. The only potential origin of non-positive conformal weights is the sector where non-contractible loops have vanishing weight. But since $\sqrt{Q}>0$,%
\footnote{We are restricting here to the ``physical part'' of the self-dual Potts model \cite{Saleur:1991npb}.}
we have necessarily $p-q<{p\over 2}$, and thus the dimension of the order parameter $h_{1/2,0}>0$.

\bigskip

To study the correlation functions, we consider the RSOS order parameters originally obtained in \cite{Pasquier:1986jc,Pasquier:1987xj}:
\begin{equation}\label{orderparameter}
\phi_r(i)=\frac{S_{(r)}^{\sigma_i}}{S_{\sigma_i}} \,,
\end{equation}
with the conformal weights given in \eqref{hdef}.
We therefore see that if $p$ is even, $p=2n$, we have $h_{p/2}=h_{1/2,0}$, i.e., the conformal weight of the operator $\phi_{p/2}$ coincides with the conformal weight of the Potts order parameter. In the case of type $D$, there are two such operators which we will denote as $\phi_{p/2}$ and $\phi_{\bar{p}/2}$. Therefore we will be mainly interested in the four-point functions of the operators $\phi_{p/2}$ and $\phi_{\bar{p}/2}$ in the RSOS model and their cluster interpretation, for the purpose of comparing with the geometric correlations in the Potts model. Notice that with our special eigenvector \eqref{speigen}, the contractible loops weight is 
\begin{equation}
\lambda_{(p-q)}=2\cos\frac{\pi(p-q)}{p}=\sqrt{Q} \,, 
\end{equation}
the same as in the Potts model. Since $h_{p-q}=0$ by \eqref{hdef}, this corresponds to the identity field.

\subsection{RSOS four-point functions}\label{RSOS4pt}

Consider now the four-point function $\langle \phi_{r_1}(i_1)\phi_{r_2}(i_2)\phi_{r_3}(i_3)\phi_{r_4}(i_4)\rangle$ on the sphere where the operators are inserted at the special sites $i_1, i_2, i_3, i_4$. Similar to the torus partition function, the four-point function can be expanded in terms of clusters/loops configurations \cite{Kostov:1989eg}. A detailed study of the RSOS weights (see appendix \ref{NptRSOS}) reveals that the weight of any loop is unchanged when it is turned inside out, i.e., wrapping around the ``point at infinity'' on the sphere punctured at the positions of the operator insertions. In particular, the loops surrounding all four insertion points are in fact contractible on the sphere, and hence they receive the usual weight $\lambda_{(p-q)}=\sqrt{Q}$ as in the Potts model. We will from now on refer to the contractible/non-contractible loops in this sense of the four-times punctured sphere.

\begin{figure}[t]
	\begin{center}
		\includegraphics[width=0.8\textwidth]{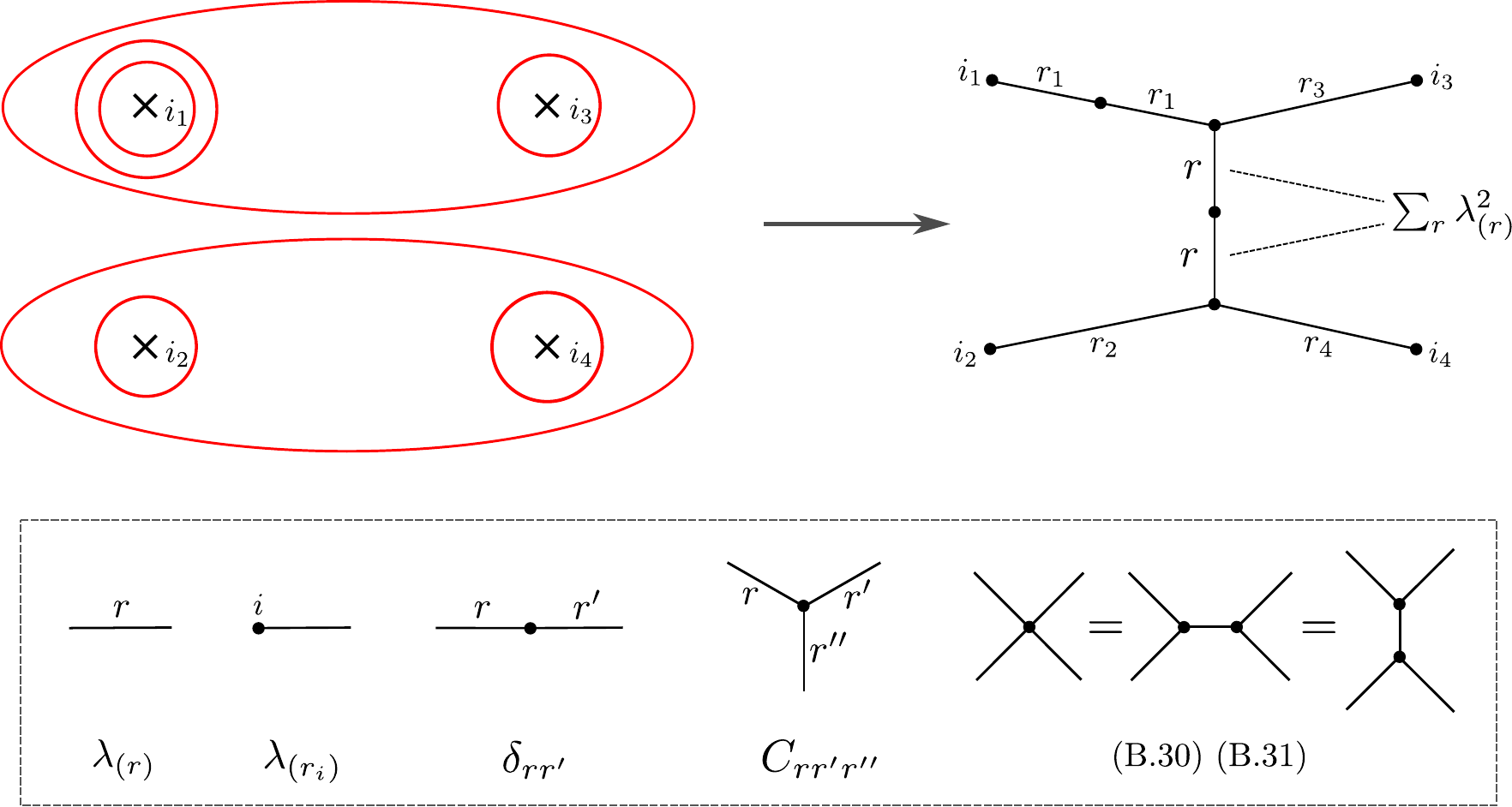}
		\caption{Determination of the weights of non-contractible loops in the four-point function $\langle \phi_{r_1}(i_1)\phi_{r_2}(i_2)\phi_{r_3}(i_3)\phi_{r_4}(i_4)\rangle$. In this example, the loop encircling points $1,3$ is represented by a three-leg vertex fusing $r_1,r_3$, and the one encircling points $2,4$ by a vertex fusing $r_2,r_4$. The weight of these two loops must be equal since they are connected by a topologically trivial domain as indicated by the two-leg vertex in the graph, and one needs to sum the loop weight $\lambda_{(r)}$ over $r\in\mathcal{D}^*$. We give in the box the general rule for assigning factors to the vertices and legs. The last diagram in the box is relevant for the configurations where all four points are within the same cluster.}\label{FigWeights}
		\label{RSOSrule}
	\end{center}
\end{figure}

For non-contractible loops, their weights in a certain configuration are given by simple rules of which we provide the detailed formulation in appendix \ref{NptRSOS} and give a brief summary here. As illustrated in figure \ref{FigWeights}, one starts by representing the domains between loops (namely, the clusters) as vertices on a graph, and loops separating the domains as legs connecting these vertices. The graph thus obtained can be evaluated by giving the legs and vertices the factors as shown in the figure. Notice that one needs to sum over $r\in\mathcal{D}^*$ for the internal legs. 

In the special cases where there are no non-contractible loops involved, i.e., all four points belong to the same big cluster, one still represents the cluster by a vertex and associate it with a four-leg vertex. As studied in appendix \ref{NptRSOS} (see \eqref{Cr1r2r3r4} and \eqref{crossing4pt}), the four-leg vertex can be decomposed into three-leg vertices \cite{Kostov:1989eg} as indicated in the last diagram in the box of figure \ref{FigWeights}. This results in the diagram's acquiring a non-trivial multiplicity, of which we will see an explicit example in the next section.\footnote{Here there is no factor associated with the internal leg, since it does not represent any loops.}

In the example of figure \ref{FigWeights}, we have the contribution of the diagram:
\begin{equation}
\lambda^2_{(r_1)}\lambda_{(r_2)}\lambda_{(r_3)}\lambda_{(r_4)}\sum_r C_{r_1r_3r}C_{rr_2r_4} \lambda_{(r)}^2 \,,
\end{equation}
as well as extra factors for the contractible loops (not shown). The three-point coupling $C_{rr'r''}$ can be calculated as we discuss in detail in Appendix \ref{NptRSOS}, and we also give the explicit expressions for type $A$ and type $D$ in Appendix \ref{3ptC}, which will be used in the next section. There, we will see that things simplify drastically for the four-point functions of $\phi_{p/2}$ (and $\phi_{\bar{p}/2}$ in type $D$) we are interested in, where we can make direct contact with the Potts correlation functions.

\bigskip

{\it Note: In our notations we shall henceforth not differentiate between the lattice correlation functions and their continuum limit, with the latter interpreted as the minimal-models correlation functions.}

\section{Geometrical  interpretation of four-point functions in minimal models}\label{cluster}

Since we are mainly interested in the comparison with the Potts model, in this section, we focus on the RSOS four-point functions $\langle\phi_{r}\phi_{r}\phi_{r}\phi_{r}\rangle$ where $\phi_r$ coincides with the Potts order parameter, i.e., $h_{r}=h_{1/2,0}$. This involves the operator $\phi_{p/2}$ in type $A$ (with $p$ even) and $\phi_{p/2},\phi_{\bar{p}/2}$ in type $D$ (with $p=\hbox{2 mod 4}$). In figure \ref{ADE}, we list the Dynkin diagrams $\mathcal{D}$ involved and the relevant conventions, which are used in appendix \ref{3ptC} for obtaining the three-point couplings $C_{r_1,r_2,r_3}$. As it turns out, the four-point functions we are interested in can be expanded in terms of clusters/loops configurations exactly like in the Potts model, but the geometric interpretation is different.

\begin{figure}[t]
	\begin{center}
		\includegraphics[width=0.8\textwidth]{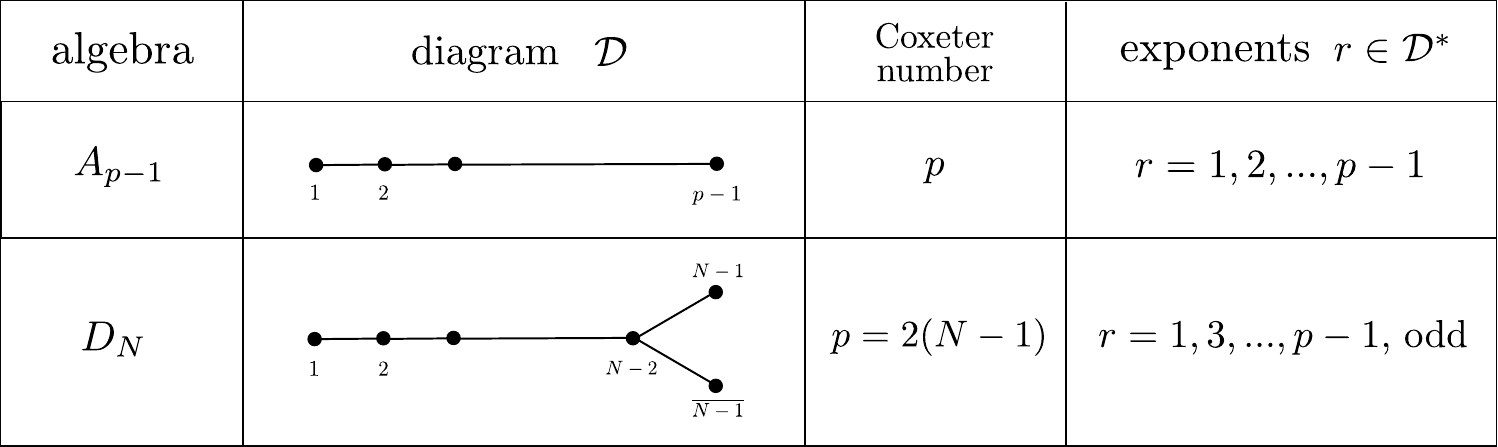}
		\caption{The type $A$ and type $D$ Dynkin diagrams $\mathcal{D}$ we are studying, where $p$ is even. In the case of type $D$ we have $N$ even, i.e., $p\equiv\hbox{2 mod 4}$, and the exponents $r=1,3,...,p-1$, odd. In the following, we will often denote $D_N$ as $D_{1+\frac{p}{2}}$.}\label{ADE}
	\end{center}
\end{figure}


\subsection{Type $A_{p-1}$}

\begin{figure}[t]
	\begin{center}
		\includegraphics[width=0.6\textwidth]{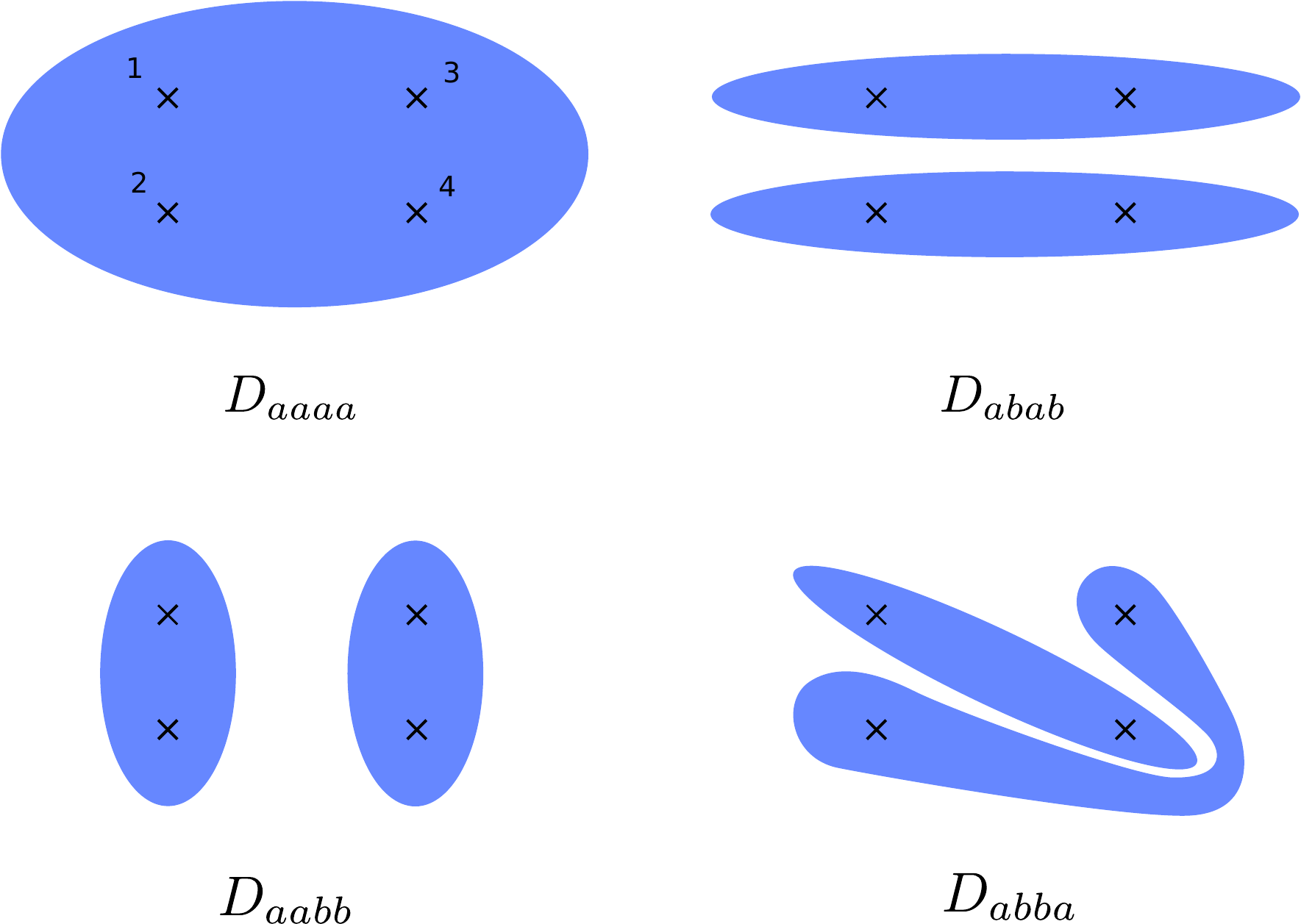}
		\caption{Four types of diagrams contributing to the four-point function $\langle\phi_{p/2}\phi_{p/2}\phi_{p/2}\phi_{p/2}\rangle$. Here we only draw the basic clusters while there can be extra contractible and non-contractible clusters surrounding them.}\label{fourdiagram}
	\end{center}
\end{figure}

In the case of type $A$, we consider the four-point function $\langle\phi_{p/2}\phi_{p/2}\phi_{p/2}\phi_{p/2}\rangle$. Since $\lambda_{p/2}=2\cos(\pi/2)=0$, any diagram with a loop encircling a single special site has weight 0 and does not contribute. The four-point function then involves four types of diagrams as shown in figure \ref{fourdiagram}. We denote them using the notations $D_{abcd}$ with the same convention as the Potts probabilities $P_{abcd}$. For example, the first type of diagrams $D_{aaaa}$ involves configurations where the four points are all within the same cluster, and the other three---$D_{aabb},D_{abab},D_{abba}$---involve two distinct clusters for the four points with $D_{abab}$, for instance, denoting the set of diagrams where 1 and 3 are within the same cluster, while 2 and 4 belong to another cluster.

The three-point couplings in this case are given in \eqref{CA} and with $r_i=p/2$ are simply:\footnote{As discussed in appendix \ref{3ptC}, the expression of the three-point coupling involves integers $\mathsf{a},\mathsf{b}$ from solving a Diophantine equation $r+\mathsf{b}p=\mathsf{a}(p-q)$ for given $r,p,q$. This can be done easily using the function \texttt{FindInstance} in {\sc Mathematica}.}
%
\begin{equation}\label{typeAc}
C_{p/2,p/2,r}=
\begin{cases}
(-1)^\mathsf{b},\; 1\leq r\leq p-1\hbox{ and $r$ odd},\; r+\mathsf{b} p=\mathsf{a}(p-q)\\
0,\; \hbox{otherwise}.
\end{cases}
\end{equation}
In the following, we will consider the weight of a diagram where all loops get the factor $\sqrt{Q}$ as its ``basic weight'' for the obvious reason to relate to the Potts model, and refer to the ratio of the weight in the graphical expansion with respect to this basic weight as the ``multiplicity''. According to the rules summarized in section \ref{RSOS4pt} (see the last diagram in the box in figure \ref{RSOSrule}), we obtain that the multiplicities in $D_{aaaa}$ are equal to 
\begin{equation}\label{Amulti1}
 M^{A_{p-1}}_{aaaa} =\sum_r (C_{p/2,p/2,r})^2=\sum_{r=1~\rm odd}^{p-1} 1={p\over 2}.
\end{equation}

The other three types of diagrams have the special sites encircled pairwise by one or more big loops respectively and connected by a topologically trivial domain where, in the case of type $A_{p-1}$, one should sum over $r=1,2,\ldots,p-1$ for the weight $\lambda_{(r)}$ of the non-contractible loops. However, thanks to the simplicity of the three-point coupling \eqref{typeAc} for the four-point function we are considering, one in fact only needs to sum over $r=1,3,\ldots,p-1,\text{odd}$. Denoting the number of non-contractible loops as $k$, we therefore see a significant simplification of the diagrammatic expansion involved: since
\begin{equation}\label{oddk}
\sum_{r=1~\rm odd}^{p-1}\lambda_{(r)}^k=\sum_{r=1~\rm odd}^{p-1}\bigg(2\cos\frac{r\pi}{p}\bigg)^k=0, \mbox{ for } k \mbox{ odd},
\end{equation}
any diagram with the two clusters separated by odd number of loops has weight zero in the four-point function $\langle\phi_{p/2}\phi_{p/2}\phi_{p/2}\phi_{p/2}\rangle$. Recalling that two RSOS clusters are mapped to Potts clusters only when they are separated by even number of loops, here we see that for the four-point function we are interested in, these are exactly the types of configurations that contribute. For this reason we henceforth suppose $k$ even and set
\begin{equation} \label{keven}
k = 2 l \,.
\end{equation}
From the remarks made at the beginning of section~\ref{PottsRSOS} this is equivalent to supposing that all clusters marked in the correlation functions
that we shall consider are of the Potts (and not dual) type.

\bigskip

In the next section, we will consider the $s$-channel spectrum involved in the RSOS four-point functions using the techniques developed in \cite{Jacobsen:2018pti} for the Potts model, where we take the four points to be on a cylinder as depicted in figure \ref{fig:cylinder} below. If we now consider the contribution from the third type of diagrams $D_{aabb}$ in figure \ref{fourdiagram} we see that, after mapping to the cylinder, this will give rise to diagrams just like for the calculation of $P_{aabb}$ in the Potts case, {\bf but} there is the important difference that in the sum over sectors, loops encircling the cylinder between points $1,2$ and $3,4$ have weight $\lambda_{(r)}$. The appearance of $\lambda_{(1)}$ is crucial. In the non-unitary case, it does not correspond to the identity field, but rather to the field with dimension
\begin{equation}
h_1={1-(p-q)^2\over 4pq}<0 \,.
\end{equation}
As mentioned before, this is in fact the field of most negative dimension in the theory, responsible for the value of the effective central charge $c_{\rm eff}=1-{6\over pq}$. In the case of the Potts model, however, as discussed in \cite{Jacobsen:2018pti}, only states with positive conformal weights propagate along the cylinder and no effective central charge appears despite the non-unitarity of the CFT. 

\bigskip

\begin{figure}[t]
	\begin{center}
		\includegraphics[width=0.8\textwidth]{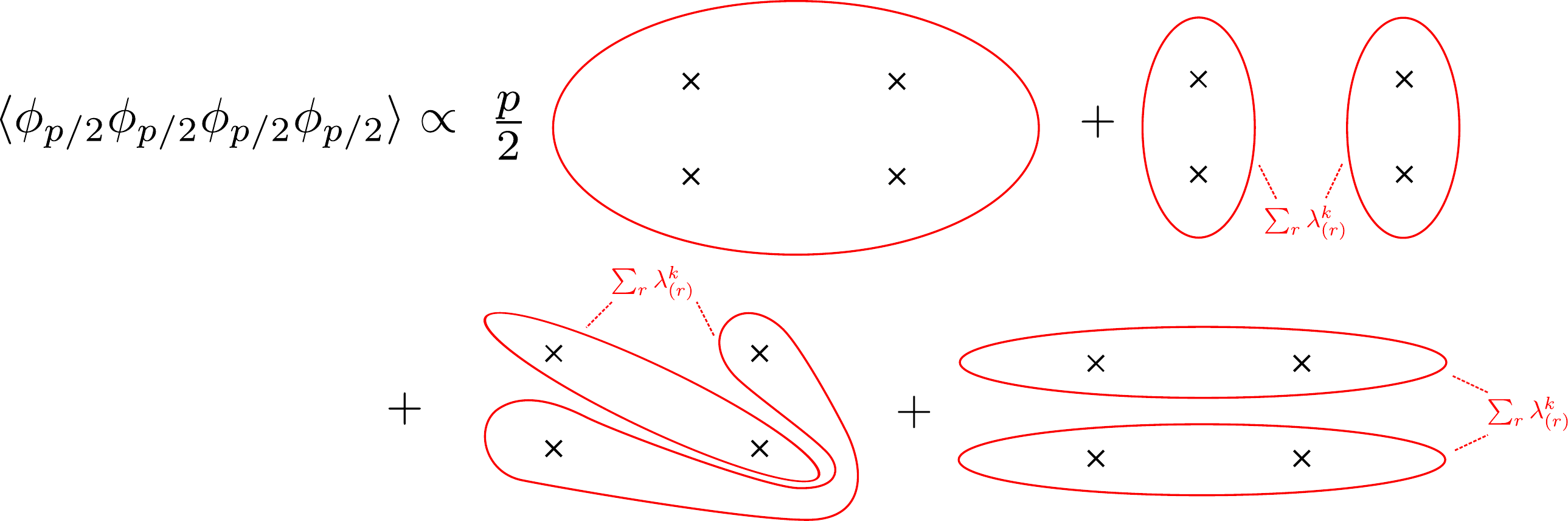}
		\caption{The diagrammatic expansion of four-point function $\langle\phi_{p/2}\phi_{p/2}\phi_{p/2}\phi_{p/2}\rangle$. The first type of diagram where all four sites are in one big loop comes with a multiplicity $\frac{p}{2}$. In the last three types of diagrams, one needs to sum over $\lambda_{(r)}^k$ $,r=1,3,...,p-1$ for even $k$ -- number of non-contractible loops encircling two of the special sites.}\label{expansion}
	\end{center}
\end{figure}

The full diagrammatic expansion of the four-point function with $\langle\phi_{p/2}\phi_{p/2}\phi_{p/2}\phi_{p/2}\rangle$ is summarized in figure \ref{expansion}. Note that in  $D_{aabb},
D_{abab},D_{abba}$, there are always two  basic clusters connecting $a$ to $a$ and $b$ to $b$ respectively, plus extra clusters encircling the basic pair, and non-contractible on the sphere. Instead of clusters we can  count their boundaries:\footnote{Although we have used so far mostly the language of loops, the mapping on the cluster formulation is obvious, simply by taking loops as cluster boundaries.} the basic pair gives rise to two boundaries, and every surrounding cluster contributes an extra pair. The total number of boundaries---namely, the number of non-contractible loops---$k$ (even) give rise to the multiplicity of the configurations
\begin{equation}\label{Amultidef}
\begin{aligned}
M^{A_{p-1}}(k)&\equiv {1\over \sqrt{Q}^k}\sum_{r=1~\rm odd}^{p-1} \lambda_{(r)}^k\\
&=\frac{1}{\sqrt{Q}^k}\sum_{\mathsf{a}=1\;\text{odd}}^{p-1} (\q^{\mathsf{a}}+\q^{-{\mathsf{a}}})^k \,,
\end{aligned}
\end{equation}
where $\mathfrak{q}=e^{i\pi \frac{p-q}{p}}$, and $\mathsf{a}$ is given in \eqref{typeAc}.
It is not hard to find a general formula for these multiplicities:
\begin{equation}\label{multiA}
M^{A_{p-1}}(k)={p\over 2\sqrt{Q}^k}\left(\begin{array}{c} k\\k/2\end{array}\right)+{p\over \sqrt{Q}^k}\sum_{n\in\mathbb{N}^*}^{\lfloor {\frac{k}{p}}\rfloor}\left(\begin{array}{c} k\\{k-np\over 2}\end{array}\right)(-1)^n.
\end{equation}
Notice when $p>k$, $\lfloor {\frac{k}{p}}\rfloor=0$ and \eqref{multiA} reduces to
\begin{equation}\label{Amulti}
M^{A_{p-1}}(k)={p\over 2\sqrt{Q}^k}\left(\begin{array}{c} k\\k/2\end{array}\right).
\end{equation}
In particular we have in this case $M^{A_{p-1}}(2)= p/Q$, $M^{A_{p-1}}(4)=3p/Q^2$, etc.  The multiplicity $p/2$ in the $D_{aaaa}$ diagram (eq. \eqref{Amulti1}) where all loops are contractible is independent of $Q$. Note that this formally coincides with $M^{A_{p-1}}(0)$ as it should.

We then introduce ``pseudo-probabilities'', such as
\begin{eqnarray}
\tilde{P}^{A_{p-1}}_{abab}=
{1\over Z_{\rm Potts}}\sum_{D\in D_{abab}} W_{\rm Potts} (D) M^{A_{p-1}}(k) \,,
\end{eqnarray}
where $k$ is the number of boundaries of the diagram $D$. We similarly define the pseudo-probabilities $\tilde{P}_{aabb}$ and $\tilde{P}_{abba}$ for the other cases of interest.
Notice that in this notation, we have the true Potts probability given by
\begin{eqnarray}
P_{abab}=
{1\over Z_{\rm Potts}}\sum_{D\in D_{abab}} W_{\rm Potts} (D).
\end{eqnarray}
We can then reexpress the four point-function in a more compact form
\begin{equation}
\langle\phi_{p/2}\phi_{p/2}\phi_{p/2}\phi_{p/2}\rangle\propto {p\over 2}P_{aaaa}+\tilde{P}^{A_{p-1}}_{abab}+\tilde{P}^{A_{p-1}}_{aabb}+\tilde{P}^{A_{p-1}}_{abba}.
\label{phi_pppp}
\end{equation}
Note that, while we use the notation $P$ for the last three terms, the sum of pseudo-probabilities is not equal to unity anymore. 

\bigskip

Let us as an application consider the  Ising model with $p=4$, $q=3$, corresponding to $\sqrt{Q}=2\cos{\pi\over 4}=\lambda_{(1)}$. In the expansion, the $D_{aaaa}$ diagrams get multiplicity ${p/2}=2$, corresponding to the two fusion channels
\begin{equation}
\sigma\times\sigma=\mathbb{I}+\epsilon \,,
\end{equation}
where $\sigma$ and $\epsilon$ denote the order parameter and energy operators, and we have used the usual simplified notation for operator product expansions (OPE). 
Meanwhile, the other geometries also get
multiplicity two, because $\lambda_{(1)}^k=\lambda_{(3)}^k=\sqrt{Q}^k$: in other words $\tilde{P}^{A_3}=2P$. Hence, in this case we find
\begin{equation}
\langle\phi_{p/2}\phi_{p/2}\phi_{p/2}\phi_{p/2}\rangle\propto P_{aaaa}+P_{aabb}+P_{abba}+P_{abab} \,,
\end{equation}
which is a well known result as can be seen directly from \eqref{Gaaaa}.

Consider now the case  $p=6$, $q=5$, corresponding to $\sqrt{Q}=2\cos{\pi\over 6}=\lambda_{(1)}$. Diagrams $D_{aaaa}$ now get multiplicity ${p/2}=3$, or since there are three fusion channels:
\begin{equation}
\sigma\times\sigma=\mathbb{I}+\sigma+\epsilon.
\end{equation}
The other diagrams still get multiplicity two since $\lambda_{(1)}^k=\lambda_{(5)}^k=\sqrt{Q}^k$ while $\lambda_{(3)}^k=0$. Hence in this case, we have 
\begin{equation}
\langle\phi_{p/2}\phi_{p/2}\phi_{p/2}\phi_{p/2}\rangle\propto 3P_{aaaa}+2(P_{aabb}+P_{abba}+P_{abab}).
\end{equation}
Meanwhile, since there is only one field with conformal weight $h_{1/2,0}=h_{33}={1\over 15}$, this four-point function should be the same as the four-point function of the spin operator in the three-state Potts model, in agreement with \eqref{Gaaaa}.

The $\phi_{p/2}$ four-point function will cease being expressed entirely in terms of the probabilities $P$ for other minimal models. Consider for  instance the case  $p=8$, $q=7$, corresponding to $\sqrt{Q}=2\cos{\pi\over 8}=\lambda_{(1)}$. In this case  we have $\lambda_{(1)}^k=\lambda_{(7)}^k=(2+\sqrt{2})^{k/2}=\sqrt{Q}^k$, whilst $\lambda_{(3)}^k=\lambda_{(5)}^k= (2-\sqrt{2})^{k/2}\neq \sqrt{Q}^k$.
So for instance, a diagram in $D_{abab}$ with one loop encircling each pair of points gets a weight $2(\lambda_{(1)}^2+\lambda_{(3)}^2)$, while a diagram with two loops encircling each pair of points gets a weight $2(\lambda_{(1)}^4+\lambda_{(3)}^4)$, etc. As soon as $\lambda_{(1)}^2\neq \lambda_{(3)}^2$, this skews the statistics compared with the pure probability/Potts problem: the $\tilde{P}$'s are {\sl not proportional} to the $P$'s.

Note that in weighing the diagrams, $p$ and $q$ play quite different roles. The weight of topologically trivial loops is $\sqrt{Q}=2\cos\pi {p-q\over p}$, while the weight of configurations with non-contractible loops depends only on $p$, since it involves a sum over all the eigenvalues of the adjacency matrix.

\subsection{Type $D_{1+\frac{p}{2}}$}\label{Dpq}

The above results can be easily extended to the type $D$ models by considering the corresponding Dynkin diagram $D_N$ where $p=2(N-1)$, as shown in figure \ref{ADE}. This is particularly interesting when $p\equiv\hbox{2 mod 4}$, corresponding to the case $N$ even.
In this case, it is known that the modular invariant partition function contains {\em two} primary fields with dimension $h_{1/2,0}=h_{m,n}$, where we recall that $m$ and $n$ are defined in \eqref{param_n_m}. Associated to these two fields are in fact two order parameters which we denote $\phi_{p/2}$ and $\phi_{\bar{p}/2}$. The existence of these two fields is related to the symmetry of the $D_N$ diagram under the exchange of the two fork nodes traditionally labeled as $N-1$ and $\overline{N-1}$, as can be seen from figure \ref{ADE}. 
Their lattice version can still be obtained using equation \eqref{orderparameter}.

Since $\lambda_{p/2}=\lambda_{\bar{p}/2}=0$, the two operators cannot be distinguished by their two-point function which, in both cases, are obtained by giving a vanishing weight to non-contractible loops on the twice punctured sphere. Four-point functions are much more interesting, and can be obtained by the same construction as for the type  $A$ models. It is easy to see that in this case the result \eqref{oddk} still holds. Also, the same four types of diagrams (figure \ref{fourdiagram}) with even number of non-contractible loops $k=2l$ participate and can be directly related to the Potts model. To expand the four-point functions in terms of diagrams, one needs the three-point couplings \eqref{CD} for $r_i=p/2,\bar{p}/2$ given by: 
\begin{subequations} \label{D3pt}
\begin{eqnarray}
C_{p/2,p/2,r} &=& (-1)^{\mathsf{a}-1\over 2},\;r+\mathsf{b} p=\mathsf{a}(p-q) \label{D3pta} \\
C_{\bar{p}/2,\bar{p}/2,r} &=& 1 \,, \label{D3ptb}\\
C_{p/2,p/2,\bar{p}/2} &=& C_{\bar{p}/2,\bar{p}/2,\bar{p}/2}=0 \,,
\end{eqnarray}
\end{subequations}
with $1\leq r\leq p-1\hbox{ and $r$ odd}$, and one has $\sum_r=\sum_\mathsf{a}$.\footnote{This comes from the fact that $\mathsf{a}$ is simply a rearrangement of $r$ with a shift $(p-q-1)/2$ and a cyclic spacing $p-q$, which results from normalizing with the $S^{\sigma}_{(p-q)}$. Taking $p=10,q=7$ for instance, $r=1,3,5,7,9$ and $\mathsf{a}=7,1,5,9,3$ where the position of 1 is shifted by $(p-q-1)/2=1$ and the spacing of consecutive odd integers is $p-q=3$. It is of course essential here that $p \wedge q = 1$, as we have supposed in \eqref{p_gcd_q}.}
Note that the vanishing of $C_{p/2,p/2,\bar{p}/2}$ follows from invariance of the $D_N$ diagram under exchange of the two fork nodes. We see that the fields of interest obey the OPEs
\begin{subequations}
\begin{eqnarray}
\phi_{p/2}\phi_{p/2} &\sim& \phi_{p/2} \,, \\
\phi_{\bar{p}/2}\phi_{\bar{p}/2} &\sim& \phi_{p/2} \,, \\
\phi_{p/2}\phi_{\bar{p}/2} &\sim& \phi_{\bar{p}/2} \,.
\end{eqnarray}
\end{subequations}
These OPEs are similar to those of the fields $V^D$, $V^N$ from \cite{Ribault:2018jdv}, as mentioned below eq. \eqref{PRSfunc}. Since there are only two fields with the correct dimensions in the CFT, we will in what follows make the identifications
\begin{equation}\label{idDN}
\phi_{p/2}\leftrightarrow V^D,\;\;\;\phi_{\bar{p}/2}\leftrightarrow V^N.
\end{equation}

\bigskip

Among the four-point functions involving $\phi_{p/2}$, $\phi_{\bar{p}/2}$, only the ones with even numbers of $\phi_{\bar{p}/2}$ are non-vanishing, as can been seen by directly carrying out the cluster expansions. This is similar to what happens for the four-point functions of $V^D$, $V^N$ in \cite{Ribault:2018jdv}. On the other hand, the cluster expansions of the four-point functions $\langle\phi_{p/2}\phi_{p/2}\phi_{p/2}\phi_{p/2}\rangle$ and $\langle\phi_{\bar{p}/2}\phi_{\bar{p}/2}\phi_{\bar{p}/2}\phi_{\bar{p}/2}\rangle$ in this case are exactly the same as the four-point function $\langle\phi_{p/2}\phi_{p/2}\phi_{p/2}\phi_{p/2}\rangle$ in the type $A$ models.
This is easily seen by recognizing that the calculation of the multiplicities in the cluster expansion of these four-point functions involves the factors
\begin{equation}
C^2_{p/2,p/2,r}=C^2_{\bar{p}/2,\bar{p}/2,r}=1 \,,
\end{equation}
so the situation reduces to the case of type $A$. Namely, the non-trivial sign difference in the three-point couplings \eqref{D3pta}, \eqref{D3ptb} and \eqref{typeAc} between type $D$ and type $A$ does not manifest itself in the four-point functions $\langle\phi_{p/2}\phi_{p/2}\phi_{p/2}\phi_{p/2}\rangle$ and $\langle\phi_{\bar{p}/2}\phi_{\bar{p}/2}\phi_{\bar{p}/2}\phi_{\bar{p}/2}\rangle$.

\bigskip

A particularly interesting case here is to consider the four-point function
\begin{equation}\label{4ptabab}
\langle \phi_{p/2}\phi_{\bar{p}/2}\phi_{p/2}\phi_{\bar{p}/2}\rangle \propto \langle V^DV^NV^DV^N\rangle \,,
\end{equation}
which is in fact the four-point function \eqref{PRSfunc}, \eqref{PRSfuncmod} studied in details in \cite{Picco:2016ilr,Migliaccio:2017dch,Ribault:2018jdv,Picco:2019dkm,Ribault:2019qrz}.
In this case, since the only three-point coupling involving $\phi_{p/2}$ and $\phi_{\bar{p}/2}$ is $C_{p/2,\bar{p}/2,\bar{p}/2}$, and since $\lambda_{\bar{p}/2}=0$, considerable simplification occurs in the cluster expansion of this correlator: diagrams where $\phi_{p/2}$ and $\phi_{\bar{p}/2}$ are in the same cluster with no other insertions are given a vanishing weight and thus disappear. Regarding \eqref{4ptabab} we are thus left with diagrams of type $D_{abab}$ and $D_{aaaa}$ only.

Diagrams of type $D_{aaaa}$ come with multiplicity
\begin{equation} \label{MDaaaa}
 M^{D_{1+\frac{p}{2}}}_{aaaa} = \sum_{r=1\;\text{odd}}^{p-1} (-1)^{\frac{\mathsf{a}-1}{2}}=\sum_{\mathsf{a}=1\;\text{odd}}^{p-1} (-1)^{\frac{\mathsf{a}-1}{2}}=1 \,.
\end{equation}
For the other three types of diagrams, we can again define the multiplicities
\begin{equation}
\begin{aligned}
M^{D_{1+\frac{p}{2}}}(k)&\equiv \frac{1}{\sqrt{Q}^{k}}\sum_{r=1\;\text{odd}}^{p-1} (-1)^{\frac{\mathsf{a}-1}{2}} \lambda_{(r)}^k, \mbox{ with } k \mbox{ even}\\
&=\frac{1}{\sqrt{Q}^k}\sum_{\mathsf{a}=1\;\text{odd}}^{p-1} (-1)^{\frac{\mathsf{a}-1}{2}} (\q^\mathsf{a}+\q^{-\mathsf{a}})^k \,, \label{DMulti}
\end{aligned}
\end{equation}
where again $\mathfrak{q}=e^{i\pi \frac{p-q}{p}}$.
It is easy to transform this expression into one that depends only on $\q$ and not on $p$ (provided 
$p\equiv2\hbox{ mod }4$). One finds
\begin{equation}
M^{D_{1+{p\over 2}}}(k=2l)={2\over Q^l}\sum_{\mathsf{m}=-l}^l \left(\begin{array}{c}
2l
\\l+\mathsf{m}
\end{array}\right){1\over \q^{2\mathsf{m}}+\q^{-2\mathsf{m}}} \,. \label{simpler}
\end{equation}
Using $\sqrt{Q}=\q+\q^{-1}$, we can express these in terms of $Q$. This is most easily done by noticing that
\begin{equation} \label{ChebyshevT}
 \q^\mathsf{j} + \q^{-\mathsf{j}} = 2 T_\mathsf{j} \left( \frac{\sqrt{Q}}{2} \right) \,,
\end{equation}
where $T_\mathsf{j}(x)$ denotes the $\mathsf{j}$'th order Chebyshev polynomial of the first kind. We obtain the explicit expressions
\begin{subequations} \label{Dweightsgen}
\begin{eqnarray}
M^{D_{1+{p\over 2}}}(k=2)&=&{2\over Q-2} \,, \\
M^{D_{1+{p\over 2}}}(k=4)&=&{2(3Q-10)\over (Q-2)(Q^2-4Q+2)} \,, \\
M^{D_{1+{p\over 2}}}(k=6)&=&{4(5Q^2-35Q+61)\over (Q-2)(Q^2-4Q+2)(Q^2-4Q+1)} \,.
\end{eqnarray}
\end{subequations}
A number of special cases of these will be discussed in section~\ref{limit}.
Note that the multiplicities in the $D$ case can be expressed only in terms of $Q$, while  in the $A$ case, an extra factor of $p$ remains (see equation (\ref{Amulti})). 
We see that the multiplicity $M^{D_{1+\frac{p}{2}}}(k)$ has poles at $\q=e^{i\pi (2\mathsf{n}+1)/4\mathsf{m}}$, for any $\mathsf{m}=-l,\ldots,l$, corresponding to
\begin{equation}\label{DQ}
\sqrt{Q}=2\cos \left( {\pi (2\mathsf{n}+1)\over 4\mathsf{m}} \right) \,.
\end{equation}

The pseudo-probabilities can then be defined as usual, for example
\begin{eqnarray} \label{pseudo_abab}
\tilde{P}^{D_{1+{p\over 2}}}_{abab}=
{1\over Z_{\rm Potts}}\sum_{D\in D_{abab}} W_{\rm Potts} (D) M^{D_{1+{p\over 2}}}(k) \,,
\end{eqnarray}
and it follows that 
\begin{subequations}
\begin{equation}\label{Dcorre1}
\langle \phi_{p/2}\phi_{\bar{p}/2}\phi_{p/2}\phi_{\bar{p}/2}\rangle\propto P_{aaaa}+\tilde{P}^{D_{1+{p\over 2}}}_{abab} \,.
\end{equation}
Other correlations with two $\phi_{p/2}$ and two $\phi_{\bar{p}/2}$ follow by braiding:
\begin{eqnarray}
&&\langle \phi_{p/2}\phi_{p/2}\phi_{\bar{p}/2}\phi_{\bar{p}/2}\rangle\propto P_{aaaa}+\tilde{P}^{D_{1+{p\over 2}}}_{aabb},\label{Dcorre2}\\
&&\langle \phi_{p/2}\phi_{\bar{p}/2}\phi_{\bar{p}/2}\phi_{p/2}\rangle\propto P_{aaaa}+\tilde{P}^{D_{1+{p\over 2}}}_{abba}.\label{Dcorre3}
\end{eqnarray}
\end{subequations}

\subsubsection{Three-state Potts model}

The simplest of all cases for the $D_N$ models of interest is with $p=6$, $q=5$, corresponding to the $D_4$ unitary CFT which is in fact the same as the $Q=3$ state Potts model \cite{Dotsenko1984,FATEEV1987644}. 
Notice that the $D_4$ Dynkin diagram is a three-star graph having the same $S_3$ symmetry as the permutations of the three Potts spins.
In this case, $r$ takes values $r=1,3,5$ and the multiplicity \eqref{DMulti} becomes simply 2, due to the symmetry $\lambda_{(1)}^k=\lambda_{(5)}^k=\sqrt{Q}^k$;
note also that $\lambda_{(3)}= 0$. In other words, 
$\tilde{P}^{D_4}_{abab}=2P_{abab}$, $\tilde{P}^{D_4}_{abba}=2P_{abba}$ and $\tilde{P}^{D_4}_{aabb}=2P_{aabb}$. We conclude that, for $Q=3$,
\begin{equation}\label{aaaaabab}
\langle \phi_{p/2}\phi_{\bar{p}/2}\phi_{p/2}\phi_{\bar{p}/2}\rangle\propto P_{aaaa}+2P_{abab}.
\end{equation}
Consider now the antisymmetric combination 
\begin{equation}
\langle \phi_{p/2}\phi_{\bar{p}/2}\phi_{p/2}\phi_{\bar{p}/2}\rangle-\langle \phi_{p/2}\phi_{\bar{p}/2}\phi_{\bar{p}/2}\phi_{p/2}\rangle\propto \tilde{P}^{D_{1+{p\over 2}}}_{abab}-\tilde{P}^{D_{1+{p\over 2}}}_{abba}
\end{equation}
which, for the $Q=3$ state Potts model becomes simply 
\begin{equation}
 \tilde{P}^{D_{4}}_{abab}-\tilde{P}^{D_{4}}_{abba}\propto P_{abab}-P_{abba} \,. \label{combo2}
 \end{equation}
In general, at $Q=3$, we expect that combinations such as (\ref{aaaaabab}) or the antisymmetric combination (\ref{combo2}) simplifies considerably. This, we believe, is in sharp contrast with the $P_{abcd}$ themselves,  whose expressions remain as complicated for $Q=3$ as in the generic case.
This is confirmed by concrete numerical evidence (eigenvalue cancellations) on finite-size cylinders.

\begin{figure}[t]
\begin{center}
    \includegraphics[width=0.4\textwidth]{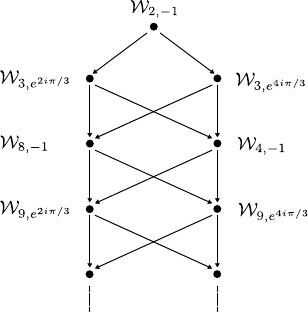}
     \caption{When $\q=e^{i\pi/6}$, the generically irreducible affine Temperley-Lieb module $\AStTL{2}{-1}$ (which contributes in its entirety to the  Potts correlations for $\q$ generic) becomes reducible, and admits a decomposition in terms of submodules as represented in this figure. Only the ``top'' contributes to the $Q=3$ correlations, leading to the disappearance of a large number of states in the $s$-channel. }\label{ModQ=3}
\end{center}
\end{figure}

This expectation is of course also in agreement with general results from representation theory of affine Temperley-Lieb algebras. Indeed, as we will discuss in more details in the following sections, from the set of all possible affine Temperley-Lieb modules $\AStTL{j}{z^2}$ appearing generically in the $Q$-state Potts model, only the simple tops $\mathcal{X}$ of $\AStTL{0}{\q^2},\AStTL{0}{-1},\AStTL{2}{-1}$ (with $\q=e^{i\pi/6}$) are relevant for the $D_4$ RSOS model \cite{Belletete:2017gwt}. The continuum limit of these modules is 
\begin{subequations}
\begin{eqnarray}
\IrrJTL{0}{\q^2}&\mapsto& \sum_{r=1}^4 |\chi_{r,1}|^2=\sum_{r=1}^4|\chi_{r,5}|^2 \,, \\
\IrrJTL{2}{-1}&\mapsto&\sum_{r=1}^4 \chi_{r,1}\overline{\chi}_{r,5}=\sum_{r=1}^4\overline{\chi}_{r,1}\chi_{r,5} \,, \\
\IrrJTL{0}{-1}&\mapsto& \sum_{r=1}^4|\chi_{r,3}|^2 \,.
\end{eqnarray}
\end{subequations}
The structure of the $\AStTL{2}{-1}$ module is shown for example in figure \ref{ModQ=3}.

Note that the particular combinations \eqref{aaaaabab} and (\ref{combo2}) at $Q=3$ could in fact be obtained from the general relationship \eqref{sumrules} between spin correlation functions in the Potts model $G_{abcd}$ and geometrical objects $P_{abcd}$. Setting $Q=3$ in these relations gives
%
\begin{subequations}
\begin{eqnarray}
4G_{abab}-G_{aaaa}&=&6(P_{aaaa}+2P_{abab}) \,, \\
G_{abab}-G_{abba}&=&3(P_{abab}-P_{abba}) \,.
\end{eqnarray}
\end{subequations}
Since the left hand sides can be  expressed strictly within the $Q=3$ Potts model, this means the same holds for the right-hand side, as we have directly established
in eqs.~\eqref{aaaaabab} and \eqref{combo2}.

\section{Pseudo-probabilities and affine Temperley-Lieb algebra}\label{pseudoATL}

\subsection{General setup} \label{setupATL}

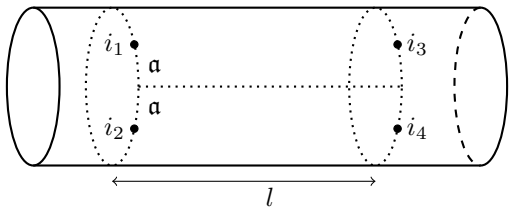
\begin{figure}
	\begin{center}
		\begin{tikzpicture}[scale=0.7]
		\draw[black,thick] (0,0) arc (0:360:0.5 and 1.5);
		\draw[black,thick] (-0.5,1.5)--(8,1.5);
		\draw[black,thick] (-0.5,-1.5)--(8,-1.5);
		\draw[black,thick,dotted] (1.5,0)--(6.5,0);
		\draw[black,thick,dotted] (1.5,0) arc (0:360:0.5 and 1.5);
		\draw[black,fill] (1.42,0.8) circle (2pt);
		\draw[black,fill] (1.42,-0.8) circle (2pt);
		\draw (1.42,0.8) node[left] {$i_1$};
		\draw (1.42,-0.8) node[left] {$i_2$};
		\draw (1.5,0.4) node[right] {$\mathfrak{a}$};
		\draw (1.5,-0.4) node[right] {$\mathfrak{a}$};
		\draw[black,thick,dotted] (6.5,0) arc (0:360:0.5 and 1.5);
		\draw[black,thick] (8.0,-1.5) arc (-90:90:0.5 and 1.5);
		\draw[black,thick,dashed] (8.0,1.5) arc (90:270:0.5 and 1.5);
		\draw[black,fill] (6.42,0.8) circle (2pt);
		\draw[black,fill] (6.42,-0.8) circle (2pt);
		\draw (6.42,0.8) node[right] {$i_3$};
		\draw (6.42,-0.8) node[right] {$i_4$};
		\draw[black,thin,<->] (1.0,-1.8)--(6.0,-1.8);
		\draw (4,-2.1) node {$l$};
		\end{tikzpicture}
	\end{center}
	\caption{Four-point functions in the cylinder geometry. The distance between the two operator insertions on each time slice is denoted $2\mathfrak{a}$.}
	\label{fig:cylinder}
\end{figure}

To proceed, we first recall the general framework discussed in \cite{Jacobsen:2018pti}. In the scaling limit, the Potts model correlation functions (\ref{sumrules}) as well as the geometrical correlations $P_{abcd}$ admit an $s$-channel expansion
\begin{equation} \label{G_s-channel-exp}
\begin{aligned}
\G(z,\bar{z})&=\sum_{\Delta,\bar{\Delta}\in {\cal S}} C_{\Phi_1\Phi_2\Phi_{\Delta\bar{\Delta}}}C_{\Phi_{\Delta\bar{\Delta}}\Phi_3\Phi_4}{\cal F}_\Delta^{(s)}(z)\overline{{\cal F}}_{\bar{\Delta}}^{(s)}(\bar{z})\\
&=\sum_{\Delta,\bar{\Delta}\in {\cal S}} A_{\Phi_{\Delta\bar{\Delta}}}{\cal F}_\Delta^{(s)}(z)\overline{{\cal F}}_{\bar{\Delta}}^{(s)}(\bar{z}) \,,
\end{aligned}
\end{equation}
where $A_{\Phi_{\Delta\bar{\Delta}}}$ denotes the amplitude of the field $\Phi_{\Delta,\bar{\Delta}}$ and the conformal blocks themselves can be expanded in (integer) powers of $z$. This full $z$ expansion is analogous to the expansion of lattice correlation functions on a cylinder in powers of eigenvalues of the geometrical transfer matrix discussed in \cite{Jacobsen:2018pti}, where the $s$-channel
geometry as shown in figure~\ref{fig:cylinder} corresponds to taking the two points $i_1, i_2$ to reside on one time slice and $i_3, i_4$ on another.
The two expansions can be matched exactly in the  limit where all lattice parameters (the width of the cylinder $L$ as well as the separation between points) are much larger than 1,%
\footnote{All measured in units of the lattice spacing.}
using the usual logarithmic mapping. 
We shall occasionally in the following also need to discuss the other two channels. For future reference, the definition of the channels is 
\begin{subequations}
	\begin{eqnarray}
	\mbox{$s$-channel} &:& i_1 \sim i_2 \mbox{ and } i_3 \sim i_4 \,, \label{s-channel-def} \\
	\mbox{$t$-channel}  &:& i_1 \sim i_4 \mbox{ and } i_2 \sim i_3 \,, \\
	\mbox{$u$-channel} &:& i_1 \sim i_3 \mbox{ and } i_2 \sim i_4 \,,
	\end{eqnarray}
\end{subequations}
where the $t$ and $u$-channels can just be obtained from the $s$-channel by relabelling the points.

A key aspect of the geometrical transfer matrix is that it can be expressed in terms of the affine Temperley-Lieb (ATL) algebra. The eigenvalues can therefore be classified in terms of (generically irreducible) representations of this algebra. The representations of interest in the case of the Potts model are of two types, denoted by $\AStTL{j}{z^2=e^{2i\pi \mathrm{p}/\mathrm{M}}}$ and $\bAStTL{0}{z^2=\q^2}$, which respectively indicate conformal weights $(h_{\mathbb{Z}+\frac{\mathrm{p}}{\mathrm{M}},j},h_{\mathbb{Z}+\frac{\mathrm{p}}{\mathrm{M}},-j})$ and $(h_{\mathbb{Z},1},h_{\mathbb{Z},1})$.
Their contributions to the correlation functions $P_{abcd}$ in the $s$-channel have been established in \cite{Jacobsen:2018pti} and are summarized in the table\footnote{We take this opportunity to correct a few misprints in\cite{Jacobsen:2018pti}: in Remark 3 of this reference,  $\AStTL{j}{z}$  (resp. $\AStTL{j}{z'}$) should read $\AStTL{j}{z^2}$ (resp. $\AStTL{j}{z'^2})$.}:
\begin{align} \label{paper1mainres}
\setlength{\arraycolsep}{6mm}
\renewcommand{\arraystretch}{1.2}
 \begin{array}{c|ll}
   & \mbox{$s$-channel} & \mbox{Parities}
   \\
   \hline
 P_{aaaa} &  \AStTL{0}{-1}\cup\AStTL{j}{e^{2i\pi \mathrm{p}/\mathrm{M}}} & j \in 2 \mathbb{N}^*, \ j\mathrm{p}/\mathrm{M}\hbox{ even}    \\
 P_{aabb} & \AStTL{0}{-1}\cup\bAStTL{0}{\q^2}\cup\AStTL{j}{e^{2i\pi \mathrm{p}/\mathrm{M}}}& j \in 2 \mathbb{N}^*, \ j\mathrm{p}/\mathrm{M}\hbox{ even}     \\
P_{abab/abba} & \AStTL{j}{e^{2i\pi \mathrm{p}/\mathrm{M}}} & j \in 2 \mathbb{N}^*, \ j\mathrm{p}/\mathrm{M}\hbox{ integer}    \\
   \hline
 \end{array}
\end{align}

We will often consider the symmetric and anti-symmetric contributions:
\begin{subequations} \label{PS_PA}
\begin{eqnarray}
 P_{S} &=& P_{abab} + P_{abba} \,, \\
 P_{A} &=& P_{abab} - P_{abba} \,.
\end{eqnarray}
\end{subequations}
Their spectra select $j\mathrm{p}/\mathrm{M}$ even and odd respectively.

The ATL representations $\AStTL{j}{z^2}$ have been discussed in details in \cite{Jacobsen:2018pti}.
They are standard modules of the algebra acting on so-called link patterns that encode the necessary information about the state of the loop model to the left
of a given timeslice of the cylinder (recall figure~\ref{fig:cylinder}), namely the pairwise connectivities between loop ends intersecting the time slice, as well as the position
of certain defect lines. More precisely,
the number $j$ corresponds to the number of clusters propagating along the cylinder: this number is half the number of cluster boundaries, often referred to as ``through lines'' in the literature. When $j=0$, modules $\AStTL{0}{z^2}$ correspond to giving to non-contractible loops wrapping around the axis of the cylinder the weight $z+z^{-1}$, and we shall need in particular the module with $z+z^{-1} = 0$ that imposes the propagation of one cluster (although no through lines are present). When $j\neq 0$, the parameter $z$ encodes the phases gathered by through lines as they wrap around the cylinder: there is a weight $z$ (resp.\ $z^{-1}$) for a through line that goes through the periodic direction in one directon (resp.\ the opposite direction). To account for these factors of $z$, it is in general necessary to keep track of whether a pairwise connectivity between loop ends straddles the periodic direction or not. However, when no cluster is propagating, the latter information is nugatory, and we shall need only the smaller quotient representation $\bAStTL{0}{z^2}$ that is devoid of this information.

For the ease of comparison with the appendices of \cite{Jacobsen:2018pti}, we recall that this reference also used the following  simpler notation:
\begin{itemize}

\item $V_0$ is the sector with no through-lines, and non-contractible loops have weight $\sqrt{Q}$: $V_0=\bAStTL{0}{z^2=\q^2}$,

\item $V_1$ is the sector with no through-lines, and non-contractible loops have weight zero: $V_1=\AStTL{0}{z^2=-1}$,

\item $V_{\ell,k}$ is the sector with $j=\ell \ge 2$ {\em pairs} of through-lines and phases $z^2=e^{2i\pi k/j}$: $V_{\ell,k}=\AStTL{j}{z^2=e^{2i\pi k/j}}$. 

\end{itemize}

\bigskip

The basic fact we want to explain now is how the complicated spectra for the $P_{abcd}$ found in \cite{Jacobsen:2018pti} can reduce to the much simpler $s$-channel spectra of minimal models where, instead of the genuine probabilities, we consider the proper combinations of pseudo-probabilities $\tilde{P}^{A,D}$ that appear in the geometrical reformulation of the four-point functions of order operators in minimal models. Note that this reduction should occur in finite size as well.

Recall that after the logarithmic conformal mapping, the $s$-channal corresponds to the cylinder geometry shown in figure~\ref{fig:cylinder}. This can be studied
in finite size by performing transfer matrix computations on an $L \times M$ lattice strip, with periodic boundary conditions in the $L$-direction, and
in the semi-infinite limit $M \gg L$. The representations acted on by the transfer matrix are those of the corresponding affine Temperley-Lieb (ATL) algebra.
Using the numerical methods described in Appendix~\ref{app:num}---and with further technical details being given in the appendices of \cite{Jacobsen:2018pti}---we can
extract, for each correlation function $P_{a_1,a_2,a_3,a_4}$ of interest, the finite-size amplitude $A_i := A(\lambda_i)$ of each participating transfer matrix
eigenvalue $\lambda_i$; see eq.\ \eqref{PAeval}. These $A_i$ are the finite-size precursors of the conformal amplitudes $A_{\Phi_{\Delta\bar{\Delta}}}$
appearing in \eqref{G_s-channel-exp}.

We have made a number of striking obversations about ratios of the amplitudes $A_i$, which, crucially, turn out
to be independent of $L$ and hence should carry over directly to their conformal counterparts, after the usual identification of representations.
Although we do not presently have complete
analytical derivations of these amplitude-ratio results in the lattice model, we wish to stress that the numerical procedures by which the observations
were made and thoroughly checked leaves no doubt that they are exact results. For the lack of a better word, we shall therefore simply
refer to them as {\bf facts} in the following.

 In full analogy with the occurence of minimal model representations of the Virasoro algebra in the continuum limit, it is well known indeed that only a small set of ``minimal'' representation of the affine Temperley-Lieb algebra appears in the correlation functions of minimal RSOS models on the lattice \cite{Pasquier:1986jc,Pasquier:1987xj,PASQUIER1990523}.
The reduction to the spectra of minimal models is made possible by virtue of the facts which we have observed.

We will now list these facts, and use them in our discussion of minimal models in the next section. 

\subsection{Facts of type 1}
\label{sec:facts1}


\noindent{\sl Whenever the same ATL module contributes to different $P_{abcd}$, the ratios of the corresponding amplitudes in these different $P_{abcd}$, depend only on the module, and are independent of the eigenvalues within this module. They also do not depend on the size $L$.}

\bigskip

To make this more explicit, consider for instance the modules $\AStTL{j}{e^{2i\pi \mathrm{p}/\mathrm{M}}}$ with $j\mathrm{p}/\mathrm{M}$ even that contribute to $P_{aaaa}$, $P_{aabb}$ and $P_S$, where we recall \eqref{PS_PA}. For such a module, consider in a certain size $L$ the eigenvalues $\lambda_i$ of the transfer matrix. The powers of these eigenvalues contribute to different probabilities with different amplitudes  $A_{aaaa}(\lambda_i),A_{aabb}(\lambda_i)$ and $A_S(\lambda_i)$. 
Our claim is that the ratios $A_{aabb}(\lambda_i)/A_{aaaa}(\lambda_i)$ and $A_S(\lambda_i)/A_{aaaa}(\lambda_i)$:
\begin{itemize}
\item{} are the same for all eigenvalues $\lambda_i$ in a given module, and thus only depend on the module;
\item{} are independent of the size $L$ of the system (provided it is big enough to allow the corresponding value of $j$
\end{itemize}
The same claim holds for eigenvalues within $\AStTL{0}{z^2=-1}$ for $A_{aaaa}/A_{aabb}$. 

We were able, by numerical fitting, to determine some of these ratios in closed form.  Defining first
\begin{equation} \label{def_alpha_coef}
\alpha_{j,z^2}\equiv {A_{aabb}\over A_{aaaa}}(\AStTL{j}{z^2}) \,,
\end{equation}
we have then
\begin{subequations}\label{alphacoeff}
\begin{eqnarray}
\alpha_{0,-1}&=&-1 \,, \\
\alpha_{2,1}&=&{1\over 1-Q} \,, \\
\alpha_{4,1}&=&-\frac{Q^5-7 Q^4+15 Q^3-10 Q^2+4 Q-2}{2 (Q^2-3Q+1)} \,, \\ 
\alpha_{4,-1}&=&{2-Q\over 2} \,.
\end{eqnarray}
\end{subequations}
Similarly defining
\begin{equation} \label{def_alpha_bar_coef}
\overline{\alpha}_{j,z^2}\equiv {A_{S}\over A_{aaaa}}(\AStTL{j}{z^2}) \,,
\end{equation}
we have then
\begin{subequations}
\begin{eqnarray}
\overline{\alpha}_{2,1}&=&2-Q \,\\
\overline{\alpha}_{4,1} &=& -\frac{ (Q^2 - 4 Q+2) (Q^2 - 3 Q -2)}{2}\,\\
\overline{\alpha}_{4,-1}&=&{(Q-1)(Q-4)\over 2} \,.
\end{eqnarray}
\end{subequations}

\bigskip

We now turn to the question of weighing differently non-contractible loops. This must be done in two quite different cases. For diagrams of type $D_{aabb}$, we can have a large number of such loops separating our two pairs of points in the $s$-channel cylinder geometry in figure \ref{fig:cylinder}. For diagrams $D_{abab}$ and $D_{abba}$ on the contrary, this number of loops---which is at least equal to two by definition---remains finite and bounded by $L/2$, and cannot increase during imaginary time propagation. Accordingly, we have two different sets of facts. 

\subsection{Facts of type 2}
\label{sec:facts2}

We focus now on the Potts probabilities involving long clusters: $P_{abab}$ and $P_{abba}$. A suitable modification of the code in
\cite{Jacobsen:2018pti}---details of which are provided in Appendix~\ref{app:algmodif}---allows us to determine, for a given eigenvalue $\lambda_i$ from $\AStTL{j}{z^2}$, the refined amplitudes corresponding to imposing a fixed number $k$ (even) of non-contractible loops. These refined amplitudes will allow us to reweigh the non-contractible loops and hence relate the pseudo-probabilities $\tilde{P}$ to the true probabilities $P$.

\begin{figure}[t]
	\begin{center}
		\includegraphics[width=0.8\textwidth]{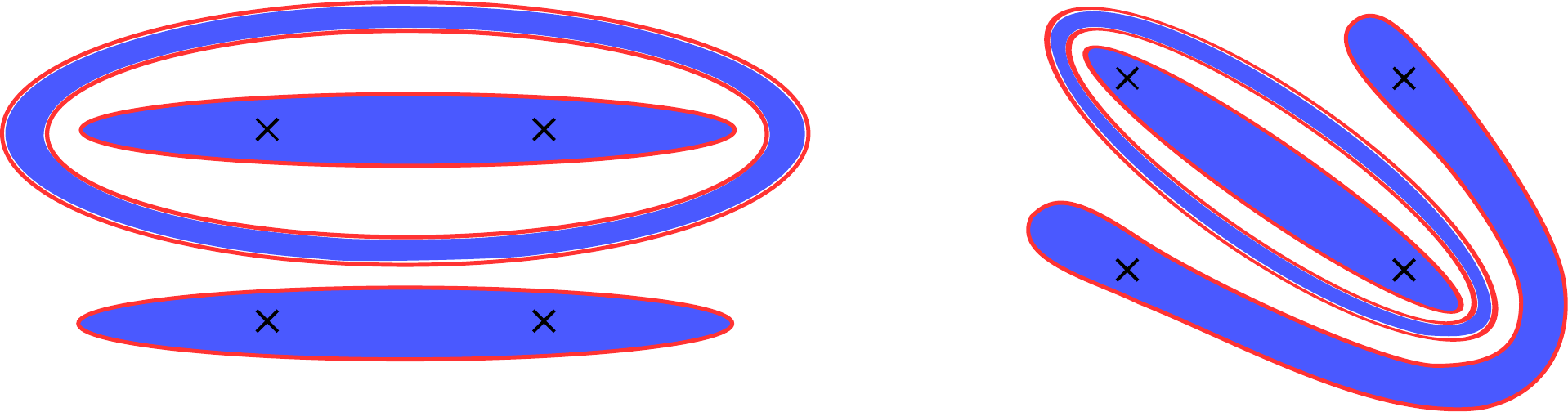}
		\caption{Diagrams of the types $D_{abab}$ and $D_{abba}$ with an extra cluster surrounding the basic clusters, i.e., the number of non-contractible loops is $k=4$.}\label{k4}
	\end{center}
\end{figure}

We first claim that the two pseudo-probabilities, $\tilde{P}_{abab}$ and $\tilde{P}_{abba}$
involve the same ATL modules exactly as their siblings $P_{abab}$ and $P_{abba}$: the only effect of the modified weights $M(k)$ is to modify the amplitudes.
To be more precise, let us consider the amplitude $A(\lambda_i)$ of some eigenvalue $\lambda_i$  occurring in the $s$-channel of the  diagram of the type $D_{abab}$ or $D_{abba}$ in finite size (for simplicity we do not indicate which type of diagram in the amplitudes). In the Potts case, this amplitude comes from summing over configurations where all loops, contractible or not, are given the same weight $\sqrt{Q}$. We now split this amplitude  into sub-amplitudes corresponding to configurations with a fixed number $k$ (even) of non-contractible loops occurring in the diagrams. Note that $k\geq 2$ since we have a least two loops each surrounding one cluster. The case $k=4$, for instance, corresponds to having, on top of these two basic clusters, an extra ``surrounding cluster'', i.e., an extra pair of loops as shown in figure \ref{k4}. Denoting by $A(\lambda_i)$ the total amplitude---that is, the one occurring in the Potts model, where no distinction is made between different values of $k$, as discussed in \cite{Jacobsen:2018pti}---we have 
\begin{equation}\label{Alambdak}
A(\lambda_i)=\sum_{k=2 ~\rm even} A^{(k)}(\lambda_i)
\end{equation}

We now state our facts of type 2:

\bigskip

\noindent{\sl For the eigenvalues in $\AStTL{j}{e^{2i\pi \mathrm{p}/\mathrm{M}}}$, the ratios of their amplitudes contributing to configurations with precisely $k$  non-contractible loops, depend only on the module and on $k$, and are  independent of the eigenvalues within this module. They  also do not depend on the size $L$.}

\bigskip

We define 
\begin{equation}
\beta^{(k)}_{j,z^2}\equiv {A_{abab}^{(k)}\over A_{abab}^{(2)}}(\AStTL{j}{z^2}).\label{betadef}
\end{equation}
Note that, since the amplitudes for the symmetric combination $P_S$ involve only $j\mathrm{p}/\mathrm{M}$ even, the amplitudes for $j\mathrm{p}/\mathrm{M}$ odd are necessarily equal and opposite in $P_{abab}$ and $P_{abba}$. Similarly, since the amplitudes for $P_A$ involve only $j\mathrm{p}/\mathrm{M}$ odd, 
the amplitudes for $j\mathrm{p}/\mathrm{M}$ even are the same for $P_{abab}$ and $P_{abba}$. The ratios $\beta^{(k)}_{j,z^2}$ are thus the same for both cases. 

Numerical determination leads to the following results (by definition, $\beta^{(2)}=1$):
\begin{subequations} \label{betas}
\begin{eqnarray}
\beta^{(4)}_{4,1}&=&  -\frac{Q^2}{3Q+2} \,, \\
\beta^{(4)}_{4,-1}&=&  -\frac{Q(Q-2)}{3Q-4} \,, \label{beta4-1} \\
\beta^{(4)}_{4,i}&=& -\frac{Q^2-4Q+2}{3Q-10} \,. 
\end{eqnarray}
\end{subequations}

\bigskip

We finally turn to the case of $P_{aabb}$, which as we will see must be handled a bit differently. 

\subsection{Facts of type 3}
\label{sec:facts3}

We now consider calculating statistical sum with $D_{aabb}$ geometries. Unlike the previous cases, the non-contractible loops are now those
that wrap around the axis of the cylinder. For a finite separation $l$ of the points along the cylinder axis (recall figure~\ref{fig:cylinder}) there can be up to
$2l$ such loops. As  $l \to \infty$, it is known that the average number of such loops in the Potts model grows like $\ln l$ \cite{PhysRevLett.100.087205}. In this case, the natural thing to do is not to focus on fixing the number of such loops,  but rather in modifying their fugacity, i.e., giving them a modified weight
\begin{equation} \label{na-modif}
n_\mathsf{a}\equiv \q^\mathsf{a}+\q^{-\mathsf{a}} \,.
\end{equation}
We denote such sums by $\tilde{P}_{aabb}^{(\mathsf{a})}$. The probability $P_{aabb}$ in the Potts model corresponds to $\mathsf{a}=1$ and involves modules $\bAStTL{0}{\q^2},\AStTL{0}{-1}$, and $\AStTL{j}{e^{2i\pi \mathrm{p}/\mathrm{M}}}$ with $j\in \mathbb{N}^*$, $j\mathrm{p}/\mathrm{M}$ even. The sums in $\tilde{P}_{aabb}^{(\mathsf{a})}$ involve the same modules, except for $\bAStTL{0}{\q^2}$ which is replaced by $\AStTL{0}{\q^{2\mathsf{a}}}$. This is expected, since such modules precisely correspond to giving to non-contractible loops the weight $n_\mathsf{a}$.

For different values of $\mathsf{a}$ (including $\mathsf{a}=1$, i.e., the case of Potts),  the $\tilde{P}_{aabb}^{(\mathsf{a})}$ involve eigenvalues from different modules. Among these  are of course the modules $\AStTL{0}{\q^{2\mathsf{a}}}$ for which, since they themselves depend on $\mathsf{a}$, there is not much point comparing amplitudes. However,  the modules $\AStTL{0}{-1}$ and $\AStTL{j}{e^{2i\pi \mathrm{p}/\mathrm{M}}}$ also contribute to the  $\tilde{P}_{aabb}^{(\mathsf{a})}$. For these, we can indeed compare the amplitudes of their  eigenvalues  contributions. Like before, another type of remarkable facts is then observed:

\bigskip

\noindent{\sl The ratios of the  amplitudes of eigenvalues from $\AStTL{0}{-1}$, and $\AStTL{j}{e^{2i\pi \mathrm{p}/\mathrm{M}}}$ that contribute to the $\tilde{P}_{aabb}^{(\mathsf{a})}$  depend only on the module and on $\mathsf{a}$, and are  independent of the eigenvalues within this module. They  also do not depend on the size $L$.}

\bigskip

Now define
\begin{equation} \label{gammadef}
\gamma^{(\mathsf{a})}_{j,z^2}\equiv{A^{(\mathsf{a})}_{aabb}\over A_{aabb}}(\AStTL{j}{z^2}) \,,
\end{equation}
where $A^{(\mathsf{a})}$ is the amplitude in $\tilde{P}_{aabb}^{(\mathsf{a})}$, and $A$ the amplitude in $P_{aabb}=\tilde{P}_{aabb}^{(1)}$. Denoting $Q_\mathsf{a}=n_\mathsf{a}^2$ such that we have $Q=n^2$ as usual, we have determined the following:
%
\begin{subequations} \label{gammacoeff}
\begin{eqnarray}
\gamma_{0,-1}^{(\mathsf{a})}&=&1 \,, \\
\gamma_{2,1}^{(\mathsf{a})}&=&{(Q_2-Q_1)Q_\mathsf{a} \over (Q_2-Q_\mathsf{a})Q_1} \,, \\
\gamma_{4,1}^{(\mathsf{a})}&=& \frac{(c_1 + Q_\mathsf{a}) Q_\mathsf{a} (Q_4 - Q_1)}{c_2 (Q_4 - Q_\mathsf{a})} \,, \label{gamma41} \\
\gamma_{4,-1}^{(\mathsf{a})}&=&{Q_\mathsf{a}\over Q_1} \,. \label{gamma4-1}
\end{eqnarray}
\end{subequations}
The expression for $\gamma_{4,1}^{(\mathsf{a})}$ involves two quantities, $c_1$ and $c_2$, which are independent of $\mathsf{a}$, but which have a complicated $Q$-dependence.
They are given by the following expressions:
\begin{subequations}
\begin{eqnarray}
 c_1 &=& \frac{8 - 26 Q + 60 Q^2 - 110 Q^3 + 112 Q^4 - 54 Q^5 + 12 Q^6 - Q^7}{Q (2 - 4 Q + Q^2)} \,, \\
 c_2 &=& \frac{(Q-4) (Q-1) (2 - 4 Q + 10 Q^2 - 15 Q^3 + 7 Q^4 - Q^5)}{2 - 4 Q + Q^2} \,.
\end{eqnarray}
\end{subequations}

\bigskip

The fact that the ratios \eqref{def_alpha_coef}, \eqref{def_alpha_bar_coef}, \eqref{betadef} and \eqref{gammadef} exist and are independent of the size of the system suggests strongly that they have a simple, algebraic origin---e.g., occurring as recoupling coefficients in quantum group representation theory. We hope to discuss this more in a forthcoming paper. For now, we use these facts (which, strictly speaking, must be considered as conjectures, since we have only checked them for a finite number of values of $L$---see Appendix \ref{app:num} for details) to discuss correlation functions in the RSOS models.

We also note that when $\mathsf{a}$ is an integer, the representation theory of ATL is not generic: the modules $\AStTL{0}{\q^{2\mathsf{a}}}$ are reducible, and contain a sub-module isomorphic to $\AStTL{\mathsf{a}}{1}$. This does not affect the coefficients in \eqref{gammacoeff}: more details can be found in appendix~\ref{app:num}.

\section{Recovering minimal model four-point functions}\label{MM}

\begin{figure}[ht]
\begin{center}
    \includegraphics[width=0.6\textwidth]{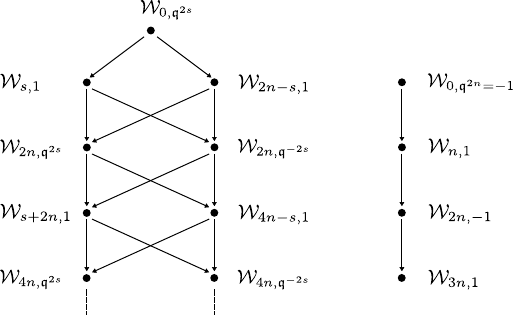}
     \caption{The structure of the standard modules involved in the RSOS model for $\q=e^{i\pi/2n}$ corresponding to $p=2n,q=2n-1$. The RSOS  model is obtained by restricting to the simple tops.}\label{ModulesRSOS}
\end{center}
\end{figure}

Recovering the $s$-channel spectrum of the minimal model transfer matrix is a subtle process. It involves not only ``throwing away'' many modules $\AStTL{j}{z^2}$, but also restricting to the irreducible tops of those which are kept. More precisely, in the continuum limit, the representation of the ATL algebra relevant for the $A_{p-1}$ RSOS minimal model is \cite{Belletete:2017gwt,ALCARAZ1989735}: 
%
\begin{equation}\label{RSOSATL}
	\rho_{\rm per} \simeq  \bigoplus_{n=1}^{p-1} \aX_{0,\q^{2n}}\ ,
\end{equation} 
with $\q=e^{i\pi {p-q\over p}}$. Here, each module $\aX_{0,\q^{2n}}$ is the irreducible top of the modules $\AStTL{0}{\q^{2n}}$, which become reducible when $\q$ is a root of unity. The structure of the some of these modules is given in figure \ref{ModulesRSOS}. 

In addition, different affine Temperley-Lieb modules $\AStTL{j}{z^2}$  may get glued in the loop model representation relevant for the Potts correlation functions. The full analysis of what happens is not our concern here, however, and will be discussed elsewhere. In this paper, we simply wish to illustrate the mechanism by which {\em unwanted eigenvalues disappear} from the $s$-channel spectrum in finite size. This turns out to be in one-to-one correspondence with the simplification of the spectrum in the continuum limit, since we have \cite{Belletete:2017gwt}:
\begin{equation}
\aX_{0,\q^{2n}}\mapsto \sum_{r=1}^{q-1} |\chi_{rn}|^2
\end{equation}
Note in particular that this only involves diagonal fields.

\subsection{The case of $A_{p-1}$}

Consider  the $A_{p-1}$ models for which we have seen in \eqref{phi_pppp} that
\begin{equation}
\langle\phi_{p/2}\phi_{p/2}\phi_{p/2}\phi_{p/2}\rangle\propto {p\over 2}P_{aaaa}+\tilde{P}^{A_{p-1}}_{aabb}+\tilde{P}^{A_{p-1}}_{abba}
+\tilde{P}^{A_{p-1}}_{abab} \,. \label{AResCom}
\end{equation}

Let us now examine, for instance, the module $\AStTL{4}{-1}$ corresponding to $j=4$ and $z^2=e^{2i\pi \mathrm{p}/\mathrm{M}}=-1$ with $\mathrm{p}/\mathrm{M}=1/2$. Using that $\q=e^{i\pi/2n}$, we can write $z^2 = \q^{2s}$ with $s=n$. For $p>4$, it is clear that the module $\AStTL{4}{-1}$ does not appear in any of the ``ladders'' (such as the ones in figure \ref{ModulesRSOS}) associated with the simple modules describing the minimal model.  Barring spurious degeneracies,\footnote{Since we study a specific Hamiltonian or transfer matrix, such degeneracies cannot be excluded {\em a priori}, though they are not observed in our numerical analysis.} this means the total amplitude for the corresponding eigenvalues in (\ref{AResCom})  should vanish. Let us now see how each term in (\ref{AResCom}) contributes to this amplitude.

While the first term in \eqref{AResCom} involves $P_{aaaa}$, all other terms involve modified weights. The total amplitude can thus be written as:
\begin{equation}\label{A4m1}
\tilde{A}(\AStTL{4}{-1})={p\over 2}A_{aaaa}(\AStTL{4}{-1})+\tilde{A}_{aabb}(\AStTL{4}{-1})+\tilde{A}_{abba}(\AStTL{4}{-1})+\tilde{A}_{abab}(\AStTL{4}{-1}) \,.
\end{equation}
Here we have introduced modified amplitudes $\tilde{A}$,
determined by the modified weights given to non-contractible loops in the RSOS correlation functions, when compared to the Potts model ones. For notational simplicity, we ignore the superscript $A_{p-1}$ for $\tilde{A}$ here---and similarly we shall omit in the next subsection the superscripts for modified amplitudes of type $D$---, while one should keep in mind that the modified amplitudes depend on the algebra in consideration due to the difference in the three-point couplings \eqref{typeAc} and \eqref{D3pt}.
We have in general
%
\begin{equation}\label{Atilde}
\tilde{A}(\lambda_i)=\sum_{k=2~\rm even}^j A^{(k)}(\lambda_i)M(k) \,,
\end{equation}
where $A$ denotes the Potts amplitudes in \eqref{Alambdak}. The sum in \eqref{Atilde} is truncated to the maximum value $j$ since for an eigenvalue $\lambda_i$ in $\mathcal{W}_{j,z^2}$, we have at most $k=j$, as is clear from the geometrical interpretation of the ATL modules in section \ref{setupATL}.
Using our facts of type 2---see eq.~(\ref{betadef})---we can therefore write 
\begin{equation}
\left.{\tilde{A}_{abab}\over A_{abab}}\right|_{\AStTL{j}{z^2}}={\sum_{k=2~\rm even}^j \beta_{j,z^2}^{(k)} M(k)\over \sum_{k=2 ~\rm even}^j \beta_{j,z^2}^{(k)}} \,.\label{amplitudesconstruction}
\end{equation}
The same holds for $A_{abba}$, since, for $\AStTL{4}{-1}$, the amplitudes for the two sectors $A_{abab}$ and $A_{abba}$ are identical. We therefore write
\begin{equation} \label{Aabab4-1}
\tilde{A}_{abab}(\AStTL{4}{-1})=A_{abab}(\AStTL{4}{-1}){M^{A_{p-1}}(2)+\beta_{4,-1}^{(4)}M^{A_{p-1}}(4)\over 
1+\beta_{4,-1}^{(4)}} \,.
\end{equation}
Now use that $\beta_{4,-1}^{(4)}=-{Q(Q-2)\over 3Q-4}$ from \eqref{beta4-1}, together with $M^{A_{p-1}}(2)={p\over Q}$ and $M^{A_{p-1}}(4)={3p\over Q^2}$ from \eqref{Amulti}. Hence 
\begin{equation}
\tilde{A}_{abab}(\AStTL{4}{-1})=-{2p\over Q(Q-1)(Q-4)}A_{abab}(\AStTL{4}{-1}) \,,
\end{equation}
and the same for $\tilde{A}_{abba}$.

Next, we have
\begin{equation}
\begin{aligned}
\tilde{A}_{aabb}(\AStTL{4}{-1})=&A_{aabb}(\AStTL{4}{-1})\sum_{\mathsf{a}=1~\rm odd}^{p-1} {A_{aabb}^{(\mathsf{a})}\over A_{aabb}}(\AStTL{4}{-1})\\
=&A_{aabb}(\AStTL{4}{-1})\sum_{\mathsf{a}=1~\rm odd}^{p-1} \gamma_{4,-1}^{(\mathsf{a})}\\
=&{p\over Q}A_{aabb}(\AStTL{4}{-1}) \,,
\end{aligned}
\end{equation}
where we used that $\gamma_{4,-1}^{(\mathsf{a})}={Q_{\mathsf{a}}\over Q_1}$ from \eqref{gamma4-1}, together with the identity
\begin{equation}
\sum_{{\mathsf{a}}=1~\rm odd}^{p-1} (\q^{\mathsf{a}}+\q^{-{\mathsf{a}}})^2=p \,,
\end{equation}
valid when $p$ is even, as we have supposed in \eqref{param_n_m}.

We therefore see that \eqref{A4m1} becomes
\begin{equation}
\begin{aligned}
\tilde{A}(\AStTL{4}{-1})&={p\over 2}A_{aaaa}(\AStTL{4}{-1})+{p\over Q}A_{aabb}(\AStTL{4}{-1})-{2p\over Q(Q-1)(Q-4)}\left(A_{abba}+A_{abab}\right)(\AStTL{4}{-1})\\
&=pA_{aaaa}(\AStTL{4}{-1})\left({1\over 2}+{\alpha_{4,-1}\over Q}-{2\over Q(Q-1)(Q-4)}\overline{\alpha}_{4,-1}\right) \,,
\end{aligned}
\end{equation}
where in the last line we have used \eqref{def_alpha_coef} and \eqref{def_alpha_bar_coef}.
Recall that $\overline{\alpha}_{4,-1}={(Q-1)(Q-4)\over 2}$ and $\alpha_{4,-1}={2-Q\over 2}$. We arrive at 
%
\begin{equation}
\left({1\over 2}+{\alpha_{4,-1}\over Q}-{2\over Q(Q-1)(Q-4)}\overline{\alpha}_{4,-1}\right)=0.
\end{equation}
We have thus established that the amplitude of eigenvalues coming from $\AStTL{4}{-1}$ all vanish in the four-point function \eqref{AResCom}. 

\bigskip

In fact, since the $s$-channel of the four-point function \eqref{AResCom} involves only diagonal fields in the type $A$ minimal models, the amplitudes of eigenvalues from all modules $\mathcal{W}_{j,z^2}$ should vanish in \eqref{AResCom} since they correspond to non-diagonal fields in the continuum limit,
leaving only the diagonal fields from $\bAStTL{0}{\q^2}$. We therefore expect, from the vanishing of the $\mathcal{W}_{j,z^2}$ contributions, to have the following relation: 
\begin{equation}
{p\over 2}+\alpha_{j,z^2}\sum_{{\mathsf{a}}=1~\rm odd}^{p-1} \gamma_{j,z^2}^{({\mathsf{a}})}+\bar{\alpha}_{j,z^2}
{\sum_{k=2~\rm even}^j \beta_{j,z^2}^{(k)} M^{A_{p-1}}(k)\over \sum_{k=2~\rm even}^j \beta_{j,z^2}^{(k)}}=0.
\end{equation}
While this can be checked numerically for $\mathcal{W}_{0,-1}$, $\mathcal{W}_{2,1}$ and $\mathcal{W}_{4,1}$ using the $\alpha$, $\bar{\alpha}$, $\beta$ and $\gamma$ we provided in the previous section, we do not have, for the moment, closed-form expressions for all the coefficients involved. Note that here $\AStTL{2}{1}$ and $\AStTL{4}{1}$ appear as submodules of some other modules when $\q$ is the relevant root of unity, so one might have feared that the overall cancellation of its contributions might involve also some of the coefficients of these other modules---this is, however, not the case.

\subsection{The case of $D_{1+\frac{p}{2}}$}

We next consider amplitudes in the $D_{1+{p\over 2}}$ case. Let us study the case of $\AStTL{2}{1}$, for example. Recall from \eqref{Dcorre1} that
\begin{equation}\label{aabbD}
\langle \phi_{p/2}\phi_{p/2}\phi_{\bar{p}/2}\phi_{\bar{p}/2}\rangle=P_{aaaa}+\tilde{P}_{aabb}^{D_{1+{p\over 2}}}.
\end{equation}
Because of \eqref{MDaaaa} the amplitude in the first term  does not depend on the modification of the weights $M$, so we have
\begin{equation}
\tilde{A}(\AStTL{2}{1})=A_{aaaa}(\AStTL{2}{1})+\tilde{A}_{aabb}(\AStTL{2}{1}),\label{combvi}
\end{equation}
and
\begin{equation}
\begin{aligned}
\tilde{A}_{aabb}(\AStTL{2}{1})=&A_{aaaa}(\AStTL{2}{1})\frac{A_{aabb}(\AStTL{2}{1})}{A_{aaaa}(\AStTL{2}{1})}\sum_{{\mathsf{a}}=1~\rm odd}^{p-1} (-1)^{{\mathsf{a}}-1\over 2}{A_{aabb}^{({\mathsf{a}})}(\AStTL{2}{1})\over A_{aabb}(\AStTL{2}{1})}\\
=&A_{aaaa}(\AStTL{2}{1})\bigg(\alpha_{2,1}\sum_{{\mathsf{a}}=1~\rm odd}^{p-1} (-1)^{{\mathsf{a}}-1\over 2} \gamma_{2,1}^{({\mathsf{a}})}\bigg) \,,
\end{aligned}
\end{equation}
where we have used \eqref{D3pt} (from which the  $(-1)^{\frac{\mathsf{a}-1}{2}}$ occurs) and \eqref{gammadef}.
Recall $\alpha_{2,1}={1\over 1-Q}$, and $\gamma_{2,1}^{({\mathsf{a}})}$ is given in \eqref{gammacoeff}, so we have
\begin{equation}
\tilde{A}(\AStTL{2}{1})=A_{aaaa}(\AStTL{2}{1})\bigg(1+\frac{Q_2-Q}{Q(1-Q)}\sum_{{\mathsf{a}}=1~\rm odd}^{p-1} (-1)^{\mathsf{a}-1\over 2}\frac{Q_{\mathsf{a}}}{Q_2-Q_{\mathsf{a}}}\bigg)=0 \,,
\end{equation}
which can be checked to vanish using {\sc Mathematica}.

In general, from the identification of \eqref{idDN}, the $s$-channel spectrum of \eqref{aabbD} involves only diagonal fields as argued in \cite{Ribault:2018jdv} and therefore we should have the following identity for modules $\mathcal{W}_{j,z^2}$:
\begin{equation}\label{aabbcancel}
1+\alpha_{j,z^2}\sum_{{\mathsf{a}}=1~\rm odd}^{p-1} (-1)^{\mathsf{a}-1\over 2}\gamma_{j,z^2}^{(\mathsf{a})}=0.
\end{equation}
This can be checked to be true for $\mathcal{W}_{0,-1}$, $\mathcal{W}_{4,-1}$, $\mathcal{W}_{4,1}$ using \eqref{alphacoeff} and \eqref{gammacoeff}.

\bigskip

%
 
Finally let us look at $\AStTL{4}{i}$. Its amplitude in $\langle\phi_{p/2}\phi_{\bar{p}/2}\phi_{p/2}\phi_{\bar{p}/2}\rangle$ comes entirely from the term $\tilde{P}_{abab}^{D_{1+{p\over 2}}}$, since by \eqref{paper1mainres} $P_{aaaa}$ has no contribution from $\AStTL{4}{i}$. By the result analogous to \eqref{Aabab4-1} we then have
\begin{equation}
\tilde{A}_{abab}(\mathcal{W}_{4,i})=A_{abab}(\mathcal{W}_{4,i})\frac{M^{D_{1+{p\over 2}}}(2)+\beta_{4,i}^{(4)}M^{D_{1+{p\over 2}}}(4)}{1+\beta_{4,i}^{(4)}} \,.\label{combi}
\end{equation}
Inserting now $\beta^{(4)}_{4,i}$ from \eqref{betas} we find
\begin{equation}
{M^{D_{1+{p\over 2}}}(4)\over M^{D_{1+{p\over 2}}}(2)}={3Q-10\over Q^2-4Q+2}=-{1\over\beta_{4,i}^{(4)}} \,, \label{miracle}
\end{equation}
so indeed (\ref{combi}) vanishes exactly.

\section{Comparison with the results of \cite{Picco:2016ilr,Migliaccio:2017dch,Ribault:2018jdv,Picco:2019dkm,Ribault:2019qrz}}\label{limit}

We now wish to return to the thread left behind in section~\ref{sec2-2}, namely the comparison between our approach and the one advocated
in \cite{Picco:2016ilr,Migliaccio:2017dch,Ribault:2018jdv,Picco:2019dkm,Ribault:2019qrz}. One of the principal ideas promoted originally in \cite{Picco:2016ilr,Migliaccio:2017dch} is to obtain the geometrical correlation functions in the generic $Q$-state Potts model by suitable analytic continuations from
correlations in the type $D$ minimal models. It has been argued in subsequent work \cite{Jacobsen:2018pti, Ribault:2018jdv,Picco:2019dkm} that such procedure is inaccurate and could at best provide an approximate description of the Potts geometrical correlations. Here we have provided an explanation of this issue, in particular why the geometrical correlation functions in the Potts model {\it cannot} be obtained this way, by explicitly reformulating the correlation functions of minimal models (i.e., their RSOS lattice realizations) to give them a geometric interpretation, and then directly comparing with the geometric correlations in the Potts model. We have seen in sections \ref{pseudoATL} and \ref{MM} that many of the ATL representations $\AStTL{j}{z^2}$
which were found in \cite{Jacobsen:2018pti} to provide contributions to the $s$-channel spectrum of the Potts geometrical correlations have, in fact, zero
net amplitude in the RSOS models and therefore in the continuum limit disappear from the minimal models spectra. Moreover, since the discussion so far have been formulated in a way that depends only on $Q$, the results apply to the spectrum first proposed in \cite{Picco:2016ilr}, which is an analytic continuation of the spectrum of minimal models obtained by taking the limit \eqref{PRS4pt}:
\begin{equation}\label{MMlimit}
\mathcal{M}(p,q):\;\;p,q\to\infty,\;\frac{q}{p-q}\to x,
\end{equation}
where $x$ is a finite number and the central charge \eqref{MMc} becomes \eqref{cx}. In this section we aim at further elucidating the nature of this limit, via the RSOS models of type $D$, for the purpose of making a direct comparison with results in \cite{Picco:2016ilr,Migliaccio:2017dch,Ribault:2018jdv,Picco:2019dkm,Ribault:2019qrz}.

In the case of $D_{1+{p\over 2}}$ models, the multiplicity $M^{D_{1+{p\over 2}}}(k)$ in \eqref{simpler} is well defined in the limit \eqref{MMlimit}---as witnessed
by its rewriting \eqref{ChebyshevT} as polynomials in $Q$---and we will denote it as $M^{D_{\infty}}(k)$. The diagrammatic expansions of 
$P_{aaaa}$, $\tilde{P}^{D_{1+{p\over 2}}}_{abab}$, $\tilde{P}^{D_{1+{p\over 2}}}_{abba}$ in the $s$-channel \eqref{s-channel-def} are also well defined.
By taking the corresponding limit of \eqref{Dcorre1} and \eqref{Dcorre3} it follows that 
\begin{subequations} \label{limitprob1}
\begin{eqnarray}
\underset{p\to\infty}{\hbox{Lim}}\langle \phi_{p/2}\phi_{\bar{p}/2}\phi_{\bar{p}/2}\phi_{p/2}\rangle &\propto& P_{aaaa}+\tilde{P}^{D_{\infty}}_{abba} \,, \\
\underset{p\to\infty}{\hbox{Lim}}\langle \phi_{p/2}\phi_{\bar{p}/2}\phi_{p/2}\phi_{\bar{p}/2}\rangle &\propto& P_{aaaa}+\tilde{P}^{D_{\infty}}_{abab} \,,
\end{eqnarray}
\end{subequations}
where on the right-hand side, the pseudo-probabilities are defined by using \eqref{simpler} and \eqref{pseudo_abab} with multiplicity $M^{D_{\infty}}$. They depend only on $Q$, so we have in (\ref{limitprob1}) two quantities that ressemble similar combinations in the Potts model. There is however an important difference with  the Potts model: while the probabilities (and thus their combinations) in the Potts model are expected to be smooth functions of $Q$, the combinations  in \eqref{limitprob1} have infinitely many poles at the values of $Q$ given by \eqref{DQ} which originate from the multiplicities $M^{D_{\infty}}(k)$. Due to \eqref{4ptabab}, we will in the following make the identifications
\begin{subequations}
\begin{eqnarray}
\underset{p\to\infty}{\hbox{Lim}}\langle\phi_{p/2}\phi_{\bar{p}/2}\phi_{\bar{p}/2}\phi_{p/2}\rangle&\leftrightarrow&\langle V^DV^NV^NV^D\rangle \,,\\
\underset{p\to\infty}{\hbox{Lim}}\langle\phi_{p/2}\phi_{\bar{p}/2}\phi_{p/2}\phi_{\bar{p}/2}\rangle&\leftrightarrow&\langle V^DV^NV^DV^N\rangle \,,
\end{eqnarray}
\end{subequations}
where the right-hand side now represent the four-point functions after taking the limit \eqref{MMlimit}, so as to extend \eqref{4ptabab} to generic central charges. We see then that the poles \eqref{DQ} in \eqref{limitprob1} obtained from direct lattice calculations exactly recover the poles \eqref{polesra0} and \eqref{polesra} from the CFT analysis in \cite{Ribault:2018jdv}. As was already argued in \cite{Jacobsen:2018pti}, on the
basis of examples, the richer $s$-channel spectrum \eqref{paper1mainres} for the Potts model has indeed the effect of cancelling these poles.

Now recall the conjecture \eqref{PRSfuncmod} inferred from Monte-Carlo simulations in \cite{Picco:2019dkm}. To be more specific, it was observed there that the four-point functions were given approximately by the combination of Potts probabilities
\begin{subequations} \label{limitprob2}
\begin{eqnarray}
\hbox{\underline{Conjecture in \cite{Picco:2019dkm}:}} \phantom{\qquad \qquad \ } &&\nonumber\\
\underset{p\to\infty}{\hbox{Lim}}\langle \phi_{p/2}\phi_{\bar{p}/2}\phi_{\bar{p}/2}\phi_{p/2}\rangle&\approx& \frac{1}{2}\bigg(P_{aaaa}+\frac{2}{Q-2}  P_{abba}\bigg) \,, \\
\underset{p\to\infty}{\hbox{Lim}}\langle \phi_{p/2}\phi_{\bar{p}/2}\phi_{p/2}\phi_{\bar{p}/2}\rangle&\approx& \frac{1}{2}\bigg(P_{aaaa}+\frac{2}{Q-2} P_{abab}\bigg)
\end{eqnarray}
\end{subequations}
and that these become exact at $Q=0,3,4$. In particular, near $Q=2$, the authors of \cite{Picco:2019dkm} conjectured:
\begin{subequations} \label{limitprobQ2}
\begin{eqnarray}
\hbox{\underline{Eqs. (3.34), (3.36)  in \cite{Picco:2019dkm}:}} \phantom{\qquad  \qquad \ }&& \nonumber\\
\underset{p\to\infty}{\hbox{Lim}}\langle \phi_{p/2}\phi_{\bar{p}/2}\phi_{\bar{p}/2}\phi_{p/2}\rangle&\overset{Q\to 2}{=}& \frac{1}{Q-2}P_{abba}+O(1)\,, \\
\underset{p\to\infty}{\hbox{Lim}}\langle \phi_{p/2}\phi_{\bar{p}/2}\phi_{p/2}\phi_{\bar{p}/2}\rangle&\overset{Q\to 2}{=}& \frac{1}{Q-2}P_{abab}+O(1)\,,
\end{eqnarray}
\end{subequations}
We now fix the coefficients in \eqref{limitprob1} to be $\frac{1}{2}$ ---same as \eqref{limitprob2}--- for the purpose of comparing our results with their claims.

For $Q=3$ we have
\begin{equation}
 T_{2\mathsf{m}} \left( \frac{\sqrt{3}}{2} \right) = \cos \left( \frac{\mathsf{m} \pi}{3} \right)
\end{equation}
in \eqref{ChebyshevT}, and using the identity
\begin{equation}
 \sum_{\mathsf{m}=-l}^l {2 l \choose l+\mathsf{m}} \cos^{-1} \left( \frac{\mathsf{m} \pi}{3} \right) = 2 \times 3^l
\end{equation}
%
the multipliticy \eqref{simpler} becomes independent of $k$:
\begin{equation}
 M^{D_{4}}(k)=2=\frac{2}{Q-2} \,.
\end{equation}
Therefore, for $Q=3$, \eqref{limitprob1} reduces to \eqref{limitprob2} exactly.
Meanwhile, for $Q=4$, we have $T_{2\mathsf{m}}(1) = 1$, so that \eqref{simpler} becomes simply:
\begin{equation}
M^{D_{\infty}}(k=2l)=\frac{1}{4^l}\sum_{\mathsf{m}=-l}^{l}\left(\begin{array}{c}
2l
\\l+\mathsf{m}
\end{array}\right)
=1=\frac{2}{Q-2} \,,
\end{equation}
and again one identifies \eqref{limitprob1} with \eqref{limitprob2}.

The situation with $Q \to 0$ is more subtle, since the Potts model partition function \eqref{Z_FK} itself vanishes in this case. As discussed in \cite{Jacobsen:2018pti}, one should renormalize the partition function by a factor of $Q$ to redefine it as the number of spanning trees. In the $Q\to 0$ limit, extra clusters disappear by the factors of $Q$ they carry, and therefore the only configuration contributing to $P_{aaaa}$ is a single spanning tree. The only configurations contributing to $P_{abab}$ and $P_{abba}$ are thus diagrams with $k=2$. Therefore, \eqref{limitprob1} is written explicitly as
\begin{subequations} \label{limitprobQ0}
\begin{eqnarray}
\underset{p\to\infty}{\hbox{Lim}}\langle \phi_{p/2}\phi_{\bar{p}/2}\phi_{\bar{p}/2}\phi_{p/2}\rangle &\overset{Q\to 0}{=}&\frac{1}{2}\bigg( P_{aaaa}+\frac{2}{Q-2}\sum_{D_{abba}}W_{\text{Potts}}(k=2)\bigg) \,, \\
\underset{p\to\infty}{\hbox{Lim}}\langle \phi_{p/2}\phi_{\bar{p}/2}\phi_{p/2}\phi_{\bar{p}/2}\rangle &\overset{Q\to 0}{=}&\frac{1}{2} \bigg(P_{aaaa}+\frac{2}{Q-2}\sum_{D_{abab}}W_{\text{Potts}}(k=2) \bigg)\,,
\end{eqnarray}
\end{subequations}
which agrees with \eqref{limitprob2}.


Near $Q=2$, we see from \eqref{Dweightsgen}, \eqref{pseudo_abab} and \eqref{limitprob1} that we have, for instance:
\begin{equation}
\underset{p\to\infty}{\hbox{Lim}}\langle \phi_{p/2}\phi_{\bar{p}/2}\phi_{p/2}\phi_{\bar{p}/2}\rangle\overset{Q\to 2}{=}\frac{1}{Q-2}\frac{1}{Z_{\text{Potts}}}\left(\sum_{D_{abab}}W_{\text{Potts}}(k=2)+2\sum_{D_{abab}}W_{\text{Potts}}(k=4)+\ldots\right)+O(1)\,.
\end{equation}
On the other hand, \eqref{limitprobQ2} reduces to:
\begin{equation}
\begin{aligned}
&\hbox{\underline{Diagrammatic expansion of eqs.\ (3.34), (3.36) in \cite{Picco:2019dkm}:}}\\
&\underset{p\to\infty}{\hbox{Lim}}\langle \phi_{p/2}\phi_{\bar{p}/2}\phi_{p/2}\phi_{\bar{p}/2}\rangle\overset{Q\to 2}{=}\frac{1}{Q-2}\frac{1}{Z_{\text{Potts}}}\left(\sum_{D_{abab}}W_{\text{Potts}}(k=2)+\sum_{D_{abab}}W_{\text{Potts}}(k=4)+\ldots\right)+O(1).\,
\end{aligned}
\end{equation}
The difference is
\begin{equation}
\frac{1}{Q-2}\frac{1}{Z_{\text{Potts}}}\left(\sum_{D_{abab}}W_{\text{Potts}}(k=4)+\ldots\right)+O(1),\
\end{equation}
still of order $\frac{1}{Q-2}$, but this is dominated by configurations with $k\ge4$, whose probabilities are small and are numerically challenging to properly sample.

\bigskip

\begin{figure}[t]
	\begin{center}
		\includegraphics[width=0.5\textwidth]{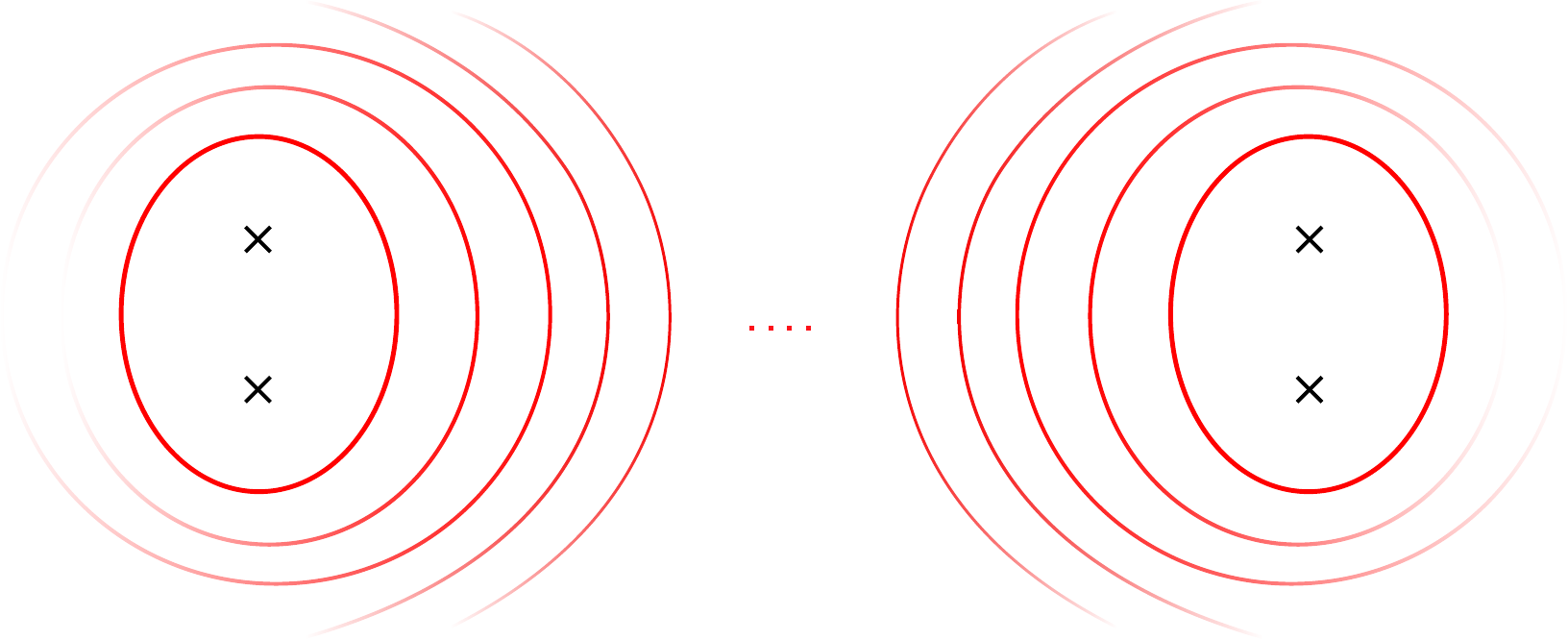}
		\caption{There are a large number of non-contractible loops surrounding the basic clusters when the distance separating them is large. Here we show this picture for $s$-channel of $\tilde{P}_{aabb}$.}\label{aabb}
	\end{center}
\end{figure}

Let us now turn to the third combination \eqref{Dcorre2}, which reads
\begin{equation}\label{limitaabb}
\underset{p\to\infty}{\hbox{Lim}}\langle \phi_{p/2}\phi_{p/2}\phi_{\bar{p}/2}\phi_{\bar{p}/2}\rangle\propto P_{aaaa}+\tilde{P}^{D_{\infty}}_{aabb} \,.
\end{equation}
While this four-point function is related to \eqref{limitprob1} by crossing, here we focus on the $s$-channel which now involves a large number of non-contractible loops separating the basic clusters in the diagrammatic expansion of $\tilde{P}_{aabb}$, as depicted in figure \ref{aabb}.
From \eqref{aabbcancel}, we have seen that only diagonal fields---i.e., the modules $\mathcal{W}_{0,\q^{2\mathsf{a}}}$---remain in the $s$-channel of this four-point function. It was claimed in \cite{Ribault:2018jdv} that in the limit \eqref{MMlimit}, the spectrum becomes continuous. Here, in terms of the ATL representations, we can formally write
\begin{equation}
\bigoplus_{\mathsf{a}=1}^{p-1} \mathcal{W}_{0,\q^{2\mathsf{a}}}
\simeq \int_0^{\pi} \mathcal{W}_{0,e^{2i\pi\theta}} \, {\rm d}\theta \,,
\end{equation} 
where the sum is replaced by an integral over a compact variable $\theta$ for generic $x$:
\begin{equation}
\theta=\frac{\mathsf{a}\pi}{x+1}.
\end{equation}
Geometrically, this corresponds to integrating over non-contractible loop weights
\begin{equation}
n_z=z+z^{-1},\;\; z=e^{i\theta}.
\end{equation}
The same picture also applies for the four-point functions
\begin{equation}\label{Alimit}
\underset{p\to\infty}{\hbox{Lim}}\langle \phi_{p/2}\phi_{p/2}\phi_{p/2}\phi_{p/2}\rangle=\underset{p\to\infty}{\hbox{Lim}}\langle \phi_{\bar{p}/2}\phi_{\bar{p}/2}\phi_{\bar{p}/2}\phi_{\bar{p}/2}\rangle\propto \frac{p}{2} P_{aaaa}+\tilde{P}^{A_{\infty}}_{abab}+\tilde{P}^{A_{\infty}}_{aabb}+\tilde{P}^{A_{\infty}}_{abba} \,,
\end{equation}
where all three channels give rise to the geometric picture of figure \ref{aabb}, with the $s$, $t$ and $u$-channels corresponding respectively to the diagrammatic expansions of $\tilde{P}_{aabb}$, $\tilde{P}_{abba}$ and $\tilde{P}_{abab}$. In the CFT, one obtains continuous spectra in all three channels. See \cite{GJS18} for a related discussion.

\section{Conclusions}
To conclude, we have first provided a graphical formulation of correlation functions  in RSOS minimal models that involves quantities which are similar but different from those in the Potts model. This formulation has allowed us to analyse in detail how the complex spectrum conjectured in  \cite{Jacobsen:2018pti} for the Potts model does, indeed, reduce to the much simpler RSOS spectrum when probabilities are replaced by ``pseudo-probabilities''. This reduction involves a series of beautiful ``facts'' (and numbers), which we do not fully understand for the moment. 

Using the geometrical formulation of correlation functions  in RSOS minimal models, we have then been able to explain what the conjecture in \cite{Picco:2016ilr} actually describes, why the ``special combinations of probabilities'' considered by these authors emerge, and to quantify how their results differ from the true Potts model result. 

We will, in our next paper \cite{bootstrappaper}, use this analysis to finally discuss the solution of the bootstrap for the Potts model itself. We obviously also plan to come back to our ``facts'' (exposed in sections \ref{sec:facts1}--\ref{sec:facts3}), which hint at  rich and largely unknown algebraic structures lurking beneath the problem of correlation functions on the lattice. It is hard not to speculate, in particular, that all the  coefficients $\alpha_{j,z^2}$, $\bar{\alpha}_{j,z^2}$, $\beta^{(k)}_{j,z^2}$ and $\gamma_{j,z^2}^{(\mathsf{a})}$ should have a natural algebraic meaning, and---especially since they can be expressed as relatively simple rational functions of $Q$---could be calculated from first principles, using maybe quantum-group \cite{PASQUIER1990523} or $S_Q$ representation theory \cite{VasseurSQ1,VasseurSQ2,CouvreurSQ3}. This, however, remains to be seen.

\section*{Acknowledgements} 

This work was supported by the ERC Advanced Grant NuQFT. We thank J. Bellet\^{e}te, A. Gainutdinov, I. Kostov, M. Kruczenski, V. Pasquier, N. Robertson, T. S. Tavares and especially S.~Ribault for many stimulating discussions. We are also grateful to S. Ribault for careful reading the manuscript and valuable comments.




\newpage

\appendix
\numberwithin{equation}{section}

\section{Proof of the partition function identity\footnote{This appendix is adapted from an unpublished work by A.D.\ Sokal and one of the authors \cite{Sokal:2008}.}}
\label{Z_Proof}

The Potts model on a connected plane graph $G=(V,E,F)$ (with vertices $V$, edges $E$ and faces $F$) is defined by the Fortuin-Kasteleyn
representation
\begin{equation}
Z_G(Q, \bv)   \;=\;
\sum_{A \subseteq E}  Q^{\kappa(A)}  \prod_{e \in A}  v_e
\;,
\label{eq1.1}
\end{equation}
where $\kappa(A)$ denotes the number of connected components in the subgraph $(V,A)$.\footnote{Here we consider the formulation in its most general form. Setting $v_e=v,\;\forall e\in A$ reduces to \eqref{Z_FK}.}

The related RSOS model is defined on the connected plane quadrangulation $\Gamma = (\scrv,\scre,\scrf)$,
where $\scrv = V \cup V^*$ and each face $f = \langle i_1 i_2 i_3 i_4 \rangle \in \scrf$ has $i_1,i_3 \in V$ and $i_2,i_4 \in V^*$
with diagonals $i_1 i_3 \in E$ and $i_2 i_4 \in E^*$. It takes values in
another finite graph $H=(X,{\bf E})$ with adjacency matrix
$\scra = (\scra_{\sigma,\sigma'})_{\sigma,\sigma'\in X}$. In the main text we focus on the case where $H$ is a Dynkin diagram $\mathcal{D}$ of type $A$ or $D$, while here in the appendix we consider the generic formulation. The RSOS partition function reads
\begin{equation}
Z^{\rm RSOS}_\Gamma   \;=\;
\sum_{\sigma \colon\, \scrv \to X}
\left( \prod_{(ij) \in \scre}  \scra_{\sigma(i) \sigma(j)} \! \right)
W(\sigma)
\;,
\label{def.Z.rsos.bis}
\end{equation}
where the sum runs over all maps $\sigma \colon\, \scrv \to X$ but the adjacency matrix restricts them to be graph
homomorphisms (neighbours map to neighbours).

The weight function $W$ is a product of local contributions
from vertices and faces:
\begin{equation}
W(\sigma)   \;=\;
\left( \prod_{i \in \scrv} W_i(\sigma_i)  \right)
\left( \prod_{F \in \scrf} W_F(\sigma_F)  \right)
\;,
\label{local.face.weights}
\end{equation}
where $\sigma_i \equiv \sigma(i)$,
and $\sigma_F$ denotes the collection of variables $\sigma_i$
for sites $i$ lying on the boundary of the face $F$.
Let $S=(S_\sigma)_{\sigma \in X}$ be an eigenvector of $\scra$ such that the entries are all nonzero,
with $\lambda$ the corresponding eigenvalue. 
Require the vertex weights to be given by
\begin{equation}
W_i(\sigma_i)  \;=\;  S_{\sigma_i}
\label{local.face.weights.Wi}
\end{equation}
and the face weights by
\begin{equation}
W_F(\sigma_{i_1}, \sigma_{i_2}, \sigma_{i_3}, \sigma_{i_4})
\;=\;
a_e \, S_{\sigma_{i_1}}^{-1} \delta(\sigma_{i_1}, \sigma_{i_3})
\,+\,
b_e \, S_{\sigma_{i_2}}^{-1} \delta(\sigma_{i_2}, \sigma_{i_4}) \;.
\label{local.face.weights.WF}
\end{equation}
In the following, we shall also need topological identity
\begin{equation}
\kappa(A)   \;=\;   |V| \,-\, |A| \,+\, c(A)
\label{eq.cyclomatic}
\end{equation}
where $c(A)$ is the cyclomatic number
(i.e., number of linearly independent cycles) of the graph $(V,A)$. Having defined our models, we now state the relation between them:

\noindent\fbox{%
	\parbox{\textwidth}{%
		\vspace*{0.1cm}
		\emph{Potts-RSOS equivalence for the partition function}
		\begin{equation}
		Z^{\rm RSOS}_{\Gamma}({\bf a}, {\bf b})
		\;=\;
		\left( \sum_{\sigma \in X}  S_{\sigma}^2 \! \right) \, \lambda^{-|V|}
		\left( \prod_{e \in E} b_e \! \right)
		Z_G(\lambda^2, \lambda {\bf a}/{\bf b})
		\;.
		\label{eq.theorem.2}
		\end{equation}
	}%
}

\proof
%
%
Insert \reff{local.face.weights.Wi}/\reff{local.face.weights.WF}
into \reff{local.face.weights}
and expand out the product over faces $F$ of $\Gamma$,
which are in one-to-one correspondence with edges $e \in E$.
Each term in this expansion can be associated
to a subset $A \subseteq E$
and the complementary subset $A^*$ as follows:
\begin{itemize}
	\item If the term contains the factor $a_e$, then $e \in A$ and
	hence $e^* \notin A^*$.
	\item If the term contains the factor $b_e$, then
	$e \notin A$ and hence $e^* \in A^*$.
\end{itemize}
This gives a formulation of the partition function in terms of cluster configurations. 
On each connected component (cluster) $\scrc$ of the graph
$(V \cup V^*, A \cup A^*)$, the $\sigma$ value must be constant
(let us call it simply $\sigma_\scrc$).
Such a configuration then gets a weight
\begin{subeqnarray}
	& &
	\left( \prod_{i \in V \cup V^*}  S_{\sigma_i} \! \right)
	\left( \prod_{e \in A} a_e \! \right)
	\left( \prod_{e^* \in A^*} b_e \! \right)
	\left( \prod_{{\rm components}\, \scrc}
	S_{\sigma_\scrc}^{-|{\rm edges}(\scrc)|}     \right)
	\\[2mm]
	& & \qquad\qquad=\;
	\left( \prod_{e \in A} a_e \! \right)
	\left( \prod_{e^* \in A^*} b_e \! \right)
	\left( \prod_{{\rm components}\, \scrc}
	S_{\sigma_\scrc}^{|{\rm vertices}(\scrc)|-|{\rm edges}(\scrc)|}
	\right)
	\\[2mm]
	& & \qquad\qquad=\;
	\left( \prod_{e \in A} a_e \! \right)
	\left( \prod_{e^* \in A^*} b_e \! \right)
	\left( \prod_{{\rm components}\, \scrc}
	S_{\sigma_\scrc}^{1 - c(\scrc)}
	\right)
	\;,\label{cluster_weight}
\end{subeqnarray}
where the last equality used \reff{eq.cyclomatic} with $k(\scrc)=1$ per component,
and $c(\scrc)$ here denotes the cyclomatic number of the chosen component $\scrc$.

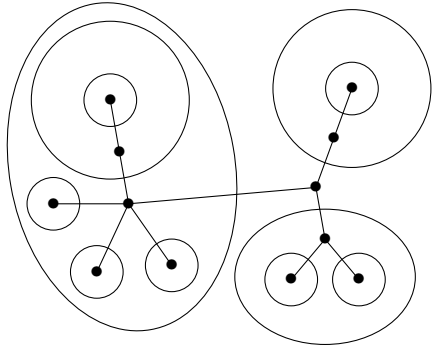
\begin{figure}
	\center
	$
	\begin{tikzpicture}
	\coordinate  (leftmid) at (0, 0) {};
	\coordinate  (leftup1) at ([shift=(100:0.7cm)]leftmid) {};
	\coordinate  (leftup2) at ([shift=(100:0.7cm)]leftup1) {};
	\coordinate  (leftdown1) at ([shift=(-115:1cm)]leftmid) {};
	\coordinate  (leftdown2) at ([shift=(-55:1cm)]leftmid) {};
	\coordinate  (leftdown3) at ([shift=(-180:1cm)]leftmid) {};
	\coordinate  (rightmid) at ([shift=(5:2.5cm)]leftmid) {};
	\coordinate  (rightup1) at ([shift=(70:0.7cm)]rightmid) {};
	\coordinate  (rightup2) at ([shift=(70:0.7cm)]rightup1) {};
	\coordinate  (rightdownmid) at ([shift=(-80:0.7cm)]rightmid) {};
	\coordinate  (rightdown1) at ([shift=(-50:0.7cm)]rightdownmid) {};
	\coordinate  (rightdown2) at ([shift=(-130:0.7cm)]rightdownmid) {};
	\node[] () at (leftmid) {\textbullet};
	\node[] () at (leftup1) {\textbullet};
	\node[] () at (leftup2) {\textbullet};
	\node[] () at (leftdown1) {\textbullet};
	\node[] () at (leftdown2) {\textbullet};
	\node[] () at (leftdown3) {\textbullet};
	\node[] () at (rightmid) {\textbullet};
	\node[] () at (rightup1) {\textbullet};
	\node[] () at (rightup2) {\textbullet};
	\node[] () at (rightdownmid) {\textbullet};
	\node[] () at (rightdown1) {\textbullet};
	\node[] () at (rightdown2) {\textbullet};
	\draw[] (leftmid) -- (rightmid);
	\draw[] (leftmid) -- (leftup1);
	\draw[] (leftup1) -- (leftup2);
	\draw[] (leftmid) -- (leftdown1);
	\draw[] (leftmid) -- (leftdown2);
	\draw[] (leftmid) -- (leftdown3);
	\draw[] (rightmid) -- (rightup1);
	\draw[] (rightup1) -- (rightup2);
	\draw[] (rightmid) -- (rightdownmid);
	\draw[] (rightdownmid) -- (rightdown1);
	\draw[] (rightdownmid) -- (rightdown2);
	\draw[] (leftup2)  circle [radius=0.35cm];
	\draw[] (leftup2)  circle [radius=1.05cm];
	\draw[] (leftdown1)  circle [radius=0.35cm];
	\draw[] (leftdown2)  circle [radius=0.35cm];
	\draw[] (leftdown3)  circle [radius=0.35cm];
	\draw[] (rightup2)  circle [radius=0.35cm];
	\draw[] (rightup2)  circle [radius=1.05cm];
	\draw[] (rightdown1)  circle [radius=0.35cm];
	\draw[] (rightdown2)  circle [radius=0.35cm];
	\draw[rotate around={100:(leftmid)}] ([shift=(0:0.5cm)]leftmid)  circle [x radius=2.2cm, y radius=1.5cm];
	\draw[] ([shift=(-90:0.5cm)]rightdownmid)  circle [x radius=1.2cm, y radius=0.9cm];
	\end{tikzpicture}
	$
	\caption{The tree ${\sf T} $ associated to a cluster configuration on the sphere. Each vertex in the tree corresponds to a cluster, each edge corresponds to a loop separating two clusters.}\label{cluster_tree}
\end{figure}

Now form the graph ${\sf T} = ({\sf V}, {\sf E})$
whose vertices are the connected components $\scrc$
of $(V \cup V^*, A \cup A^*)$
and which puts an edge between $\scrc_1$ and $\scrc_2$
whenever at least one vertex of $\scrc_1$ is adjacent in $\Gamma$ to at least one vertex of $\scrc_2$. One observes that ${\sf T}$ is a tree, and that a component $\scrc$ of cyclomatic number $c$ is adjacent in ${\sf T}$ to exactly $c+1$ other components (namely its exterior and $c$ cycles on the interior). Therefore, $ S_{\sigma_\scrc}^{1 - c(\scrc)}
=  S_{\sigma_\scrc}^{2 - d_{\sf T}(\scrc)}$
where $d_{\sf T}(\scrc)$ is the degree of $\scrc$ in ${\sf T}$. An example of a tree ${\sf T}$  associated to a cluster configuration is shown in figure \ref{cluster_tree}.

\bigskip

\begin{absolutelynopagebreak}
	To proceed we need the following lemma:
	\begin{lemma}
		\label{lemma.tree}
		Let ${\sf T} = ({\sf V}, \vec{{\sf E}})$ be a rooted tree
		whose edges are directed towards the root vertex $\rho \in {\sf V}$.
		For each $i \in {\sf V}$, let $d_{\rm in}(i)$ (resp.\ $d_{\rm out}(i)$)
		denote the in-degree (resp.\ out-degree) of $i$ in ${\sf T}$.
		(Thus, $d_{\rm out}(i) = 1$ for all $i \neq \rho$, and $d_{\rm out}(\rho) = 0$.)
		Let $M$ be a matrix indexed by a finite set $X$,
		and let $S$ be an eigenvector of $M$ with eigenvalue $\lambda$.
		Then, for each $\tilde{\sigma} \in X$, we have
		\begin{equation}
		\sum_{\begin{scarray}
			\sigma \colon\, \mathsf{V} \to X \\
			\sigma_\rho = \tilde{\sigma}
			\end{scarray}}
		\left( \prod_{(ij) \in \vec{{\sf E}}} M_{\sigma_j,\sigma_i} \right)
		\left( \prod_{i \in {\sf V}}  S_{\sigma_i}^{d_{\rm out}(i) - d_{\rm in}(i)}
		\right)
		\;=\;
		\lambda^{|{\sf V}|-1}  \;.
		\label{eq.lemma.tree.1}
		\end{equation}
	\end{lemma}
\end{absolutelynopagebreak}

\proof
The proof of \reff{eq.lemma.tree.1} is by induction on
the cardinality of ${\sf V}$.
If $|{\sf V}| = 1$ (i.e., ${\sf T}$ consists of the root vertex and no edges),
then \reff{eq.lemma.tree.1} is trivial.
If $|{\sf V}| > 1$, then ${\sf T}$ contains
at least one leaf vertex $i \neq \rho$,
for which $d_{\rm out}(i) = 1$ and $d_{\rm in}(i) = 0$.
Letting $j$ be the parent of $i$
we can perform the sum over $\sigma_i$ using $M S = \lambda  S$,
yielding $\lambda  S_{\sigma_j}$.
This extra factor of $ S_{\sigma_j}$ is exactly what we need
to apply the inductive hypothesis to the tree ${\sf T} \setminus i$,
in which $j$ has in-degree one lower than it does in ${\sf T}$.
\qed

In particular, if $M$ is a symmetric matrix, as is the case for our adjacency matrix, we can ignore the orientations of the edges. We then have
\begin{equation}
\sum_{\begin{scarray}
	\sigma \colon\, \mathsf{V} \to X \\
	\sigma_\rho = \tilde{\sigma}
	\end{scarray}}
\left( \prod_{(ij) \in {\sf E}} M_{\sigma_i,\sigma_j} \right)
\left( \prod_{i \in {\sf V}}  S_{\sigma_i}^{2-d(i)} \right)
\;=\;
S_{\tilde{\sigma}}^2  \lambda^{|{\sf V}|-1}
\label{eq.lemma.tree.2}
\end{equation}
where $d(i)$ is the total degree of the vertex $i$;
the result is independent of the choice of the root vertex $\rho$. This result follows immediately from Lemma \ref{lemma.tree},
since $d_{\rm out}(i) = 1$ for all $i \neq \rho$
and $d_{\rm out}(\rho) = 0$.

We now resume the proof of the main result \reff{eq.theorem.2}.
Using \reff{eq.lemma.tree.2}
to sum over RSOS configurations satisfying $\sigma_\rho = \tilde{\sigma}$
we obtain
\begin{equation}
S_{\tilde{\sigma}}^2 \,  \lambda^{\kappa(A) + \kappa(A^*) - 1}
\left( \prod_{e \in A} a_e \! \right)
\left( \prod_{e^* \in A^*} b_e \! \right)
\;.
\end{equation}
But by \reff{eq.cyclomatic} we have
\begin{equation}
\kappa(A) + \kappa(A^*) - 1
\;=\;
\kappa(A) + c(A)
\;=\;
2 \kappa(A) + |A| - |V|
\;,
\end{equation}
which proves \reff{eq.theorem.2} by summing over $\tilde{\sigma} \in X$.
\qed

\section{RSOS $N$-point functions}\label{NptRSOS}

In this appendix we shall focus on the equivalence between the RSOS model and the loop model defined on the medial
graph ${\cal M}(G) = \Gamma^*$, i.e. the dual of the plane quandrangulation. 
%
%
%
The loops are shown in figure \ref{cluster_tree} together with the tree ${\sf T} = ({\sf V}, {\sf E})$, which shall play an important role in the following. In terms of loops and trees, the essential part of the result (\ref{eq.theorem.2}) is that
\begin{itemize}
	\setlength\itemsep{0em}
        \item The expansion of the local weights in the RSOS model followed by the summation over heights, subject to the constraints imposed by the adjacency matrix $\scra$, leads to a corresponding formulation in terms of clusters on $G$, or equivalently to a completely packed loop model on ${\cal M}(G)$.
	\item Each loop gets a weight $\lambda$ equal to the eigenvalue of the chosen eigenvector $S$ of the adjacency matrix $\scra$.
	These weights are due to the recurrence relation on the tree ${\sf T}$ that serves to eliminate it starting from the leaves. 
	\item There is an extra factor $\sum_{\sigma \in X}  S_{\sigma}^2$ coming from the summation over the root vertex. Henceforth we choose to normalize all eigenvectors of $\scra$, so that this factor is 1.
\end{itemize}

We label the different eigenvectors and eigenvalues of $\scra$ as $ S^\sigma_{(r)}$ and $\lambda_{(r)}$, with $r = 1,2,\ldots,{\rm dim}\,\scra$.
%
Below, we shall also refer to the $S_{(r)}$ as states, calling $S\equiv S_{(r_{\text{id}})}$ the identity state.\footnote{In the RSOS lattice formulation of minimal models $\mathcal{M}(p,q)$, we have $r_{\text{id}}=p-q$ (see section~\ref{PottsRSOS}).}
$\scra$ is real and symmetric, so the matrix ${\cal O}$ formed by its normalized eigenvectors is orthogonal. Both the rows and columns of ${\cal O}$ provide an orthonormal basis of $\R^{{\rm dim}\,\scra}$:
%
\be
\sum_\sigma  S_{(r_1)}^\sigma S_{(r_2)}^\sigma = \delta_{r_1,r_2} \label{ortho_k}
\ee
and
\be
\sum_r  S^{\sigma_1}_{(r)}  S^{\sigma_2}_{(r)} = \delta_{\sigma_1,\sigma_2} \label{ortho_x} \,.
\ee

The definition of order parameters from the normalized eigenvectors extends that of \cite{Pasquier:1986jc,Pasquier:1987xj} (in which $S_{(r_{\text{id}})}$ is the Perron-Frobenius vector) to any $r_{\text{id}}$ such that $S_{\sigma} := S^{\sigma}_{(r_{\text{id}})}\ne0$:
\begin{equation} \label{def_order_param}
\phi_r(i)=\frac{S^{\sigma_i}_{(r)}}{S_{\sigma_i}}\,.
\end{equation}
We mark $N$ vertices $i \in V$ by a label $r_i \in \{1,2,\ldots,{\rm dim}\,\scra\}$. The corresponding $N$-point correlation functions are given by 
insertions of $\phi_{r_i}(i)$, which amounts to replacing the vertex weights $S_{\sigma_i}$ of \eqref{local.face.weights.Wi} by $ S^{\sigma_i}_{(r_i)}$ at each marked vertex. 
The corresponding weights in the loop model will depend on how the marked vertices are situated in the tree ${\sf T}$. In particular, in the edge subset expansion, it will be possible for a given vertex in ${\sf V}$ to
be marked several times, if the corresponding marked vertices in $V$ are situated in the same cluster.

In the section \ref{UpTo3} below we shall revisit the inductive argument on ${\sf T}$, first for the partition function, and then generalising it to all $N$-point
correlation functions in the RSOS model with $N \le 3$. 
We shall then describe the case of $N>3$ from a slightly different perspective in section \ref{Higher_N}.

\subsection{Explicit computation on ${\sf T}$ up to $N=3$}\label{UpTo3}

A common feature of the proofs in this section is that we have the liberty to chose the root of ${\sf T} = ({\sf V}, {\sf E})$ at any vertex in ${\sf V}$. Certain calculations can be done in different ways, depending on the choice of the root, but the result will of course be independent of that choice. This independence is guaranteed by certain identities that we shall establish along the way.

\subsubsection*{Partition function}

Chose any $\rho \in {\sf V}$ as the root of ${\sf T}$. To sum out a leaf $i \in {\sf V}$, let $j$ denote its (unique) parent.
The leaf has degree $d_i = 1$, and let $d_j$ denote the degree of the parent vertex before the summation. The inductive argument made in Lemma~\ref{lemma.tree} then
hinges on the eigenvalue identity for the adjacency matrix $\scra$
\be
\sum_{\sigma_i} \scra_{\sigma_j,\sigma_i}  S_{\sigma_i}  S_{\sigma_j}^{2-d_j} = \lambda  S_{\sigma_j}^{2-(d_j-1)} \,, \label{usual_induction}
\ee
where $d_j-1$ is now the degree of $j$ after the leaf has been summed out. This produces a weight $\lambda$ per loop. After summing out inductively all the leaves, only the root vertex $\rho$ will remain. Since $d_\rho = 0$ the corresponding sum
produces
\be
\sum_{\sigma_\rho}  S_{\sigma_\rho}^2 = 1 \,,
\ee
where we have used the normalisation of the eigenvectors.

\subsubsection*{One-point function}

Take the marked point $i_1 := \rho \in {\sf V}$ as the root of ${\sf T}$, and let $r_1$ denote the corresponding label of $ S_{(r_1)}$.
The argument for the leaves can be taken over from the computation of the partition function, producing again a factor $\lambda^{|{\sf V}|}$.
At the root we get an extra factor
\be
\sum_{\sigma_\rho} S_{\sigma_\rho}^2 \phi_{r_1}(i_1) = \sum_{\sigma_\rho}  S_{\sigma_\rho}  S^{\sigma_\rho}_{(r_1)} = \delta_{r_1,r_{\text{id}}} \,,
\ee
where we have used the definition \eqref{def_order_param} of the order parameters, followed by the orthogonality (\ref{ortho_k}) of the eigenvectors. (Recall $S\equiv S_{(r_{\text{id}})}$.)

\subsubsection*{Two-point function}

With more than one point, the regrouping of marked vertices in $V$ into connected components in $(V \cup V^*,A \cup A^*)$
will induce a set partition of the marked vertices. Specifically, with $N=2$ marked vertices, we shall denote by $\{12\}$ the situation in which the two marked vertices $i_1,i_2 \in V$ correspond to the same vertex of the tree ${\sf T}$
, and by
$\{1\}\{2\}$ the situation in which they correspond to two distinct vertices. In either case, the corresponding labels
of the eigenvectors are denoted $r_1$ and $r_2$.

We first treat the case of the partition $\{12\}$. We take the marked points to be at the root
, a choice that we write for short as $\rho=\{12\}$. As before we get a factor $\lambda^{|{\sf V}|}$ from the summation over the leaves, while at the root we obtain
\be
\sum_{\sigma_\rho} S_{\sigma_\rho}^2 \phi_{r_1}(i_1) \phi_{r_2}(i_2) = \sum_{\sigma_\rho}  S^{\sigma_\rho}_{(r_1)}  S^{\sigma_\rho}_{(r_2)} = \delta_{r_1,r_2} \,. \label{2ptrootsum}
\ee

In the case of the partition $\{1\}\{2\}$ we take $i_2 := \rho \in {\sf V}$ to be the root of ${\sf T}$. The other marked point $i_1$ corresponds to
a different vertex in ${\sf V}$. Since ${\sf T}$ is a tree, there is a unique path ${\sf P}$ from $i_1$ to $i_2$. We denote by $\ell$ the number of edges in ${\sf P}$.
All the vertices not in ${\sf P}$ can be summed out using (\ref{usual_induction}), giving rise to a total factor 
of $\lambda^{|{\sf V}|-|{\sf P}|}$.
Once this has been done, we must sum over the vertices remaining in ${\sf P}$. We start by summing over $i_1$. Let $j$ denote its parent
in ${\sf T}$, of degree $d_j$. 
We get that \eqref{usual_induction} must be replaced by
%
\begin{eqnarray}
 \sum_{\sigma_{i_1}} \scra_{\sigma_j,\sigma_{i_1}} S_{\sigma_{i_1}} S_{\sigma_j}^{2-d_j} \phi_{r_1}(i_1) &=& \sum_{\sigma_1} \scra_{\sigma_j,\sigma_1} S^{\sigma_1}_{(r_1)} S_{\sigma_j}^{2-d_j} \nonumber \\
 &=& \lambda_{(r_1)} \left( S^{\sigma_j}_{(r_1)} / S_{\sigma_j} \right) S_{\sigma_j}^{2-(d_j-1)} \nonumber \\
 &=& \lambda_{(r_1)} \phi_{r_1}(j) S_{\sigma_j}^{2-(d_j-1)} \,.
\label{2ptreduc}
\end{eqnarray}
This has the effect of producing a factor $\lambda_{(r_1)}$ corresponding to the summed-out vertex $i \in {\sf P}$, and moving the marked weight to the parent vertex $j$. Therefore the inductive argument can be continued until we have reduced ${\sf P}$ to the root vertex $\rho = i_2$, and summing
over this provides the same factor (\ref{2ptrootsum}) as before. 
In total we obtain
\be
\lambda^{|{\sf V}|-\ell} \lambda_{(r_1)}^{\ell} \delta_{r_1,r_2} \,. \label{2ptres}
\ee
We could divide by the partition function $Z$ to write the correlation function as
\be
\left( \frac{\lambda_{(r_1)}}{\lambda} \right)^\ell \delta_{r_1,r_2} \,;
\ee
however, in what follows we prefer to keep the correlation functions un-normalised as in (\ref{2ptres}).

The result in (\ref{2ptres}) can be summarised by saying that any loop that separates the two marked vertices $i_1,i_2 \in V$ has its weight modified from $\lambda$ to $\lambda_{(r_1)}$. In addition there is a factor $\delta_{r_1,r_2}$, so it is equivalent to say that the weight is modified to $\lambda_{(r_2)}$. This equivalence agrees naturally with the possibility to turn a loop inside out on the Riemann sphere. We also notice that the case of the partition $\{12\}$ emerges as a particular case of the $\{1\}\{2\}$ computation; it suffices to set $\ell = 0$ in \eqref{2ptres}. This is a general observation that will carry over to appropriate $N$-point functions with $N>2$.

\subsubsection*{Three-point functions}\label{Three-point}


We begin by considering the case of the partition $\{123\}$. Take the root $\rho = \{123\}$. We get the factor $\lambda^{|{\sf V}|}$ as usual,
meaning the all loop weights are unchanged. At the root we obtain the factor
\be
\sum_{\sigma_\rho}  S^{\sigma_\rho}_{(r_1)}  S^{\sigma_\rho}_{(r_2)}  S^{\sigma_\rho}_{(r_3)}  S_{\sigma_\rho}^{-1} =: C_{r_1,r_2,r_3} \,. \label{structfac}
\ee
By definition the structure constant $C_{r_1,r_2,r_3}$ is symmetric in all three indices.
Note also that \eqref{structfac} correctly contains the two-point function as a special case, since
\be
 C_{r_1,r_2,r_{\rm id}} = \sum_{\sigma_\rho}  S_{(r_1)}^{\sigma_\rho}  S_{(r_2)}^{\sigma_\rho}  = \delta_{r_1,r_2} \,,
 \label{three_to_two_reduc}
\ee
where we have used (\ref{ortho_k}).

Next consider the partition $\{12\}\{3\}$. The easiest way to compute this correlation function is to take $\rho = \{12\}$ and $i = \{3\}$.
We strip off the leaves of ${\sf T}$ as usual, ending up with the path graph ${\sf P}$ with $\ell$ edges and extremities $i$ and $r$. To perform the
sum over $\sigma_i$, we can take over the inductive argument (\ref{2ptreduc}) from the computation of the two-point function to get a factor $\lambda_{(r_3)}^\ell$ by undoing ${\sf P}$. Finally at the root we get $C_{r_1,r_2,r_3}$ by the same
calculation as above.

We can redo this computation the other way around by taking the root $\rho = \{3\}$ and $i = \{12\}$. We shall use
the following lemma:
\begin{lemma} \label{lemma.identity1}
	Let $\scra$ be a real and symmetric matrix, $S_{(r)}$ (with $r=1,2,\ldots,{\rm dim}\,\scra$) its normalized eigenvectors with corresponding eigenvalues $\lambda_{(r)}$, and $S$ a distinguished eigenvector whose entries are all non-zero. For $C_{r_1,r_2,r_3}$ as defined by \eqref{structfac} we have
	\be
	\sum_{\sigma_i} \scra_{\sigma_j,\sigma_i}  S^{\sigma_i}_{(r_1)}  S^{\sigma_i}_{(r_2)}  S_{\sigma_i}^{-1} =
	\sum_r \lambda_{(r)} C_{r_1,r_2,r}  S^{\sigma_j}_{(r)} \,. \label{identity1}
	\ee
\end{lemma}
\proof
Since the eigenvectors $ S_{(\tilde{r})}$ form a basis of $\R^{{\rm dim}\,\scra}$, this identity can be proven by showing the the left-hand and right-hand
sides have the same projections on each of these vectors. First consider the left-hand side:
\be
\sum_{\sigma_j}  S^{\sigma_j}_{(\tilde{r})} [{\rm l.h.s.}] =
\lambda_{(\tilde{r})} \sum_{\sigma_i}  S^{\sigma_i}_{(\tilde{r})}  S^{\sigma_i}_{(r_1)}  S^{\sigma_i}_{(r_2)}  S_{\sigma_i}^{-1} = \lambda_{(\tilde{r})} C_{r_1,r_2,\tilde{r}} \,,
\ee
where the first equality uses the symmetry of $\scra$. Similarly, the projection of the right-hand side reads:
\be
\sum_{\sigma_j}  S^{\sigma_j}_{(\tilde{r})} [{\rm r.h.s.}] =
\sum_r \lambda_{(r)} C_{r_1,r_2,r} \sum_{\sigma_j}  S^{\sigma_j}_{(r)}  S^{\sigma_j}_{(\tilde{r})} =
\sum_r \lambda_{(r)} C_{r_1,r_2,r} \delta_{r,\tilde{r}} = \lambda_{(\tilde{r})} C_{r_1,r_2,\tilde{r}} \,,
\ee
proving (\ref{identity1}).\qed

Lemma \ref{lemma.identity1} is exactly what is needed in the inductive proof in order to replace the marking $\{12\}$ from vertex $i$ by a
marking $(r)$ of the parent vertex $j$. At the same time we obtain a sum over all $r$, a structure constant $C_{r_1,r_2,r}$, and a factor
$\lambda_{(r)}$. This can be physically interpreted as the fusion of the two states $ S_{(r_1)}$ and $ S_{(r_2)}$ into the superposition
of all intermediate channels $ S_{(r)}$, as will be discussed further in section \ref{Higher_N}.

Undoing successive vertices of ${\sf P}$ we get more factors of $\lambda_{(r)}$, and at the root we end up with
\be
\sum_r \lambda_{(r)}^\ell C_{r_1,r_2,r} \sum_{\sigma_\rho}  S^{\sigma_\rho}_{(r)}  S^{\sigma_\rho}_{(r_3)} =
\sum_r \lambda_{(r)}^\ell C_{r_1,r_2,r} \delta_{r,r_3} = \lambda_{(r_3)}^\ell C_{r_1,r_2,r_3} \,, \label{undo_P}
\ee
which is the same result as obtained by the first, easy computation. We shall need Lemma \ref{lemma.identity1} further below.

We finally consider the partition $\{1\}\{2\}\{3\}$. The three marked points can be positioned in various ways on the tree ${\sf T}$.
Once all unmarked leaves have been undone (giving rise to factors of $\lambda$), all arrangements are special cases of
the situation where ${\sf T}$ has been reduced to a three-star graph ${\sf S}_{\ell_1,\ell_2,\ell_3}$
\begin{equation}
\begin{tikzpicture}
\newcommand\circrad{1.2}
\newcommand\scaling{2}
\coordinate (A) at (0, 0) {};
\coordinate (B) at (1*\scaling, 0) {};
\coordinate (C) at (0.5*\scaling, 0.866*\scaling) {};
\coordinate (D) at (0.5*\scaling, 0.289*\scaling) {};
\node(A1) at (A) {$\times$};
\node(B1) at (B) {$\times$};
\node(C1) at (C) {$\times$};
\filldraw (A) circle (\circrad pt);
\filldraw (B) circle (\circrad pt);
\filldraw (C) circle (\circrad pt);
\filldraw (D) circle (\circrad pt);
\node(D1) at ([shift=(-90:8pt)]D) {$_{i_0}$};
\node(A2) at ([shift=(-180:8pt)]A) {$_{r_1}$};
\node(B2) at ([shift=(-0:8pt)]B) {$_{r_3}$};
\node(C2) at ([shift=(90:8pt)]C) {$_{r_2}$};
\node(l1) at ([shift=(90:6pt)]$(A)!0.45!(D)$) {$_{\ell_1}$};
\node(l3) at ([shift=(90:6pt)]$(B)!0.4!(D)$) {$_{\ell_3}$};
\node(l2) at ([shift=(0:7pt)]$(C)!0.5!(D)$) {$_{\ell_2}$};
\draw[] (A) -- ($(A)!0.33!(D)$);
\draw[dotted] ($(A)!0.33!(D)$) -- ($(A)!0.67!(D)$);
\draw[] ($(A)!0.67!(D)$) -- (D);
\draw[] (B) -- ($(B)!0.33!(D)$);
\draw[dotted] ($(B)!0.33!(D)$) -- ($(B)!0.67!(D)$);
\draw[] ($(B)!0.67!(D)$) -- (D);
\draw[] (C) -- ($(C)!0.33!(D)$);
\draw[dotted] ($(C)!0.33!(D)$) -- ($(C)!0.67!(D)$);
\draw[] ($(C)!0.67!(D)$) -- (D);
\end{tikzpicture} \label{three-star}
\end{equation}
with marked points $\{1\}$, $\{2\}$ and
$\{3\}$ positioned at each extremity of the branches which have respective lengths $\ell_1$, $\ell_2$ and $\ell_3$ as indicated in \eqref{three-star}. The three branches
meat at a central vertex $i_0$ that we take as the root, $\rho := i_0$.
In this configuration, it is simple to undo the branches, giving rise
to a factor $\lambda_{(r_1)}^{\ell_1} \lambda_{(r_2)}^{\ell_2} \lambda_{(r_3)}^{\ell_3}$. At the end, we sum over the root, which reduces to the computation (\ref{structfac}) done for the $\{123\}$ partition, and leads to a contribution $C_{r_1,r_2,r_3}$.

We see that all cases of three-point functions are special cases of the ${\sf S}_{\ell_1,\ell_2,\ell_3}$ arrangement, provided we allow some or
all of the branch lengths to be zero. The general result for the three-point function can be summarised as
\be\label{three-point}
\lambda_{(r_1)}^{\ell_1} \lambda_{(r_2)}^{\ell_2} \lambda_{(r_3)}^{\ell_3} C_{r_1,r_2,r_3} \,.
\ee
In other words, apart from the structure constant, there is a factor $\lambda_{(r_j)}$ for each loop that separates point $j \in \{1,2,3\}$
from the other two points. Each loop that surrounds none or all of the points meanwhile gets the usual weight $\lambda$. This is
very similar to the setup in \cite{Ikhlef:2015eua} for the non-unitary loop model with generic loop weights. We note once again that for any given loop on the Riemann sphere, we may freely choose which of the two regions separated by the loop to consider as the inside.

\subsection{Higher $N$-point functions and Feynman rules for the trees}\label{Higher_N}

To consider general $N$-point functions it is convenient to shift perspective, making links with the formulation in \cite{Kostov:1989eg} in terms of ``Feynman rules'' for the relevant trees. To obtain these rules, let us consider weight of a single cluster in the cluster expansion of the partition function. One important feature of the arguments below can be summarized as follows: As seen in \eqref{cluster_weight}, any cluster ${\cal C}$ comes with a weight that depends on its cyclomatic number $c({\cal C})$ as $ W_{\cal C} = S^{1-c({\cal C})}$. An $n$-vertex---i.e., a vertex $i \in {\sf T}$ with $d_i = n$---corresponds to a cluster that is adjacent to $n$ other clusters: the cluster surrounding it and $c({\cal C}) = n-1$ cycles on the interior. We shall call the latter circuits in the following. 
When considering the tree corresponding to a given cluster configuration, we can decompose an $n$-vertex ($n>3$) into $S^{2-n}=S^{-1}\times S^{-1}\times...\times S^{-1}$, where each of the $(n-2)$ factor $S^{-1}$ corresponds to a 3-vertex. We will represent this decomposition with ``symbolic loops'', as shown below. Such loops are also used to handle other situations, such as when several marked points are in the same cluster. After taking care of these details, any tree for any value of $N$ will be computed from the Feynman rules stated in List \ref{Feynman}, which generalize those in figure~5 of \cite{Kostov:1989eg} to the case with marked points where order parameters $S_{(r)}/S$ are inserted.


\begin{rules}
	\begin{minipage}{0.45\textwidth} 
		\textbf{Position space:}
		\begin{itemize}
			\setlength\itemsep{0em}
			\item each edge corresponds to a propagator $\scra_{\sigma,\sigma'}$
			\item each 1-vertex (leaf) gives
			\begin{enumerate}
				\setlength\itemsep{0em}
				\item $S^{\sigma}_{(r_{\text{id}})}$ if it has no marked points
				\item $S^{\sigma}_{(r)}$ if it has one marked point corresponding to the state $S_{(r)}$
				\item if there are several marked points we first fuse the states, see Sections \ref{fusion_of_states}, \ref{n_vertices}
			\end{enumerate}
			\item each 2-vertex gives 1
			\item each 3-vertex gives $S_\sigma^{-1}$
			\item we sum over any internal lines
			\item any marked point that is not on a leaf will get fused into the tree, see Sections \ref{fusion_of_states}, \ref{n_vertices}
		\end{itemize}
	\end{minipage}%
	\hspace*{0.5cm}
	\begin{minipage}{0.45\textwidth} 
		\textbf{Momentum space:}
		\begin{itemize}
			\setlength\itemsep{0em}
			\item each edge carries a label $r'$ and corresponds to a propagator $\lambda_{(r')}$
			\item each 1-vertex (leaf) gives
			\begin{enumerate}
				\setlength\itemsep{0em}
				\item $\delta_{r',r_{\text{id}}}$ if it has no marked points
				\item $\delta_{r',r}$ if it has one marked point corresponding to the state $S_{(r)}$
				\item if there are several marked points we first fuse the states, see Sections \ref{fusion_of_states}, \ref{n_vertices}
			\end{enumerate}
			\item each 2-vertex gives $\delta_{r,r'}$ 
			\item each 3-vertex gives $C_{r_1,r_2,r_3}$
			\item we sum over any internal lines
			\item any marked point that is not on a leaf will get fused into the tree, see Sections \ref{fusion_of_states}, \ref{n_vertices}
		\end{itemize} 
	\end{minipage}
	\vspace*{0.1cm}
	\caption{Feynman rules for RSOS models}\label{Feynman}
\end{rules}

\subsubsection{Fusion of states}\label{fusion_of_states}
Let us consider a cluster configuration on a sphere where we take $N$ marked vertices, some of these possibly belonging to the same cluster. We first establish a convenient pictorial reformulation of some results already seen in the sections above. We have seen that \eqref{2ptreduc} lets us recursively sum out clusters, starting at the leaves and gaining a factor $\lambda_{(r)}$ any time we cross a loop. We repeat this equation here for convenience:
%
%
\begin{equation}
\sum_{\sigma'} \scra_{\sigma,\sigma'} S^{\sigma'}_{(r)} S_{\sigma}^{2-d} =\lambda_{(r)} \left( S^{\sigma}_{(r)} / S_{\sigma} \right) S_{\sigma}^{2-(d-1)} \,.
\end{equation}
%
%
%
%
Within any given cluster of $1-\tilde{c}$ circuits, that includes any number $m$ of marked points, we can use the similar looking but trivial identity
%
\begin{equation}
\sum_{\sigma'} \delta_{\sigma,\sigma'}  \prod_{i=1}^k \left( S^{\sigma'}_{(r_i)} / S_{\sigma'} \right)  S_{\sigma'}^{c_1} \prod_{j=k+1}^{m} \left( S^{\sigma}_{(r_j)} / S_{\sigma} \right)  S_{\sigma}^{c_2} = \prod_{i=1}^m \left( S^{\sigma}_{(r_k)} / S_{\sigma} \right) S_\sigma^{\tilde{c} }
\end{equation}
to formally split this cluster into two, one inside the other, such that $c_1+c_2=\tilde{c}$. Comparing the two expressions above, we represent the latter pictorially as inserting a ``symbolic loop'' where instead of a factor $\scra_{\sigma,\sigma'}$ at the boundary, we have a factor $\delta_{\sigma,\sigma'}$, and where we do not get a weight $\lambda_{(r)}$ when removing the loop. We draw this symbolic loop as a dashed line to distinguish it from the ordinary cluster boundaries, as in the following example of a leaf with one marked point $\times$:
\begin{equation}
\begin{tikzpicture}
\coordinate (A) at (0, 0) {};
\coordinate (B) at (-0.1, 0.1) {};
\filldraw[fill opacity=0.2,fill=white] ([shift=(0:0.7cm)]A) arc (0:360:0.7cm);
\filldraw[fill opacity=1,fill=white,draw=black,dashed] ([shift=(0:0.4cm)]B) arc (0:360:0.4cm);
\node (B1) at (B) {$\times$};
\end{tikzpicture}
\end{equation}
Let us now insert such loops around two marked points sitting in the same cluster with respective RSOS variables $\sigma', \sigma'' \in X$:
\begin{equation}
\vcenter{\hbox{
		\begin{tikzpicture}
		\coordinate (A) at (0, 0) {};
		\coordinate (B) at (1, 0) {};
		\node(A1) at (A) {$\times$};
		\node(B1) at (B) {$\times$};
		\end{tikzpicture}
}}
\vcenter{\hbox{\hspace*{1cm}$\rightarrow$\hspace*{1cm}}}
\vcenter{\hbox{
		\begin{tikzpicture}
		\coordinate (A) at (0, 0) {};
		\coordinate (B) at (1, 0) {};
		\filldraw[fill opacity=0.2,fill=white,draw=black,dashed] ([shift=(0:0.4cm)]A) arc (0:360:0.4cm);
		\filldraw[fill opacity=0.2,fill=white,draw=black,dashed] ([shift=(0:0.4cm)]B) arc (0:360:0.4cm);
		\node(A1) at (A) {$\times$};
		\node(B1) at (B) {$\times$};
		\end{tikzpicture}
}}
\label{twopointinsert}
\end{equation}
%
%
If we consider the \emph{surrounding} cluster, it contributes a weight of $S_\sigma^{1-c} = S_\sigma^{-1}$ due to its two circuits, while the two symbolic loops will insert a factor of $\delta_{\sigma,\sigma'}\delta_{\sigma,\sigma''}$. 
With some rewriting of \eqref{structfac} using \eqref{ortho_x} this gives
\begin{equation}\label{C123}
\delta_{\sigma,\sigma'}\delta_{\sigma,\sigma''}S_\sigma^{-1} = \sum_r\sum_{r'}\sum_{r''} C_{r,r',r''} S_{(r)}^\sigma S_{(r')}^{\sigma'} S_{(r'')}^{\sigma''}.
\end{equation}
We can now express the fusion of the two states $S_{(r_1)}$ and $S_{(r_2)}$, by which we mean that each of the marked points
in \eqref{twopointinsert} carries an additional factor $S_{(r_1)}^{\sigma'}$ or $S_{(r_2)}^{\sigma''}$, respectively. Using now \eqref{ortho_k},
this simplifies as
\begin{equation}\label{fusion}
\sum_{\sigma'}\sum_{\sigma''} \sum_r\sum_{r'}\sum_{r''} C_{r,r',r''}  S_{(r)}^\sigma S_{(r')}^{\sigma'} S_{(r'')}^{\sigma''}  S_{(r_1)}^{\sigma'} S_{(r_2)}^{\sigma''}  =  \sum_rC_{r,r_1,r_2} S^\sigma_{(r)}.
\end{equation}
That is: any time we have two marked points of labels $r_1,r_2$ within the same cluster, we can replace them by a sum over the possible fusion products. To recover the 3-point function discussed above we introduce a third marked vertex, with a corresponding $\sum_\sigma S_{(r_3)}^\sigma$, which will take care of the last sum and single out $C_{r_1,r_2,r_3}$ as the only surviving term. With only two marked vertices, we must use $C_{r_1,r_2,{r_{\text{id}}}}=\delta_{r_1,r_2}$, as in \eqref{three_to_two_reduc}. We then recover the 2-point function. Similarly if two of the vertices are unmarked we use $C_{r_1,r_{\text{id}},r_{\text{id}}}=\delta_{r_1,r_{\text{id}}}$ to recover the 1-point function.

When we encounter a marked point that is not on a leaf, we can use a symbolic loop to treat if as if it were, constructing a 3-vertex and fusing it into the tree. We have seen a similar idea already in the general three-point result of \eqref{three-point}, where we could allow branch lengths to be zero. We note that fusing any state $S_{(r)}$ with the identity state $S_{(r_{\text{id}})}$ will give back $S_{(r)}$ as the only output. Here is a sample figure (with the loop weights indicated on the r.h.s.):
\begin{equation}
\newcommand\circrad{1.2}
\vcenter{\hbox{
		\begin{tikzpicture}
		\coordinate (A) at (0, -0.3) {};
		\node (A1) at (0, -0.3) {$\times$};
		\node (A2) at (0, -0.6) {${}_{r_1}$};
		\coordinate (B) at (0.5, 1) {};
		\node (B1) at (0.5, 1) {$\times$};
		\node (B2) at (0.5, 0.7) {${}_{r_2}$};
		\coordinate (E) at (2.2, 0.3) {};
		\node () at ([shift=(-90:1pt)]E) {\hspace*{0.7cm}$\cdots$};
		\coordinate (F) at (0.9, 0.3) {};
		\filldraw (A) circle (\circrad pt);
		\filldraw (F) circle (\circrad pt);
		\draw[] (A) -- (F);
		\draw[] (F) -- (E);
		\draw[] ([shift=(0:0.45cm)]A) arc (0:360:0.45cm);
		\draw[] ([shift=(20:1.8cm)]A) arc (0:360:1.3cm);
		\end{tikzpicture}
}}
\vcenter{\hbox{\hspace*{1cm}$\leftrightarrow$\hspace*{1cm}}}
\vcenter{\hbox{
		\begin{tikzpicture}
		\coordinate (A) at (0, -0.3) {};
		\node (A1) at (0, -0.3) {$\times$};
		\node (A2) at (0, -0.6) {${}_{r_1}$};
		\coordinate (B) at (0.5, 1) {};
		\node (B1) at (0.5, 1) {$\times$};
		\node (B2) at (0.5, 0.7) {${}_{r_2}$};
		\coordinate (E) at (2.2, 0.3) {};
		\node () at ([shift=(-90:1pt)]E) {\hspace*{0.7cm}$\cdots$};
		\coordinate (F) at (0.9, 0.3) {};
		\filldraw (A) circle (\circrad pt);
		\filldraw (B) circle (\circrad pt);
		\filldraw (F) circle (\circrad pt);
		\draw[] (A) -- (F);
		\draw[] (B) -- (F);
		\draw[] (F) -- (E);
		\draw[] ([shift=(0:0.45cm)]A) arc (0:360:0.45cm);
		\draw[] ([shift=(20:1.8cm)]A) arc (0:360:1.3cm);
		\draw[dashed] ([shift=(0:0.45cm)]B) arc (0:360:0.45cm);
		\node at (E) [below] {$\hspace*{1.5cm} \sum_r C_{r_1,r_2,r}\lambda_{(r)}$};
		\node at (A) [right] {$\;\;\, \lambda_{(r_1)}$};
		\end{tikzpicture}
}}
\end{equation}

\subsubsection{$n$-vertices, $n>3$}\label{n_vertices}
An $n$-vertex corresponds to a cluster with $n-1$ circuits, giving a weight $S_\sigma^{1-c} = S_\sigma^{2-n}$ in position space. 
%
With the results established above, the aforementioned idea of factorising vertices as $S^{2-n}=S^{-1}\times S^{-1}\times...\times S^{-1}$ is made rigorous. Consider for instance $n=4$, as in e.g.:
\begin{equation}
\newcommand\scaling{0.7}
\vcenter{\hbox{
		\begin{tikzpicture}
		\node (A) at (0, 0) {$\times$};
		\node (B) at (1*\scaling, 1.3*\scaling) {$\times$};
		\node (C) at (3*\scaling, 1.5*\scaling) {$\times$};
		\node (D) at (2*\scaling, 0) {$\times$};
		\coordinate (E) at (1*\scaling, 0.2*\scaling) {};
		\draw[] ([shift=(4:1.7*\scaling cm)]E) arc (0:360:1.7*\scaling cm);
		\draw[] ([shift=(0:0.3cm)]A) arc (0:360:0.3cm);
		\draw[] ([shift=(0:0.3cm)]B) arc (0:360:0.3cm);
		\draw[] ([shift=(0:0.3cm)]D) arc (0:360:0.3cm);
		\node() at (1*\scaling,0.4*\scaling) {$_{S^{-2}}$};
		\end{tikzpicture}}}
\vcenter{\hbox{\hspace*{0cm}$\leftrightarrow$\hspace*{0.5cm}}}
\vcenter{\hbox{
		\begin{tikzpicture}
		\node (A) at (0, 0) {$\times$};
		\node (B) at (1*\scaling, 1.3*\scaling) {$\times$};
		\node (C) at (3*\scaling, 1.5*\scaling) {$\times$};
		\node (D) at (2*\scaling, 0) {$\times$};
		\coordinate (E) at (1*\scaling, 0.2*\scaling) {};
		\draw[] ([shift=(4:1.7*\scaling cm)]E) arc (0:360:1.7*\scaling cm);
		\draw[] ([shift=(0:0.3cm)]A) arc (0:360:0.3cm);
		\draw[] ([shift=(0:0.3cm)]B) arc (0:360:0.3cm);
		\draw[] ([shift=(0:0.3cm)]D) arc (0:360:0.3cm);
		\draw[dashed, rotate around={-35:(0,0)}] ([shift=(175:0.6*\scaling cm)]E)  circle [x radius=0.9*\scaling cm, y radius=1.4*\scaling cm];
		\node() at (0.6*\scaling,0.6*\scaling) {$_{S^{-1}}$};
		\node() at (1.4*\scaling,-0.7*\scaling) {$_{S^{-1}}$};
		\end{tikzpicture}}}
\vcenter{\hbox{\hspace*{0cm}$\leftrightarrow$\hspace*{0.5cm}}}
\vcenter{\hbox{
		\begin{tikzpicture}
		\node (A) at (0, 0) {$\times$};
		\node (B) at (1*\scaling, 1.3*\scaling) {$\times$};
		\node (C) at (3*\scaling, 1.5*\scaling) {$\times$};
		\node (D) at (2*\scaling, 0) {$\times$};
		\coordinate (E) at (1*\scaling, 0.2*\scaling) {};
		\draw[] ([shift=(4:1.7*\scaling cm)]E) arc (0:360:1.7*\scaling cm);
		\draw[] ([shift=(0:0.3cm)]A) arc (0:360:0.3cm);
		\draw[] ([shift=(0:0.3cm)]B) arc (0:360:0.3cm);
		\draw[] ([shift=(0:0.3cm)]D) arc (0:360:0.3cm);
		\draw[dashed] ([shift=(-90:0.15cm)]E)  circle [x radius=1.55*\scaling cm, y radius=0.8*\scaling cm];
		\node() at (2*\scaling,1.1*\scaling) {$_{S^{-1}}$};
		\node() at (1*\scaling,0*\scaling) {$_{S^{-1}}$};
		\end{tikzpicture}}}
\end{equation}
In terms of trees, the above figures correspond to:
\begin{equation}
\newcommand\circrad{1.2}
\vcenter{\hbox{
		\begin{tikzpicture}
		\coordinate (A) at (0, 0) {};
		\coordinate (B) at (1, 1.2) {};
		\coordinate (C) at (3, 1.5) {};
		\coordinate (D) at (2, 0) {};
		\node (A1) at (A) {$r_2\hspace*{20pt}$};
		\node (B1) at (1, 1.2) {$r_1\hspace*{20pt}$};
		\node (C1) at (3, 1.5) {$\hspace*{20pt}r_3$};
		\node (D1) at (2, 0) {$\hspace*{20pt}r_4$};
		\coordinate (E) at (1.2, 0.4) {};
		\filldraw (A) circle (\circrad pt);
		\filldraw (B) circle (\circrad pt);
		\filldraw (C) circle (\circrad pt);
		\filldraw (D) circle (\circrad pt);
		\filldraw (E) circle (\circrad pt);
		\draw[] (A) -- (E);
		\draw[] (B) -- (E);
		\draw[] (C) -- (E);
		\draw[] (D) -- (E);
		\end{tikzpicture}}}
\vcenter{\hbox{\hspace*{0cm}$\leftrightarrow$\hspace*{0.5cm}}}
\vcenter{\hbox{
		\begin{tikzpicture}
		\coordinate (A) at (0, 0) {};
		\coordinate (B) at (1, 1.2) {};
		\coordinate (C) at (3, 1.5) {};
		\coordinate (D) at (2, 0) {};
		\node (A1) at (A) {$r_2\hspace*{20pt}$};
		\node (B1) at (1, 1.2) {$r_1\hspace*{20pt}$};
		\node (C1) at (3, 1.5) {$\hspace*{20pt}r_3$};
		\node (D1) at (2, 0) {$\hspace*{20pt}r_4$};
		\coordinate (E) at (1.5, 0.5) {};
		\coordinate (F) at (1.2, 0.5) {};
		\filldraw (A) circle (\circrad pt);
		\filldraw (B) circle (\circrad pt);
		\filldraw (C) circle (\circrad pt);
		\filldraw (D) circle (\circrad pt);
		\filldraw (E) circle (\circrad pt);
		\filldraw (F) circle (\circrad pt);
		\draw[] (A) -- (F);
		\draw[] (B) -- (F);
		\draw[] (C) -- (E);
		\draw[] (D) -- (E);
		\draw[] (F) -- (E);
		\end{tikzpicture}}}
\vcenter{\hbox{\hspace*{0cm}$\leftrightarrow$\hspace*{0.5cm}}}
\vcenter{\hbox{
		\begin{tikzpicture}
		\coordinate (A) at (0, 0) {};
		\coordinate (B) at (1, 1.2) {};
		\coordinate (C) at (3, 1.5) {};
		\coordinate (D) at (2, 0) {};
		\node (A1) at (A) {$r_2\hspace*{20pt}$};
		\node (B1) at (1, 1.2) {$r_1\hspace*{20pt}$};
		\node (C1) at (3, 1.5) {$\hspace*{20pt}r_3$};
		\node (D1) at (2, 0) {$\hspace*{20pt}r_4$};
		\coordinate (E) at (1.1, 0.3) {};
		\coordinate (F) at (1.3, 0.6) {};
		\filldraw (A) circle (\circrad pt);
		\filldraw (B) circle (\circrad pt);
		\filldraw (C) circle (\circrad pt);
		\filldraw (D) circle (\circrad pt);
		\filldraw (E) circle (\circrad pt);
		\filldraw (F) circle (\circrad pt);
		\draw[] (A) -- (E);
		\draw[] (B) -- (F);
		\draw[] (C) -- (F);
		\draw[] (D) -- (E);
		\draw[] (F) -- (E);
		\end{tikzpicture}}}
\end{equation}
In the two trees on the right, \eqref{fusion} applies at the new 3-vertices such that we have one sum $\sum_r$ along the internal line, which we can interpret as an $s/t$-channel. The result must be the same, showing the notion of crossing symmetry mentioned in \cite{Kostov:1989eg}. 

It is clear that we can follow the same scheme for any vertex with $n>3$, as well as for any case of several marked points sitting in the same cluster.
Any states $S_{(r)}$ that are ``close'' (by which we that they would sit in the same cluster after taking away any loop surrounding only one state) can be fused with each other, and crossing symmetry makes the result independent of in which order we perform the fusion. Having established this final result, we see that we can write all $N$-point functions in terms of the Feynman rules stated before. 
%
%
\subsubsection{Expressions for the $N$-point functions}
For the 4-point function, we can consider a H-shaped tree ${\sf H}_{\ell_1,\ell_2; \ell; \ell_3, \ell_4}$
\begin{equation}
\begin{tikzpicture}
\newcommand\circrad{1.2}
\newcommand\scaling{2}
\coordinate (A) at (0, -1) {};
\coordinate (B) at (0, 1) {};
\coordinate (C) at (2, 1) {};
\coordinate (D) at (2, -1) {};
\coordinate (E) at (0, 0) {};
\coordinate (F) at (2, 0) {};
\node(A1) at (A) {$\times$};
\node(B1) at (B) {$\times$};
\node(C1) at (C) {$\times$};
\node(D1) at (D) {$\times$};
\filldraw (A) circle (\circrad pt);
\filldraw (B) circle (\circrad pt);
\filldraw (C) circle (\circrad pt);
\filldraw (D) circle (\circrad pt);
\filldraw (E) circle (\circrad pt);
\filldraw (F) circle (\circrad pt);
\node(A2) at ([shift=(-180:8pt)]A) {$_{r_2}$};
\node(B2) at ([shift=(-180:8pt)]B) {$_{r_1}$};
\node(C2) at ([shift=(0:8pt)]C) {$_{r_3}$};
\node(D2) at ([shift=(0:8pt)]D) {$_{r_4}$};
\node(l1) at ([shift=(-180:5pt)]$(A)!0.5!(E)$) {$_{\ell_2}$};
\node(l2) at ([shift=(-180:5pt)]$(B)!0.5!(E)$) {$_{\ell_1}$};
\node(l3) at ([shift=(0:7pt)]$(C)!0.5!(F)$) {$_{\ell_3}$};
\node(l4) at ([shift=(0:7pt)]$(D)!0.5!(F)$) {$_{\ell_4}$};
\node(l) at ([shift=(90:5pt)]$(E)!0.5!(F)$) {$_{\ell}$};
\draw[] (A) -- ($(A)!0.33!(E)$);
\draw[dotted] ($(A)!0.33!(E)$) -- ($(A)!0.67!(E)$);
\draw[] ($(A)!0.67!(E)$) -- (E);
\draw[] (B) -- ($(B)!0.33!(E)$);
\draw[dotted] ($(B)!0.33!(E)$) -- ($(B)!0.67!(E)$);
\draw[] ($(B)!0.67!(E)$) -- (E);
\draw[] (C) -- ($(C)!0.33!(F)$);
\draw[dotted] ($(C)!0.33!(F)$) -- ($(C)!0.67!(F)$);
\draw[] ($(C)!0.67!(F)$) -- (F);
\draw[] (D) -- ($(D)!0.33!(F)$);
\draw[dotted] ($(D)!0.33!(F)$) -- ($(D)!0.67!(F)$);
\draw[] ($(D)!0.67!(F)$) -- (F);
\draw[] (E) -- ($(E)!0.33!(F)$);
\draw[dotted] ($(E)!0.33!(F)$) -- ($(E)!0.67!(F)$);
\draw[] ($(E)!0.67!(F)$) -- (F);
\end{tikzpicture}
\label{H-graph}
\end{equation}
where upper left, lower left, upper right and lower right vertical branches have respective lengths $\ell_1,\ell_2,\ell_3,\ell_4$, while the connecting horizontal
branch has length $\ell$. Up to factors of $\lambda_{(r_{\text{id}})}$ from summing out the loops not separating the marked points, we get the weight 
\be\label{4pt_fun}
\left( \prod_{j=1}^4 \lambda_{(r_j)}^{\ell_j} \right) C^{(\ell)}_{r_1,r_2,r_3,r_4} \,,
\ee
with
\be \label{Cr1r2r3r4}
C^{(\ell)}_{r_1,r_2,r_3,r_4} := \sum_r C_{r_1,r_2,r} \lambda_{(r)}^\ell C_{r,r_3,r_4} \,.
\ee
That is, each loop separating one of the points points $j \in \{1,2,3,4\}$ from the other three points provides a factor $\lambda_{(r_j)}$, whereas
the $\ell$ loops that separate the fused group $\{12\}$ from the other block $\{34\}$ of the set partition provide a total contribution of $C^{(\ell)}_{r_1,r_2,r_3,r_4}$. 
Allowing for some of the $\ell_j$ and/or $\ell$ to be zero, any other type of tree contributing to the 4-point function leads to a special case of this result. Taking the positions of the marked points into account we note that there are three possible configurations when all $\ell,\ell_j>0$: the $s,t$ and $u$-channel trees.\footnote{If we consider $\ell=0$ to be a separate case, we can compare the resulting four types of trees to the four diagrams in figure \ref{fourdiagram}.}
The diagram in \eqref{H-graph} illustrates the $s$-channel.

When $\ell = 0$, the $s$, $t$ and $u$-channels coincide. We therefore have
\be\label{crossing4pt}
 \sum_r C_{r_1,r_2,r} C_{r,r_3,r_4} = \sum_r C_{r_1,r_3,r} C_{r,r_2,r_4} = \sum_r C_{r_1,r_4,r} C_{r,r_2,r_3} \,,
\ee
a statement referred to as crossing symmetry in \cite{Kostov:1989eg}. An equivalent statement is that $C^{(0)}_{r_1,r_2,r_3,r_4}$, defined by \eqref{Cr1r2r3r4},
is symmetric in all its four indices.

\bigskip

Let us briefly remark on higher-point correlation functions.
For the 5-point function, we can consider a tree 
\begin{equation}
\begin{tikzpicture}
\newcommand\circrad{1.2}
\newcommand\scaling{2}
\coordinate (A) at (0, -1) {};
\coordinate (B) at (0, 1) {};
\coordinate (C) at (4, 1) {};
\coordinate (D) at (4, -1) {};
\coordinate (E) at (0, 0) {};
\coordinate (F) at (4, 0) {};
\coordinate (midup) at (2, 1) {};
\coordinate (midmid) at (2, 0) {};
\node(A1) at (A) {$\times$};
\node(B1) at (B) {$\times$};
\node(C1) at (C) {$\times$};
\node(D1) at (D) {$\times$};
\node(midup1) at (midup) {$\times$};
\node(l2) at ([shift=(-180:5pt)]$(A)!0.5!(E)$) {$_{\ell_2}$};
\node(l1) at ([shift=(-180:5pt)]$(B)!0.5!(E)$) {$_{\ell_1}$};
\node(l4) at ([shift=(0:7pt)]$(C)!0.5!(F)$) {$_{\ell_4}$};
\node(l5) at ([shift=(0:7pt)]$(D)!0.5!(F)$) {$_{\ell_5}$};
\node(l3) at ([shift=(0:7pt)]$(midup)!0.5!(midmid)$) {$_{\ell_3}$};
\node(l) at ([shift=(90:5pt)]$(E)!0.5!(midmid)$) {$_{\ell}$};
\node(lprime) at ([shift=(90:5pt)]$(F)!0.5!(midmid)$) {$_{\ell'}$};
\filldraw (A) circle (\circrad pt);
\filldraw (B) circle (\circrad pt);
\filldraw (C) circle (\circrad pt);
\filldraw (D) circle (\circrad pt);
\filldraw (E) circle (\circrad pt);
\filldraw (F) circle (\circrad pt);
\filldraw (midup) circle (\circrad pt);
\node(A2) at ([shift=(-180:8pt)]A) {$_{r_2}$};
\node(B2) at ([shift=(-180:8pt)]B) {$_{r_1}$};
\node(C2) at ([shift=(0:8pt)]C) {$_{r_4}$};
\node(D2) at ([shift=(0:8pt)]D) {$_{r_5}$};
\node(midup2) at ([shift=(0:8pt)]midup) {$_{r_3}$};
\draw[] (A) -- ($(A)!0.33!(E)$);
\draw[dotted] ($(A)!0.33!(E)$) -- ($(A)!0.67!(E)$);
\draw[] ($(A)!0.67!(E)$) -- (E);
\draw[] (B) -- ($(B)!0.33!(E)$);
\draw[dotted] ($(B)!0.33!(E)$) -- ($(B)!0.67!(E)$);
\draw[] ($(B)!0.67!(E)$) -- (E);
\draw[] (C) -- ($(C)!0.33!(F)$);
\draw[dotted] ($(C)!0.33!(F)$) -- ($(C)!0.67!(F)$);
\draw[] ($(C)!0.67!(F)$) -- (F);
\draw[] (D) -- ($(D)!0.33!(F)$);
\draw[dotted] ($(D)!0.33!(F)$) -- ($(D)!0.67!(F)$);
\draw[] ($(D)!0.67!(F)$) -- (F);
\draw[] (midup) -- ($(midup)!0.33!(midmid)$);
\draw[dotted] ($(midup)!0.33!(midmid)$) -- ($(midup)!0.67!(midmid)$);
\draw[] ($(midup)!0.67!(midmid)$) -- (midmid);
\draw[] (E) -- ($(E)!0.33!(midmid)$);
\draw[dotted] ($(E)!0.33!(midmid)$) -- ($(E)!0.67!(midmid)$);
\draw[] ($(E)!0.67!(midmid)$) -- (midmid);
\draw[] (midmid) -- ($(midmid)!0.33!(F)$);
\draw[dotted] ($(midmid)!0.33!(F)$) -- ($(midmid)!0.67!(F)$);
\draw[] ($(midmid)!0.67!(F)$) -- (F);
\end{tikzpicture}
\end{equation}
Let $\ell,\ell'$ be the lengths of the horizontal branches, and let $\ell_j,\;j=1,\ldots,5$ be the lengths of the vertical branches, as shown. Up to factors of $\lambda_{(r_{\text{id}})}$ from summing out the loops not separating the marked points, we get the weight 
\be\label{5pt_fun}
\left( \prod_{j=1}^5 \lambda_{(r_j)}^{\ell_j} \right) C^{(\ell,\ell')}_{r_1,r_2,r_3,r_4,r_5} \,,
\ee
with
\be
C^{(\ell,\ell')}_{r_1,r_2,r_3,r_4,r_5} := 
\sum_{r,r'} C_{r_1,r_2,r} \lambda_{(r)}^\ell 
C_{r,r_3,r'} \lambda_{(r')}^{\ell'} C_{r',r_4,r_5}
\,.
\ee
As in the cases of $N\leq 4$ we recover all possible shapes of trees when we allow some or all of $\ell,\ell',\ell_j$ to be zero. Taking the positions of the marked points into account we need to consider $4\times 3$ possible configurations; starting from any of the three 4-point trees ($s,t$ or $u$-channel), we can let any of the four branches  $j=1,\ldots,4$ split into two branches $j$ and $5$. 

The recursive method of finding all $N$-trees by splitting branches of the $(N-1)$-trees extends to $N>5$. As before we consider trees with 3-vertices only, seeing the other trees as special cases with some branch lengths set to zero. As soon as there are more than two internal lines, it is important to keep in mind that different trees may be non-isomorphic, for instance at $N=6$:
\begin{equation}
\vcenter{\hbox{
		\begin{tikzpicture}
		\newcommand\circrad{1.2}
		\coordinate (r1) at (-1, 0) {};
		\coordinate (r2) at (0, 1) {};
		\coordinate (r3) at (1, 1) {};
		\coordinate (r4) at (2, 1) {};
		\coordinate (r5) at (3, 1) {};
		\coordinate (r6) at (4, 0) {};
		\coordinate (c2) at (0, 0) {};
		\coordinate (c3) at (1, 0) {};
		\coordinate (c4) at (2, 0) {};
		\coordinate (c5) at (3, 0) {};
		\filldraw (r1) circle (\circrad pt);
		\filldraw (r2) circle (\circrad pt);
		\filldraw (r3) circle (\circrad pt);
		\filldraw (r4) circle (\circrad pt);
		\filldraw (r5) circle (\circrad pt);
		\filldraw (r6) circle (\circrad pt);
		\filldraw (c2) circle (\circrad pt);
		\filldraw (c3) circle (\circrad pt);
		\filldraw (c4) circle (\circrad pt);
		\filldraw (c5) circle (\circrad pt);
		\node(phantom) at (3, 1.5) { };
		\draw[] (r1) -- (c2);
		\draw[] (r2) -- (c2);
		\draw[] (r3) -- (c3);
		\draw[] (r4) -- (c4);
		\draw[] (r5) -- (c5);
		\draw[] (r6) -- (c5);
		\draw[] (c2) -- (c5);
		\end{tikzpicture}}}
\vcenter{\hbox{\hspace*{1cm}$\not\simeq$\hspace*{1cm}}}
\vcenter{\hbox{
		\begin{tikzpicture}
		\newcommand\circrad{1.2}
		\coordinate (r1) at (-1, 0) {};
		\coordinate (r2) at (0, 1) {};
		\coordinate (r3) at (0.5, 1.5) {};
		\coordinate (r6) at (1.5, 1.5) {};
		\coordinate (r4) at (2, 1) {};
		\coordinate (r5) at (3, 0) {};
		\coordinate (c2) at (0, 0) {};
		\coordinate (c3) at (1, 0) {};
		\coordinate (c4) at (2, 0) {};
		\coordinate (c5) at (1, 1) {};
		\filldraw (r1) circle (\circrad pt);
		\filldraw (r2) circle (\circrad pt);
		\filldraw (r3) circle (\circrad pt);
		\filldraw (r4) circle (\circrad pt);
		\filldraw (r5) circle (\circrad pt);
		\filldraw (r6) circle (\circrad pt);
		\filldraw (c2) circle (\circrad pt);
		\filldraw (c3) circle (\circrad pt);
		\filldraw (c4) circle (\circrad pt);
		\filldraw (c5) circle (\circrad pt);
		\draw[] (r1) -- (c2);
		\draw[] (r2) -- (c2);
		\draw[] (r4) -- (c4);
		\draw[] (r5) -- (c4);
		\draw[] (r3) -- (c5);
		\draw[] (r6) -- (c5);
		\draw[] (c2) -- (c4);
		\draw[] (c3) -- (c5);
		\end{tikzpicture}}}
\end{equation}
For any given tree, the corresponding weight will follow the general pattern seen in \eqref{2ptres}, \eqref{three-point}, \eqref{4pt_fun}, \eqref{5pt_fun}, encoded in the Feynman rules of List \ref{Feynman}.

\section{Three-point couplings $C_{r_1,r_2,r_3}$ in type $A_{p-1}$ and type $D_{1+\frac{p}{2}}$}\label{3ptC}
In this appendix, we give the three-point couplings $C_{r_1,r_2,r_3}$ for RSOS models of type $A$ and $D$ which we used in the main text for studying the cluster expansions of the RSOS four-point functions. Here we consider generic $(p,q)$ with $p-q\ge1$ and $p\wedge q=1$ associated with a Dynkin diagram of type $A$ or $D$ with Coxeter number $p$ as in figure \ref{ADE} and use the formula \eqref{structfac}, where the special vector corresponding to identity field is $S_{\sigma}=S^{\sigma}_{(p-q)}$ and becomes $S^{\sigma}_{(1)}$ in the unitary case.

\subsection*{Type $A$}
In $A_{p-1}$, the eigenvectors of the adjacency matrix $\mathcal{A}$ are
\begin{equation}
S^\sigma_{(r)}=\sqrt{\frac{2}{p}}\sin\bigg(\frac{r\pi}{p}\sigma\bigg),\;r=1,\ldots,p-1 \,.
\end{equation}
We thus obtain the following three-point couplings:
\begin{equation}\label{CA}
C^{A_{p-1}}_{r_1,r_2,r_3}=\begin{cases}
(-1)^{\mathsf{b}_1+\mathsf{b}_2+\mathsf{b}_3}\frac{1-(-1)^{\mathsf{a}_1+\mathsf{a}_2+\mathsf{a}_3}}{2},&|\mathsf{d}_1-\mathsf{d}_2|+1\le \mathsf{d}_3\le\;(\mathsf{d}_1+\mathsf{d}_2-1),\;\mathsf{d}_1+\mathsf{d}_2+\mathsf{d}_3\le 2p-1 \,, \\
0,&\text{otherwise} \,,
\end{cases}
\end{equation}
where $\mathsf{a}_i$, $\mathsf{b}_i$ solve the Diophantine equation
\begin{equation}
r_i+\mathsf{b}_ip=\mathsf{a}_i(p-q)
\end{equation}
and $\mathsf{d}_i$ is given by
\begin{equation}
\mathsf{d}_i=\frac{1+(-1)^{\mathsf{b}_i}}{2}\mathsf{a}_i+\frac{1-(-1)^{\mathsf{b}_i}}{2}(p-\mathsf{a}_i) \,.
\end{equation}
Notice that in the unitary case, $\mathsf{d}_i=\mathsf{a}_i=r_i$ and $\mathsf{b}_i=0$, and we recover the well-known unitary minimal models fusion rules. 

\subsection*{Type $D$}
In the case of $D_{N}=D_{1+\frac{p}{2}}$ with $p\equiv\hbox{2 mod 4}$, the Dynkin diagram has a fork labeled by $N-1$ and $\overline{N-1}$, as shown in figure \ref{ADE}. The eigenvectors of $\mathcal{A}$ read
\begin{eqnarray}
S^\sigma_{(r)}&=&\frac{2}{\sqrt{p}}\cos\frac{(N-1-\sigma) r\pi}{p},\; \sigma\ne N-1, \overline{N-1}\\
S^{N-1}_{(r)}&=&S^{\overline{N-1}}_{(r)}=\frac{1}{\sqrt{p}},\; \text{for} \;r=1,3,\ldots,p-1,\;\text{odd}
\end{eqnarray}
and the last eigenvector corresponding to $r=\frac{\bar{p}}{2}$ is
\begin{equation}
S^{\sigma}_{(\bar{p}/2)}=(0,...,0,\frac{1}{\sqrt{2}},-\frac{1}{\sqrt{2}}) \,.
\end{equation}
The three-point couplings are
\begin{equation}\label{CD}
C^{D_{1+\frac{p}{2}}}_{r_1,r_2,r_3}=\begin{cases}
(-1)^{\frac{\mathsf{a}_1+\mathsf{a}_2+\mathsf{a}_3+1}{2}},&|\mathsf{d}_1-\mathsf{d}_2|+1\le \mathsf{d}_3\le\;(\mathsf{d}_1+\mathsf{d}_2-1),\;\mathsf{d}_1+\mathsf{d}_2+\mathsf{d}_3\le 2p-1 \,, \\
0,&\text{otherwise}\\
\end{cases}
\end{equation}
for $r\ne \bar{p}/2$ and the only non-vanishing $C$'s involving $\bar{p}/2$ are $C^{D_{1+\frac{p}{2}}}_{\bar{p}/2,\bar{p}/2,r}=1$, for $r=1,\ldots,p/2,\ldots,p-1$.

\section{Numerical computation of exact amplitude ratios} \label{app:num}

\subsection{General setup}
\label{num-gen-setup}

Our numerical transfer matrix computations of four-point functions take place on a cylinder of circumference $L$, as shown in figure~\ref{fig:cylinder}.
The four points are inserted on two different time slices, with the group consisting of $i_1$ and $i_2$ on the first slice, and the second group of
$i_3$ and $i_4$ on another slice, $l$ lattice spacings distant from the first one. In both groups, the distance between the two points is $2\mathfrak{a}$
lattice spacings (or $L-2\mathfrak{a}$ when going around the periodic direction), and we take $2\mathfrak{a} \le \frac{L}{2}$. The distance to the free
boundary conditions at either length of the cylinder is taken sufficiently large in order not to influence the results, to within the chosen numerical precision.

The relevant probabilities $P_{\cal P}$ of \eqref{P_corr_def}, for any given partition ${\cal P} = a_1,a_2,a_3,a_4$, as well as the transfer matrix
eigenvalues $\Lambda_0 > \Lambda_1 > \cdots > \Lambda_i > \cdots$ are then computed to very high precision (4\,000 digits) in order to be able 
to accurately determine even contributions to  $P_{\cal P}$ which are exponentially small (in $l$) with respect to the leading term.

For generic values of $Q$ the transfer matrix is diagonalisable---i.e., no non-trivial Jordan cells appear---and the probabilities take the form 
\begin{equation}
 P_{a_1,a_2,a_3,a_4} = \sum_i A_i \left( \frac{\Lambda_i}{\Lambda_0} \right)^l \,, \label{PAeval}
\end{equation}
where the amplitudes $A_i = A_i(\mathfrak{a},L)$ can be determined precisely, provided that we have at our disposal data with as many different
values of $l$ as the number of eigenvalues appearing in the sum. This can be done in practice for $L \le 7$.

Setting up this transfer matrix computation involves dealing with a fairly large number of technical aspects. These include specifying the affine 
Temperley-Lieb representations on which the transfer matrix acts in order to be able to compute the desired probabilities, how to diagonalise
it efficiently to the required precision, how to set up the necessary data structures and computational schemes, and more. For the basic
probabilities $P_{\cal P}$ this has already been described in much detail in the extensive appendix A of \cite{Jacobsen:2018pti}, to which the
interested reader is referred. We however do describe below the modifications of the basic method which are necessary to compute certain
refined and modified probabilities required by the present article.

Since $Q$ is generic, each of the eigenvalues appearing in \eqref{PAeval} can be assigned to a definite affine Temperley-Lieb (ATL) module $\AStTL{j}{z^2}$.
Our main conclusions---namely, the facts exposed in sections \ref{sec:facts1}--\ref{sec:facts3}---are that whenever the same ATL module contributes
to two different probabilities $P_{\cal P}$, the ratios between the corresponding amplitudes are the same for every eigenvalue within that module.
Moreover, the ratios do not depend on the size $L$, nor on the separation $2\mathfrak{a}$, provided that both are
sufficiently large to accommodate the representation $\AStTL{j}{z^2}$---in practice
this means that ${\rm min}(2\mathfrak{a},L-2\mathfrak{a}) \ge j$.

We have verified these statements for many different ratios, different sizes $L$ and $2\mathfrak{a}$, and many different values of $Q$, finding them to be exact to
hundreds of digits of numerical precision. But we can go further yet. Conducting the computations for rational values%
\footnote{In these computations we eschew integer values of $Q$, since they are not generic (in the sense that the quantum group parameter
$\q$ would be a root of unity). It however turns out that the amplitude ratios
are in fact continuous functions of $Q$, except for the presence of poles, so in most cases taking $Q$ integer would actually do no harm.}
of $Q=\frac{1}{10},\frac{2}{10},\ldots$, our numerical precision is such that we can use {\sc Mathematica}'s function {\tt Rationalize} to establish that
the amplitude ratios are, in fact, themselves rational numbers. We have carefully checked that the same fraction is obtained for any eigenvalue within
a given ATL module---and also for different sizes $L$---and so we are fully confident that this method, albeit numerical, does in fact produce
exact results.

The resulting amplitude ratios could of course still be complicated functions of $Q$. But we find, remarkably, that comparing the same amplitude ratio
for a sufficiently large number of $Q$-values, it can invariably be produced as the ratio between two integer-coefficient polynomials in $Q$. Once
this expression has been established---which in some complicated cases, such as \eqref{gamma41}, required assembling a dozen of values of $Q$---, we have
double-checked it by repeating the computations for several more values.

\subsection{Modification of the probability $P_{aabb}$}
\label{app:algmodif}

For the purpose of establishing the facts of type 3 in section~\ref{sec:facts3} we need to
compute a modified version of the probability of $P_{aabb}$, denoted $P_{aabb}^{(\mathsf{a})}$, in which each loop separating the
two ``short'' clusters has a weight given by \eqref{na-modif}, viz.
\begin{equation}
n_\mathsf{a}\equiv \q^\mathsf{a}+\q^{-\mathsf{a}} \,,
\end{equation}
different from the weight $n = \sqrt{Q} = \mathfrak{q} + \mathfrak{q}^{-1}$ of the usual contractible loops.
In the cylinder geometry (see figure~\ref{fig:cylinder}) both types of loops may or may not wrap the periodic direction, but since the model should be considered
on the Riemann sphere, the only distinction is whether the loops separate the two clusters or not.

The ATL modules contributing to $P_{aabb}^{(\mathsf{a})}$ are the same as those contributing to $P_{aabb}$, except that
the quotient module $\bAStTL{0}{\q^2}$ is replaced by non-quotient module $\AStTL{0}{\q^{2\mathsf{a}}}$.
This is significant for the transfer matrix approach, since in the sector with no through-lines one now
needs to distinguish the contractibility (on the cylinder) of a loop, in
order to determine whether its weight is $n_\mathsf{a}$ or $n$.

\begin{figure}
	\begin{center}
		\begin{tikzpicture}[scale=0.7]
		\draw[black,thick] (0,0) arc (0:360:0.5 and 1.5);
		\draw[black,thick] (-0.5,1.5)--(8,1.5);
		\draw[black,thick] (-0.5,-1.5)--(8,-1.5);
		\draw[black,thick,dotted] (1.5,0.8)--(6.5,0.8);
		\draw (4.0,0.8) node[above] {${\cal S}_2$};
		\draw[black,thick,dotted] (1.5,-0.8)--(6.5,-0.8);
		\draw (4.0,-0.8) node[below] {${\cal S}_3$};
		\draw[black,thick,dotted] (1.5,0.8)--(1.5,-0.8);
		\draw (1.5,0.0) node[left] {${\cal S}_1$};
		\draw[black,fill] (1.42,0.8) circle (2pt);
		\draw[black,fill] (1.42,-0.8) circle (2pt);
		\draw (1.42,0.8) node[left] {$i_1$};
		\draw (1.42,-0.8) node[left] {$i_2$};
		\draw[black,thick] (8.0,-1.5) arc (-90:90:0.5 and 1.5);
		\draw[black,thick,dashed] (8.0,1.5) arc (90:270:0.5 and 1.5);
		\draw[black,fill] (6.42,0.8) circle (2pt);
		\draw[black,fill] (6.42,-0.8) circle (2pt);
		\draw (6.42,0.8) node[right] {$i_3$};
		\draw (6.42,-0.8) node[right] {$i_4$};
		\end{tikzpicture}
	\end{center}
	\caption{Cylinder geometry with three seams, ${\cal S}_1$, ${\cal S}_2$ and ${\cal S}_3$, ensuring the correct weighting of non-contractible loops for the computation of $P_{aabb}^{(\mathsf{a})}$.}
	\label{fig:cylinder-seam}
\end{figure}
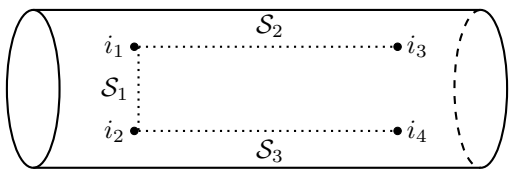

To achieve this, position the four points $i_1$, $i_2$, $i_3$ and $i_4$ as
usual on the cylinder (see figure \ref{fig:cylinder-seam}), and let ${\cal S}_1$, ${\cal S}_2$ and ${\cal S}_3$ denote three different seams running respectively from $i_1$ to $i_2$,
from $i_1$ to $i_3$, and from $i_2$ to $i_4$. There are eight topological types of loops, according to whether they traverse each
of the seams an even or an odd number of times. Let the triplet of signs $((-1)^{N_1},(-1)^{N_2},(-1)^{N_3})$ be associated with the type of loop that traverses seam
${\cal S}_i$ a number $N_i$ of times (for $i=1,2,3$). To compute $P_{aabb}^{(\mathsf{a})}$ we give a weight $n$ to the topologically trivial loops
of type $(+,+,+)$, the modified weight $n_\mathsf{a}$ to separating loops of type $(+,-,-)$, and the weight zero to the remaining six types of loops.%
\footnote{Obviously the same transfer matrix construction, with different choices for the eight weights, can also compute various modifications of
other probabilities $P_{a_1,a_2,a_3,a_4}$.}
Each state acted on by the transfer matrix is endowed with the three binary variables $(-1)^{N_i}$, in addition to the usual connectivity information for the Temperley-Lieb
loop representation \cite{Jacobsen:2018pti}.
Moreover, there is a fourth binary variable
that registers whether at least one loop of type $(+,-,-)$ has been closed; once the lattice has been entirely built up
(i.e., when one reaches the far right boundary of the cylinder) the weight of any
configuration with no loop of type $(+,-,-)$ must be set to zero, since otherwise the $D_{aabb}$ constraint is not fulfilled.

\medskip

In the limit of a very long cylinder---obtained formally by pushing the groups of points $i_1$, $i_2$ and $i_3$, $i_4$ towards respectively the extreme left and the
extreme right of the cylinder---, it is obvious that any loop of type $(+,-,-)$ is almost surely non-contractible on the cylinder, meaning that it wraps the periodic
direction. It is of course a simple matter to write a transfer matrix $T_\mathsf{a}$ in which {\em all} such wrapping loops get a modified weight $n_\mathsf{a}$:
for this it suffices to disregard
the points $i_1$, $i_2$, $i_3$, $i_4$, draw a single seam ${\cal S}$ all along the cylinder, and attribute the weight $n$ (resp.\ $n_\mathsf{a}$) to loops that
traverse ${\cal S}$ an even (resp.\ odd) number of times. What is less obvious, however, is that the probability $P_{aabb}^{(\mathsf{a})}$
can be expressed for a {\em finite} cylinder in terms of the eigenvalues of $T_\mathsf{a}$. This is nevertheless what we observe. In terms of ATL representations,
the spectrum of $T_\mathsf{a}$ is that of the module $\AStTL{0}{\q^{2\mathsf{a}}}$. So indeed, to describe $P_{aabb}^{(\mathsf{a})}$ rather than
$P_{aabb}$, we must replace $\bAStTL{0}{\q^2}$ by $\AStTL{0}{\q^{2\mathsf{a}}}$, as stated initially.

\medskip

The amplitude ratios $\gamma_{j,z^2}^{(\mathsf{a})}$ covered by the facts of type 3 (see section~\ref{sec:facts3}) are more involved than
the remaining ones, since they depend on both $Q$ and $\mathsf{a}$ (through the weight $Q_\mathsf{a}$). Just as we took care to keep
$Q$ generic---by taking $Q = (\q+\q^{-1})^2$ with $\q$ not being a root of unity---we should beware of non-generic values of $\mathsf{a}$.
Indeed, for $\mathsf{a} \ge 2$ integer, the ATL module $\AStTL{\mathsf{a}}{1}$ is a proper submodule of $\AStTL{0}{\q^{2\mathsf{a}}}$,
and when $\mathsf{a} \le L$ this may lead to coincident eigenvalues and hence potentially Jordan cells. However, when $\mathsf{a} \le L$ is odd,
$\AStTL{\mathsf{a}}{1}$ actually does not contribute to $P_{aabb}^{(\mathsf{a})}$ by the general result on the $s$-channel spectra
\cite{Jacobsen:2018pti}, and accordingly we find no Jordan cells.
Instead, the eigenvalues in $\AStTL{\mathsf{a}}{1} \subset \AStTL{0}{\q^{2\mathsf{a}}}$ have in fact zero amplitude.
On the other hand, when $\mathsf{a} \le L$ is even, we observe rank-two Jordan cells
for the eigenvalues in the intersection $\AStTL{\mathsf{a}}{1} \cap \AStTL{0}{\q^{2\mathsf{a}}}$. Finally, when $\mathsf{a} > L$ there are neither Jordan cells,
nor vanishing amplitudes. This latter case thus exhibits the generic behaviour, that should also be observed
when $\mathsf{a}$ is non-integer.

In our numerical work it turns out practical to study first integer values of $\mathsf{a}$. Based on a sufficient number of generic cases, with $\mathsf{a} > L$,
we have established the $\mathsf{a}$-dependence of the ratios $\gamma_{j,z^2}^{(\mathsf{a})}$ given by the facts of type 3, and subsequently
double-checked the expressions for non-integer $\mathsf{a}$. The final expressions \eqref{gammacoeff} have in some cases poles when $j=\mathsf{a}$,
but apart from that they do not exhibit any exceptional behaviour for $\mathsf{a}$ integer.

\subsection{Refinements of the probabilities $P_{abab}$ and $P_{abba}$}

We have also written a transfer matrix in the loop representation that computes a refined version of the probabilities $P_{abab}$ and $P_{abba}$,
denoted $P_{abab}^{(p)}$ and $P_{abba}^{(p)}$, in which the two imposed ``long'' clusters are separated by exactly $2p$ loops (with $p \ge 1$).
All loops, including the separating ones, have weight $n = \sqrt{Q}$.
We concentrate on the symmetric and antisymmetric combinations, as in \eqref{PS_PA}, namely
\begin{subequations}
\begin{eqnarray}
 P_S^{(p)} &=& P_{abab}^{(p)} + P_{abba}^{(p)} \,, \\
 P_A^{(p)} &=& P_{abab}^{(p)} - P_{abba}^{(p)} \,.
\end{eqnarray}
\end{subequations}

The trick for constructing this transfer matrix is to endow each state with an extra integer variable that counts the number of separating loops having been
closed at any stage in the transfer process. This variable takes values $0,1,\ldots,2 \, {\rm min}(2\mathfrak{a},L-2\mathfrak{a})$.

We find that $P_S^{(p)}$ and $P_A^{(p)}$ have non-zero contributions from eigenvalues coming from the sectors with the number of clusters $j'$ being even,
taking values in the range $2p \le j' \le 2 \, {\rm min}(2\mathfrak{a},L-2\mathfrak{a})$, and having the specified symmetry ($S$ or $A$).
These are precisely the sectors corresponding to the ATL
representations ${\mathcal W}_{j',z^2}$ with $z^2 = {\rm e}^{2 i \pi k / j'}$, and the parity of $k$ is even (resp.\ odd) for $P_S^{(p)}$ (resp.\ $P_A^{(p)}$).

Consider now an eigenvalue $\Lambda_i$ belonging to a definite ATL representation ${\mathcal W}_{j,z^2}$---i.e., with $j$ chosen among the possible values $j'$---and which
contributes to $P^{(p)}$ with amplitude $A_i^{(p)}$, and to
$P^{(p+1)}$ with amplitude $A_i^{(p+1)}$. Define the ratio
\begin{equation}
 \rho_{j,z^2}^{(p)} = \frac{A_i^{(p)}}{A_i^{(p+1)}} \,.
\end{equation}
As indicated by the notation, we find that the ratio is the same for all eigenvalues within ${\mathcal W}_{j,z^2 = {\rm e}^{2 i \pi k / j}}$
and hence only depends on the integer labels $j,k$. The translation to the notation used in section \ref{sec:facts2} for the ``facts of type 2''
is
\begin{equation}
\beta^{(2)}_{j,z^2}=1 \,, \qquad \beta^{(4)}_{j,z^2} = \frac{1}{\rho^{(1)}_{j,z^2}} \,, \qquad \beta^{(6)}_{j,z^2}=\frac{1}{\rho^{(1)}_{j,z^2}\rho^{(2)}_{j,z^2}} \,.
\end{equation}

Computations for a sufficient number of rational values of $Q$, as described in section \ref{num-gen-setup} has allowed us to obtain full results
for the $\rho_{j,z^2}^{(p)}$ with $j=4$, and partial results with $j=6$.
For $P_S^{(p)}$ we find:
\begin{equation}
 \rho_{4,1}^{(1)} = -\frac{2+3Q}{Q^2} \,, \qquad
 \rho_{4,-1}^{(1)} = -\frac{3Q-4}{Q(Q-2)} \,.
\end{equation}
and
\begin{equation}
 \rho_{6,1}^{(2)} = -\frac{-2-8 Q+5 Q^2}{Q^2(-2+Q)} \,, \qquad
 \rho_{6,{\rm e}^{\pm 2i\pi/3}}^{(2)} = -\frac{2-52 Q+86 Q^2-38 Q^3+5 Q^4}{Q(Q-1)(Q-3)(Q^2-4 Q+1)} \,.
\end{equation}
Similarly, for $P_A^{(p)}$ we find:
\begin{equation}
 \rho_{4,\pm i}^{(1)} = -\frac{3Q-10}{Q^2-4Q+2}
\end{equation}
and
\begin{equation}
 \rho_{6,{\rm e}^{\pm i\pi/3}}^{(2)} = -\frac{-10-52 Q+86 Q^2-38 Q^3+5 Q^4}{Q (1-16 Q+20 Q^2-8 Q^3+Q^4)} \,, \qquad
 \rho_{6,-1}^{(2)} = -\frac{4-18 Q+5 Q^2}{Q(Q^2-4Q+2)} \,.
\end{equation}

Notice that if we take $2 \mathfrak{a} < \frac{L}{2}$ (with a strict inequality in order to avoid the too symmetric situation in which the marked points
are inserted at opposite positions on the cylinder), we can only observe such $\rho_{j,z^2}^{(p)}$ for which $p \le \frac{j}{2}$ and $j < L$.
For the series $\rho_{4,z^2}^{(1)}$ we have therefore made the computations for $L=5$ and $2 \mathfrak{a} = 2$, and subsequently verified the size-independence
by double checking for $L=7$ and $2 \mathfrak{a} = 3$. In this case, the correlation probabilities pick up contributions where the eigenvalues belong
to ATL modules  ${\mathcal W}_{j',z^2}$ with $j' = 2, 4$. Those with $j=4$, for which we need to compute the amplitudes, are hence burried beneath
those with $j'=2$, so we need to delve rather deep into the transfer matrix spectrum.

For the series $\rho_{6,z^2}^{(2)}$, the computations were made for $L=7$ and $2 \mathfrak{a} = 3$. In that case the participating eigenvalues have
$j'=4,6$, and those with $j=6$ are burried below those with $j'=4$. This is a demanding but still doable computation. However, we have not been able
to obtain the results for $\rho_{6,z^2}^{(1)}$, despite considerable effort (some 30 years of CPU time was wasted on an unsuccessful attempt). Namely,
in this case we have contributions from $j'=2,4,6$ and to ``see past'' the $\simeq 500$ eigenvalues with $j'=2,4$, in order to access the amplitudes of
those with $j=6$ with sufficient numerical precision, turned out to be impossible, given our numerical methods.

\bibliographystyle{ieeetr}
\bibliography{RSOSreferences}

\end{document}